\documentclass[authoryear]{article}

\usepackage{amsmath}

\usepackage{times}
\usepackage{natbib}

\usepackage[plain,noend]{algorithm2e}

\usepackage{threeparttablex}
\usepackage[font=small, justification=centering,labelsep=period, labelfont=sc, singlelinecheck=off]{caption}
\usepackage{colortbl}
\definecolor{lightgray}{gray}{0.95}
\usepackage{booktabs}
\usepackage{enumitem}
\usepackage{placeins}
\usepackage{mathptmx}
\usepackage{amsfonts}
\usepackage{amssymb}
\usepackage{multirow}
\usepackage{supertabular}
\usepackage{longtable}
\usepackage{lscape}
\usepackage{upgreek}

\usepackage{appendix}

\usepackage[multiple]{footmisc}
\usepackage{titling}
\usepackage[runin]{abstract}

\newcommand{\keyword}{\normalfont\textsc{Keywords. }\enspace}
\newcommand{\JEL}{\normalfont\textsc{JEL classification. }\enspace}

\usepackage{titlesec}
\titleformat{\section}{\sc\center}{\normalfont\thesection.}{1em}{}
\titleformat{\subsection}{\it\center}{\normalfont\thesection.\thesubsection}{1em}{}
\titleformat{\subsubsection}{\it\raggedright}{}{0em}{}
\renewcommand{\thesection}{\arabic{section}}
\renewcommand{\thesubsection}{\arabic{subsection}}

\usepackage{setspace}

\usepackage[cmyk]{xcolor}
\usepackage[colorlinks=true,allcolors=blue]{hyperref}
\usepackage[all]{hypcap}
\usepackage{footnotebackref}
\usepackage[all]{hypcap}

\usepackage{authblk}

\begin{document}

%\onehalfspacing

\setlength{\abovedisplayskip}{2.1pt}
\setlength{\belowdisplayskip}{2.1pt}

\title{\large\textbf{Heteroscedastic stratified two-way EC models\\of single equations and SUR systems}}

\author[1]{\normalsize\sc{Silvia Platoni}\thanks{Address correspondence to Silvia Platoni, Dipartimento di Scienze economiche sociali, Universit\`{a} Cattolica del Sacro Cuore, via Emilia Parmense 84, 29122 Piacenza, Italy; silvia.platoni@unicatt.it; tel. +390523599337; fax +390523599303.}}
\author[1]{\normalsize\sc{Laura Barbieri}}
\author[2]{\normalsize\sc{Daniele Moro}}
\author[2]{\normalsize\sc{Paolo Sckokai}}

\affil[1]{\textit{Dipartimento di Scienze economiche e sociali} and}
\affil[2]{\textit{Dipartimento di Economia agro-alimentare,}}
\affil[ ]{\textit{Universit\`{a} Cattolica del Sacro Cuore, Piacenza, Italy}}

%\author{\normalsize\sc Silvia Platoni\thanks{Silvia Platoni: \href{mailto:silvia.platoni@unicatt.it}{silvia.platoni@unicatt.it}}
%    \vspace{-1.5em}\\
%        \small{Dipartimento di Scienze economiche e sociali, Universit\`{a} Cattolica del Sacro Cuore}
%    \vspace{1em}\\
%        \normalsize\sc Laura Barbieri\thanks{Laura Barbieri: \href{mailto:laura.barbieri@unicatt.it}{laura.barbieri@unicatt.it}}
%    \vspace{-0.5em}\\
%        \small{Dipartimento di Scienze economiche e sociali, Universit\`{a} Cattolica del Sacro Cuore}
%    \vspace{1em}\\
%        \normalsize\sc Daniele Moro\thanks{Daniele Moro: \href{mailto:daniele.moro@unicatt.it}{daniele.moro@unicatt.it}}
%    \vspace{-0.5em}\\
%        \small{Dipartimento di Economia agro-alimentare, Universit\`{a} Cattolica del Sacro Cuore}
%    \vspace{1em}\\
%        \normalsize\sc Paolo Sckokai\thanks{Paolo Sckokai: \href{mailto:paolo.sckokai@unicatt.it}{paolo.sckokai@unicatt.it}}
%    \vspace{-0.5em}\\
%        \small{Dipartimento di Economia agro-alimentare, Universit\`{a} Cattolica del Sacro Cuore}}

%\author{\vspace{-2em}}

\date{\vspace{-2em}}

\renewcommand\footnotemark{}
\renewcommand{\footnoterule}{\vspace*{0.3cm}\noindent\rule{2cm}{0.4pt}\vspace*{0.3cm}}

\maketitle

\begin{abstract}
A relevant issue in panel data estimation is heteroscedasticity, which often occurs when the sample is large and individual units are of varying size. Furthermore, many of the available panel data sets are unbalanced in nature, because of attrition or accretion, and micro-econometric models applied to panel data are frequently multi-equation models. This paper considers the general least squares estimation of the heteroscedastic stratified two-way error component (\textit{EC}) models of both single equations and seemingly unrelated regressions (\textit{SUR}) systems (with cross-equations restrictions) on unbalanced panel data. The derived heteroscedastic estimators of both single equations and \textit{SUR} systems improve the estimation efficiency.\vspace{0.5em}\\
\keyword{Unbalanced panel, EC model, SUR, heteroscedasticity.} \vspace{0.5em}\\
\JEL{C13, C23, C33.}
\end{abstract}

%%%%%%%%%%%%%%%%%%%%%%%%%%%%%%%%%%%%%%%%%%%%%%%%%%%%%%%%%%%%%%%%%%%%%%%%%%%%%%%%%%%%%%%%%%%%%%%%%%%%%%%%%%%%%%%%%%%%%%%%%%%%%%%%%%%%

\section{Introduction}\label{INTRODUCTION}

In applied econometrics, there is an increasing use of panel data, that \citet[page 1]{Baltagi(2013)} defines as `the pooling of observations on a cross-section of households, countries, firms, etc. over several time periods'. The reason for this increasing use is that panel data sets are more informative, since they often provide richer and more disaggregated information. Furthermore, they allow to model individual heterogeneity and to address aggregation issues. Finally, since they span over several time periods, they also allow to describe the dynamics of the phenomena under study.

The error component (\textit{EC}) model is the standard approach to the estimation of individual and time effects in econometric single-equation models based on panel data \citep[see][for a review of the methods]{Baltagi(2013)}. Many of the available data sets are unbalanced in nature, that is, not all the individuals are observed over the whole time period. Several and different reasons, such as attrition or accretion, may produce an incomplete panel data set. Therefore, standard single-equation \textit{EC} models have been extended to the econometric treatment of unbalanced panel data: \citet{Biorn(1981)} and \citet{Baltagi(1985)} discussed the single-equation one-way \textit{EC} model, \citet{WansbeekKapteyn(1989)} and \citet{Davis(2002)} extended such estimation method to the two and multi-way cases.

Although often discarded in empirical applications, a relevant issue in panel data estimation is heteroscedasticity, which often occurs when the sample is large and observations differ in ``size characteristic'' (i.e., the level of the variables). Under this perspective, heteroscedasticity arises from the fact that the degree to which a relationship may explain actual observations is likely to depend on individual specific characteristics. On the other hand, the error variance may also systematically vary across observations of similar size and, in practice, the two different sources of heteroscedasticity may be simultaneously present \citep[see][]{Lejeune(1996),Lejeune(2004)}. This means that heteroscedasticity is the rule rather than the exception when dealing with individual data concerning households or firms. Assuming homoscedastic disturbances when heteroscedasticity is present will still result in consistent estimates of the regression coefficients, but these estimates will not be efficient. Also, the standard errors of the fixed-effect (\textit{FE}) estimates will be biased and robust standard errors should be computed in order to correct for the possible presence of heteroscedasticity.

Several authors have analyzed the problem of heteroscedasticity in balanced panel data, usually considering a single-equation regression model with one-way disturbances $\varepsilon_{it}=\mu_{i}+u_{it}$\footnote{While all these papers assume constant slope coefficients, \citet{BressonHsiaoPirotte(2006),BressonHsiaoPirotte(2011)} allow variations in parameters across cross-sectional units in order to take into account the between individual heterogeneity. Hence, these authors derive a hierarchical Bayesian panel data estimator for a random coefficient model (\textit{RCM}), where heteroscedasticity is modeled following both the \textit{RCM}s on panel data proposed by \citet{HsiaoPesaran(2004)} and \citet{Chib(2008)} and the general heteroscedastic one-way \textit{EC} model proposed by \citet{Randolph(1988)}, who assumes that both the individual-specific term $\mu_{i}$ and the remainder error term $u_{it}$ are heteroscedastic.}. \citet{BaltagiGriffin(1988)} are concerned with the estimation of a random-effect (\textit{RE}) model allowing for heteroscedasticity on the individual-specific error term $\text{var}\left(\mu_{i}\right)=\varphi_{i}^2$. In contrast, \citet{RaoKaplanCochran(1981)}, \citet{Magnus(1982)}, \citet{Baltagi(1988)}, and \citet{Wansbeek(1989)} adopt a symmetrically opposite specification allowing for heteroscedasticity on the remainder error term $\text{var}\left(u_{it}\right)=\psi_{i}^2$.

As \citet{MazodierTrognon(1978)} pointed out, if the $\varphi_{i}^2$'s are unknown, then there is no hope to estimate them from the data: even if the $\mu_{i}$'s were observed, it would be impossible to estimate their variances from only one observation on each individual disturbance. Therefore, the model proposed by \citet{BaltagiGriffin(1988)} suffers from the incidental parameters problem\footnote{\citet{NeymanScott(1948)} study maximum likelihood (\textit{ML}) estimation of models having both structural and incidental parameters: while the structural parameters can be consistently estimated, the incidental parameters cannot be consistently estimated. These authors show that the estimation of the \textit{ML} model is inconsistent (or partially inconsistent) if the model contains nuisance or incidental parameters which increase in number with the sample size.} \citep[see][]{Phillips(2003),Baltagi(2013)}. Furthermore, also the models allowing for heteroscedasticity on the remainder error term $u_{it}$ suffer from the incidental parameters problem when the time dimension of the panel is short.

There are two possible solutions to avoid the incidental parameters problem \citep[see][]{Baltagi(2013)}: either to allow the variances to change across strata (i.e., stratified \textit{EC} models) or, if the variables that determine heteroscedasticity are known, to specify parametric variance functions (i.e., adaptive estimation of heteroscedasticity of unknown form).

\citet{MazodierTrognon(1978)} proposed a stratified two-way \textit{EC} model, i.e., $\varepsilon_{it}=\mu_{i}+\nu_{t}+u_{it}$, on balanced panels in which both the individual-specific effect $\mu_{i}$ and time-specific effect $\nu_{t}$ variances are constant within subsets of observations (or strata), but are allowed to change across strata. More recently, \citet{Phillips(2003)} considers a stratified one-way \textit{EC} model, again on
balanced panels, where the variances of the individual-specific effect $\mu_{i}$ are allowed to change not across individuals but across strata, and provides an expectation-maximization (\textit{EM}) algorithm to estimate the model's parameters.

\citet{LiStengos(1994)} derive an adaptive estimator for the heteroscedastic one-way \textit{EC} model using balanced panel data where heteroscedasticity is placed on the remainder error term, and hence, $\text{var}\left(u_{it}|\mathrm{x}_{it}\right)=\psi\left(\mathrm{x}_{it}\right)\equiv\psi_{it}^2$.\footnote{Throughout the paper, all vectors and matrices are in non-italics.} Later, \citet{Roy(2002)} derives a similar adaptive estimator where heteroscedasticity is placed on the individual-specific term rather than the remainder disturbance, and hence, $\text{var}\left(\mu_{i}|\bar{\mathrm{x}}_{i \centerdot}\right)=\varphi\left(\bar{\mathrm{x}}_{i \centerdot}\right)\equiv \varphi_{i}^2$. \citet{BaltagiBressonPirotte(2005)} check the sensitivity of these two adaptive heteroscedastic estimators to misspecification of the form of heteroscedasticity, showing that misleading inference may occur when heteroscedasticity is present in both components. Therefore, accounting for both sources of heteroscedasticity seems to be very important in empirical work.

Indeed, if heteroscedasticity is due to differences in size characteristic across statistical units (i.e., individuals, households, firms or countries), then both error components are expected to be heteroscedastic, and it may be difficult to argue that only one component of the error term is heteroscedastic but not the other \citep[see][]{BressonHsiaoPirotte(2006),BressonHsiaoPirotte(2011)}. To this end, \citet{Randolph(1988)}, working on unbalanced panel data, allows for a more general heteroscedastic single-equation one-way \textit{EC} model, assuming that both the individual-specific and remainder error terms are heteroscedastic, i.e., $\text{var}\left(\mu_{i}\right) = \varphi_i^2$ and $\textit{E}\left(\mathrm{u}\mathrm{u}^{\textrm{T}}\right) = \textrm{diag}(\psi_{it}^2)$. \citet{Lejeune(1996),Lejeune(2004)} is concerned with the estimation and specification testing of a full heteroscedastic one-way \textit{EC} model, in the spirit of \citet{Randolph(1988)} and \citet{BaltagiBressonPirotte(2005)}, and specifies parametrically the variance functions. \citet{BaltagiBressonPirotte(2006)}, in the spirit of \citet{Randolph(1988)} and \citet{Lejeune(1996),Lejeune(2004)}, derive a joint Lagrange multiplier (\textit{LM}) test for homoscedasticity against the alternative of heteroscedasticity both in the individual-specific term $\mu_{i}$ and in the remainder error term $u_{it}$.

Micro-econometric models applied to panel data are often multi-equation models. Primal and dual production models are a common case, when systems of input demands and/or output supply equations have to be estimated; the same is true for systems of demand equations in consumer analysis. \citet{Baltagi(1980)} and \citet{Magnus(1982)} extended the estimation procedure of the single-equation model to the case of seemingly unrelated regressions (\textit{SUR}s) for balanced panels; \citet{Biorn(2004)} proposed a parsimonious technique to estimate one-way \textit{SUR} systems on unbalanced panel data; \citet{PlatoniSckokaiMoro(2012)} extended the procedure suggested by \citet{Biorn(2004)} to the two-way case. Although heteroscedasticity is a frequent and relevant issue also in the multi-equation models applied to (unbalanced) panel data, to our knowledge very few papers concerning heteroscedastic \textit{SUR} systems have been published. A relevant exception is \citet{Verbon(1980)}, who derived a \textit{LM} test for heteroscedasticity in a model of \textit{SUR} equations for balanced panels.

In order to fill this gap in the literature, this paper extends previous results to the estimation of the heteroscedastic stratified two-way \textit{EC} model, i.e., $\varepsilon_{it}=\mu_i+\nu_t+u_{it}$, on unbalanced panel data\footnote{The estimation procedures proposed here can definitely be applied also to balanced panel data.} in case of both single equations and \textit{SUR} systems (with cross-equations restrictions). The individual-specific effect $\mu_{i}$ and remainder error term $u_{it}$ variances and covariances are constant within strata, but they are allowed to change across strata. Indeed, the variance and covariance estimations in two-way \textit{SUR} systems are implemented, starting from the extension of the two-way single-equation \textit{EC} model from the homoscedastic to the heteroscedastic stratified case. Moreover, the estimation is implemented by two methods: the quadratic unbiased estimation (\textit{QUE}) procedure suggested by \citet{WansbeekKapteyn(1989)} and the within-between (\textit{WB}) procedure proposed by \citet{Biorn(2004)}.

The remainder of the paper proceeds as follows. While Section \ref{SINGLE-EQUATION} describes the heteroscedastic two-way estimation for single equations, Section \ref{SUR-SYSTEMS} extends the analysis to the corresponding estimation for \textit{SUR} systems. Finally, Section \ref{CONCLUSIONS} draws some conclusions.

\section{Heteroscedastic single-equation two-way EC model}\label{SINGLE-EQUATION}

We start by considering an unbalanced panel characterized by a total of $n$ observations, with $N$ individuals (indexed $i=1,\ldots,N$) observed over $T$ periods (indexed $t=1,\ldots,T$). Let $T_{i}$ denote the number of times the individual $i$ is observed and $N_{t}$ the number of individuals observed in period $t$. Hence, $\sum_{i} T_{i} = \sum_{t} N_{t} = n$.

In the following we consider the regression model:
\begin{equation}
      \begin{array}{l}
            y_{it} = \mathrm{x}_{it}^{\textrm{T}} {\upbeta} + \mu_{i} + \nu_{t} + u_{it} = \mathrm{x}_{it}^{\textrm{T}} {\upbeta} + \varepsilon_{it} ,
      \end{array}
\end{equation}
where $\mathrm{x}_{it}$ is a $k \times 1$ vector of explanatory variables and ${\upbeta}$ a $k \times 1$ vector of parameters, $\mu_{i}$ is the individual-specific effect, $\nu_{t}$ the time-specific effect, and $u_{it}$ the remainder error term; in the \textit{RE} model $\varepsilon_{it}$ is the composite error term.

Using the $n \times N$ matrix ${\Updelta}_{\mu}$ and the $n \times T$ matrix ${\Updelta}_{\nu}$, that are matrices of indicator variables denoting observations on individuals and time periods respectively, we can define the $N \times N$ diagonal matrix ${\Updelta}_{N} \equiv {\Updelta}_{\mu}^{\textrm{T}} {\Updelta}_{\mu}$ (diagonal elements correspond to the $T_i$'s) and the $T\times T$ diagonal matrix ${\Updelta}_{T} \equiv {{\Updelta}_{\nu}^{\textrm{T}}} {\Updelta}_{\nu}$ (diagonal elements correspond to the $N_{t}$'s), as well as the $T \times N$ matrix of zeros and ones ${\Updelta}_{TN} \equiv {\Updelta}_{\nu}^{\textrm{T}} {\Updelta}_{\mu}$, indicating the absence or presence of an individual in a certain time period. Hence, using matrix notation, we can write:
\begin{equation}
      \begin{array}{l}
            \mathrm{y} = \mathrm{X} {\upbeta} + {\Updelta}_{\mu} {\upmu} + {\Updelta}_{\nu} {\upnu} + \mathrm{u} = \mathrm{X} {\upbeta} + {\upvarepsilon},
      \end{array}
\end{equation}
where $\mathrm{X}$ is a $n \times k$ matrix of explanatory variables.

Let us assume there exists a meaningful stratification of observations\footnote{In empirical work the number of strata is unidentified. Therefore, it is necessary to use a selection procedure, such as the \citet{Akaike(1974)} information criterion, to determine the number of strata.}. Hence, the unbalanced panel can also be characterized by $A$ strata (indexed $a=1,\ldots,A$), with $N_a$ the number of individuals belonging to stratum $a$. %(indexed $\hat{\imath}_a=\hat{1}_a,\ldots,N_a$) and $I_a$ the set of individuals belonging to stratum $a$.
Moreover, the number of observations related to stratum $a$ is $n_a= \sum_{i \in I_a} T_{i}$, with $I_a$ the set of individuals belonging to stratum $a$.% = \sum_{\hat{\imath}_a = \hat{1}_a}^{N_a} T_{\hat{\imath}_a}$.
\footnote{Note that $\sum_{a=1}^{A} N_a = N$ and $\sum_{a=1}^{A} n_a = n$.}

Using the $n \times A$ matrix ${\Updelta}_{\alpha}$ of indicator variables denoting observations on strata, we can define the $A \times A$ diagonal matrix ${\Updelta}_{A} \equiv {\Updelta}_{\alpha}^{\textrm{T}} {\Updelta}_{\alpha}$ (diagonal elements correspond to the $n_a$'s) and the $A \times N$ matrix of zeros and ones ${\Updelta}_{AN} \equiv {\Updelta}_{\alpha}^{\textrm{T}} {\Updelta}_{\mu} {\Updelta}_{N}^{-1}$, indicating the absence or presence of an individual in a certain stratum (notice that ${\Updelta}_{\alpha}^{\textrm{T}} {\Updelta}_{\mu}$ is a matrix of zeros and $T_{i}$'s for $i \in I_a$).

As \citet{MazodierTrognon(1978)} and \citet{Phillips(2003)}, we assume the individual-specific error and remainder error variances are constant within stratum but change across strata. Hence, heteroscedasticity on the individual-specific disturbance implies $\mu_{i} \sim \left( 0 , \varphi_{a}^2 \right)$, while heteroscedasticity on the remainder error term implies $u_{it} \sim \left( 0 , \psi_{a}^2 \right)$.

\subsection{Robust two-way FE} \label{sect:FE}

In the \textit{FE} model the individual-specific term $\mu_i$ and the time-specific term $\nu_t$ are parameters to be estimated. Therefore, heteroscedasticity is placed only on the remainder error $u_{it}$ by assuming $u_{it} \sim \left( 0 , \psi_{a}^2 \right)$. The Within (\textit{W}) estimator\footnote{The number of explanatory variables, obviously without the intercept, is $k-1$.} is:
\begin{equation} \label{Westimator}
      \begin{array}{l}
            \hat{{\upbeta}}^{W} =
            \left( \mathrm{X}^{\textrm{T}} \mathrm{Q}_{\Updelta} \mathrm{X} \right) ^ {-1}
                   \mathrm{X}^{\textrm{T}} \mathrm{Q}_{\Updelta} \mathrm{y},
      \end{array}
\end{equation}
where the $n \times n$ matrix $\mathrm{Q}_{\Updelta}$ on which the two-way \textit{EC} model transformation is based is:
\begin{equation}
      \begin{array}{l}
              \mathrm{Q}_{\Updelta}
            = \mathrm{Q}_{A} - \mathrm{P}_{B}
            = \mathrm{Q}_{A} - \mathrm{Q}_{A}{\Updelta}_{\nu} \mathrm{Q}^{-} {\Updelta}^{\textrm{T}}_{\nu} \mathrm{Q}_{A},
      \end{array}
\end{equation}
with $\mathrm{Q}_{A}=\mathrm{I}_n-\mathrm{P}_{A}$, $\mathrm{P}_{A}={\Updelta}_{\mu} {\Updelta}_{N}^{-1} {\Updelta}_{\mu}^{\textrm{T}}$, $\mathrm{Q} = {\Updelta}^{\textrm{T}}_{\nu} \mathrm{Q}_{A} {\Updelta}_{\nu}$, and $\mathrm{Q}^{-}$ the generalized inverse \citep[see][]{WansbeekKapteyn(1989),Davis(2002)}.\footnote{For a \textit{FE} model the number of fixed-effect parameters $\mu_{1},\dots,\mu_{N}$ and $\nu_{1},\dots,\nu_{T}$ increases with the number of individuals $N$ and periods $T$, respectively. Hence, the conventional asymptotic result cannot be applied: if $N \rightarrow \infty$, then estimates of the parameters $\mu_{1},\dots,\mu_{N}$ are necessarily inconsistent for a fixed $T$ \citep[see][]{WangHo(2010)}, and if $T \rightarrow \infty$, then estimates of the parameters $\nu_{1},\dots,\nu_{T}$ are necessarily inconsistent for a fixed $N$. Therefore, when the time dimension of the panel is short, the noise in the estimation of the incidental parameters $\mu_{i}$ contaminates the \textit{ML} estimates of the structural parameters \citep[see][]{BesterHansen(2016)}. The literature proposes some solutions to the incidental parameters problem for some of the models, usually relying on removing the incidental parameters before estimations \citep[see][]{WangHo(2010)}. One popular approach, widely used in linear models, is to transform the model by the \textit{W} transformation (i.e., $y_{it}$ and the $\left(k-1\right) \times 1 $ vector $\mathrm{x}_{it}$ are demeaned), as we have done in deriving our estimation.}

Under the assumptions of strict exogeneity, consistency, homoscedasticity and no serial correlation (see assumptions FE.1-FE.3 in Appendix A of \citealp{PlatoniSckokaiMoro(2012)}), the \textit{W} estimator is consistent and asymptotically normal \citep[see][]{Wooldridge(2010)} with
\begin{equation}
\label{WvarOMO}
      \begin{array}{l}
            \text{var}\left(\hat{{\upbeta}}^{W}\right) =
            \hat{\sigma}_u^2 \left( \mathrm{X}^{\textrm{T}} \mathrm{Q}_{\Updelta} \mathrm{X} \right)^{-1},
      \end{array}
\end{equation}
where $ \hat{\sigma}_u^2$ is the estimator of $ \sigma_u^2$. However, relaxing the homoscedasticity assumption (see assumption FE.3 in Appendix \ref{appFE}), the expression \eqref{WvarOMO} gives an improper variance-covariance matrix estimator \citep[see][]{Wooldridge(2010)}.

To obtain robust standard errors we follow the simple method suggested by \citet{Arellano(1987)} for the one-way \textit{EC} model, and proposed also by \citet{Baltagi(2013)}. If we stack the observations for each individual $i$, we can write:
\begin{equation}
      \begin{array}{rl}
            \widetilde{\mathrm{y}}_{i} & \hspace{-0.1em} =
            \left( \mathrm{E}_{T_i} - \mathrm{E}_{T_i} \mathrm{D}_{i} \mathrm{Q}^{-} \mathrm{D}^{\textrm{T}}_{i} \mathrm{E}_{T_i} \right) \mathrm{y}_{i}, \\
            \widetilde{\mathrm{X}}_{i} & \hspace{-0.1em} =
            \left( \mathrm{E}_{T_i} - \mathrm{E}_{T_i} \mathrm{D}_{i} \mathrm{Q}^{-} \mathrm{D}^{\textrm{T}}_{i} \mathrm{E}_{T_i} \right) \mathrm{X}_{i} ,
      \end{array}
\end{equation}
where $\mathrm{E}_{T_i} = \mathrm{I}_{T_i} - \bar{\mathrm{J}}_{T_i}$, with $\mathrm{I}_{T_i}$ an identity matrix of dimension $T_i$, $\bar{\mathrm{J}}_{T_i} = \frac{\mathrm{J}_{T_{i}}}{T_i}$, and $\mathrm{J}_{T_{i}}$ a matrix of ones of dimension $T_i$, and $\mathrm{D}_{i}$ is the $T_i \times T$ matrix obtained from the $T \times T$ identity matrix $\mathrm{I}_{T}$ by omitting the rows corresponding to periods in which the individual $i$ is not observed. Therefore, we can compute the $T_i \times 1$ vector $\widetilde{\mathrm{e}}_{i} = \widetilde{\mathrm{y}}_{i} - \widetilde{\mathrm{X}}_{i} \hat{{\upbeta}}^{W}$ and the robust asymptotic variance-covariance matrix of $\hat{{\upbeta}}^{W}$ is:
\begin{equation}
      \begin{array}{l}
            \text{var}\left(\hat{{\upbeta}}^{W}\right) =
            \left( \mathrm{X}^{\textrm{T}} \mathrm{Q}_{\Updelta} \mathrm{X} \right)^{-1}
            \underset{i=1}{\overset{N}{\textstyle\sum}} \widetilde{\mathrm{X}}^{\textrm{T}}_{i} \widetilde{\mathrm{e}}_{i} \widetilde{\mathrm{e}}^{\textrm{T}}_{i} \widetilde{\mathrm{X}}_{i}
            \left( \mathrm{X}^{\textrm{T}} \mathrm{Q}_{\Updelta} \mathrm{X} \right)^{-1} .
      \end{array}
\end{equation}

However, since $u_{it} \sim \left( 0 , \psi^2_{a} \right)$, it is possible to obtain robust standard errors also by stacking the observations for each stratum $a$, as described later in Appendix \ref{appFEROB}.

\subsection{GLS estimation}

In the \textit{RE} model, not only the remainder error $u_{it}$, but also the individual-specific error $\mu_i$ and the time-specific error $\nu_t$ are random variables.

If we assume that the variances of $\mu_i$, $\nu_t$, and $u_{it}$ are known, then the general least squares (\textit{GLS}) estimator\footnote{Note that the number of explanatory variables, obviously including the intercept, is $k$.} for ${\upbeta}$, obtained by minimizing ${\upvarepsilon}_{it}^{\textrm{T}} {\Upomega}^{-1} {\upvarepsilon}_{it}$ where ${\Upomega}$ is the $n \times n$ variance-covariance matrix, is given by:
\begin{equation} \label{GLSestimator}
      \begin{array}{l}
            \hat{{\upbeta}}^{GLS} = \left( \mathrm{X}^{\textrm{T}} {\Upomega}^{-1} \mathrm{X} \right) ^ {-1} \mathrm{X}^{\textrm{T}} {\Upomega}^{-1} \mathrm{y}.
      \end{array}
\end{equation}

%The following assumptions are made\footnote{Details on the assumptions RE.1 and RE.2 can be found in Appendix B of \citet{PlatoniSckokaiMoro(2012)}.}.
%\begin{description}[font=\normalfont\sffamily]
%      \item [RE.1.a] {\sffamily Strict exogeneity} Same definition as assumption FE.1.
%      \item [RE.1.b and RE.1.c] {\sffamily Orthogonality conditions} Both $\mu_{i}$ and $\nu_{t}$ are orthogonal to the corresponding sets of explanatory variables, that is the $k\cdot T_{i}$ explanatory variables for
%each individual $\mathrm{x}_{i \circ }$ and the $k\cdot N_{t}$ explanatory variables in each time period $\mathrm{x}_{\circ t}$:
%          \begin{equation*}
%                \begin{array}{l}
%                      \textit{E} \left( \mu_{i} \left\vert \mathrm{x}_{i \circ } \right. \right) = \textit{E} \left( \mu_{i} \right) = 0 \text{ and }
%                      \textit{E} \left( \nu_{t} \left\vert \mathrm{x}_{\circ t} \right. \right) = \textit{E} \left( \nu_{t} \right) = 0.
%                \end{array}
%          \end{equation*}
%      \item [RE.2] {\sffamily Rank condition} The $k \times k$ weighted outer product matrix $ \mathrm{X}^{\prime} {\Upomega}^{-1} \mathrm{X}$ has the appropriate rank, ensuring the \textit{GLS} estimator in
%(\ref{GLSestimator}) is consistent:
%          \begin{equation*}
%                \begin{array}{l}
%                      \text{rank} \left( \mathrm{X}^{\textrm{T}} {\Upomega}^{-1} \mathrm{X} \right) = k.
%                \end{array}
%          \end{equation*}
%\end{description}

Assuming homoscedasticity and no serial correlation \citep[i.e., the assumption RE.3 in Appendix B of][]{PlatoniSckokaiMoro(2012)}, the variance-covariance matrix ${\Upomega}$ has the following form:
\begin{equation}\label{omegaHomo}
      \begin{array}{l}
            {\Upomega} = \sigma_{u}^{2} \mathrm {I}_n + \sigma_{\mu}^{2} {\Updelta}_{\mu} {\Updelta}_{\mu}^{\textrm{T}} + \sigma_{\nu}^{2} {\Updelta}_{\nu}
            {\Updelta}^{\textrm{T}}_{\nu},
      \end{array}
\end{equation}
and the \textit{GLS} estimator in (\ref{GLSestimator}) is efficient. However, assuming homoscedastic $\mu_{i}$ and/or $u_{it}$ when heteroscedasticity is present will still result in consistent estimates of the regression coefficients, but these estimates will not be efficient.

With general heteroscedasticity (see assumption RE.3 in Appendix \ref{appRE}), that is $\mu_{i} \sim \left( 0 , \varphi_{a}^2 \right)$ and $u_{it} \sim \left( 0 , \psi_{a}^2 \right)$, the matrix ${\Upomega}$ in (\ref{omegaHomo}) is modified to:
\begin{equation}\label{omegaHetero}
      \begin{array}{l}
            {\Upomega} =  {\Uppsi} + {\Updelta}_{\mu} {\Upphi} {\Updelta}_{\mu}^{\textrm{T}} + \sigma_{\nu}^{2} {\Updelta}_{\nu}
            {\Updelta}^{\textrm{T}}_{\nu},
      \end{array}
\end{equation}
with the $n \times n$ matrix\footnote{This matrix and the related vector ${\uppsi}$ have been already defined in Appendix \ref{appFE}.} ${\Uppsi} = \textrm{diag} \left({\Updelta}_\mu {\Updelta}_{AN}^{\textrm{T}} {\uppsi} \right)$, and the ${A\times 1}$ vector ${\uppsi} = (\psi_{1}^2,\psi_{2}^2,\dots, \psi_{A}^2)^{\textrm{T}}$, the $N \times N$ matrix ${\Upphi} = \textrm{diag} \left( {\Updelta}_{AN}^{\textrm{T}} {\upvarphi} \right)$, and the ${A\times 1}$ vector ${\upvarphi} = (\varphi_{1}^2,\varphi_{2}^2,\dots, \varphi_{A}^2)^{\textrm{T}} $.

The ANOVA-type quadratic unbiased estimator of the variance components based on the \textit{W} residuals in the homoscedastic case (\ref{omegaHomo}) is determined in \citet{WansbeekKapteyn(1989)} and \citet{Davis(2002)}. The estimation of the components of the variance-covariance matrix ${\Upomega}$ in the heteroscedastic case (\ref{omegaHetero}) can be obtained modifying the \textit{QUE} procedure suggested by \citet{WansbeekKapteyn(1989)}.

This latter procedure considers the $n \times 1$ residuals $\mathrm{e} \equiv \mathrm{y} - \mathrm{X} \hat{{\upbeta}}^{W}$ from the \textit{W} estimator in \eqref{Westimator}, where $\mathrm{X}$ is a matrix of dimension $n \times \left(k-1\right)$, since it does not include the intercept. Given that the $n \times k$ matrix $\mathrm{X}$ in (\ref{GLSestimator}) contains a vector of ones, we have to define the $n \times 1$ consistent centered residuals $\mathrm{f} \equiv \mathrm{E}_{n} \mathrm{e} = \mathrm{e} - \bar{e}$, where $\mathrm{E}_{n} = \mathrm{I}_{n} - \bar{\mathrm{J}}_{n}$, with $\mathrm{I}_{n}$ being an identity matrix of dimension $n$, $\bar{\mathrm{J}}_{n} = \frac{\mathrm{J}_{n}}{n}$, and $\mathrm{J}_{n}$ a matrix of ones of dimension $n$. Moreover, we have to define also the $n_a \times 1$ consistent centered residuals $\mathrm{f}_{a} = \mathrm{H}_{a} \mathrm{f}$, with $\mathrm{H}_{a}$ the $n_a \times n $ matrix obtained from the identity matrix $\mathrm{I}_{n}$ by omitting the rows referring to observations not related to stratum $a$, and the matrix $\bar{\mathrm{J}}_{n_a} = \frac{\mathrm{J}_{n_a}}{n_a}$, with $\mathrm{J}_{n_a}$ a matrix of ones of dimension $n_a$.

The adapted \textit{QUE}s for ${\Uppsi}$, ${\Upphi}$, and $\sigma_{\nu}^{2}$ is obtained by equating:
\begin{equation}\label{q_QUE}
      \begin{array}{lll}
            q_{n_a} & \hspace{-0.5em} \equiv \mathrm{f}^{\textrm{T}} \mathrm{Q}_{\Updelta} \mathrm{H}_{a}^{\textrm{T}} \mathrm{H}_{a} \mathrm{Q}_{\Updelta} \mathrm{f} & \rightarrow \underset{a=1}
            {\overset{A} {\textstyle\sum}} q_{n_a} = q_{n} \equiv \mathrm{f}^{\textrm{T}} \mathrm{Q}_{\Updelta} \mathrm{f} , \\
            q_{N_a} & \hspace{-0.5em} \equiv  \mathrm{f}_{a}^{\textrm{T}} \bar{\mathrm{J}}_{n_a} \mathrm{f}_{a} & \rightarrow \underset{a=1} {\overset{A} {\textstyle\sum}} q_{N_a} = q_{N} \equiv \mathrm{f}^{\textrm{T}}
            {\Updelta}_{\mu} {\Updelta}_{N}^{-1} {\Updelta}_{\mu}^{\textrm{T}} \mathrm{f} , \\
            q_{T} & \hspace{-0.5em} \equiv \mathrm{f}^{\textrm{T}} {\Updelta}_{\nu} {\Updelta}_{T}^{-1} {\Updelta}_{\nu}^{\textrm{T}} \mathrm{f} , &
      \end{array}
\end{equation}
to their expected values. For more details on the identities in (\ref{q_QUE}), see the formula \eqref{q_QUE_details} in Appendix \ref{appTEC_QUE}.

Hence, the estimator of $\psi_{a}^2$ is:
\begin{equation}
      \begin{array}{l}
            \hat{\psi}_{a}^2 = \dfrac{q_{n_a} + k_{a} \hat{\sigma}_{u}^{2}}{n_a - N_a - \tau_a}.
      \end{array}
\end{equation}
where $k_{a} \equiv \text{tr} [ ( \mathrm{X}^{\textrm{T}}\mathrm{Q}_{\Delta}\mathrm{X})^{-1} \mathrm{X}^{\textrm{T}} \mathrm{Q}_{\Delta} \mathrm{H}_{a}^{\textrm{T}} \mathrm{H}_{a} \mathrm{Q}_{\Delta} \mathrm{X} ]$, with $\sum_{a=1}^{A} k_{a}= k-1$, $\tau_a \equiv n_a-N_a-\text{tr} (\mathrm{H}_{a} \mathrm{Q}_{\Updelta} \mathrm{H}_{a}^{\textrm{T}})$, with $\sum_{a=1}^{A} \tau_a = T-1$. The estimated variance $\hat{\sigma}_{u}^{2}$ is obtained by equating $q_{n}$ to its expected value \citep[see][]{WansbeekKapteyn(1989)}.
Furthermore, the estimator of $\varphi_{a}^2$ is:
\begin{equation}
      \begin{split}
            \hat{\varphi}_{a}^2 = &
                \dfrac{q_{N_a} - \left( N_a - 2 \frac{n_a}{n} \right) \hat{\psi}_{a}^2 - \left( k_{N_{a}} - k_{0_{a}} + \frac{n_a}{n} k_{0} + \frac{n_a}{n} \right) \hat{\sigma}_{u}^{2}}{n_a - 2
                \lambda_{\mu_a}} \\
            & + \dfrac{ - \frac{n_a}{n} \lambda_{\mu} \hat{\sigma}_{\mu}^{2} - \left( N_a - 2 \lambda_{\nu_a} + \frac{n_a}{n} \lambda_{\nu} \right) \hat{\sigma}_{\nu}^{2} } {n_a - 2 \lambda_{\mu_a}} .
      \end{split}
\end{equation}
where $k_{N_{a}} \equiv \text{tr} [ ( \mathrm{X}^{\textrm{T}}\mathrm{Q}_{\Updelta}\mathrm{X})^{-1} \mathrm{X}_{a}^{\textrm{T}} \bar{\mathrm{J}}_{n_a} \mathrm{X}_{a} ]$, $k_{0} \equiv \frac{ {\upiota}_{n}^{\textrm{T}} \mathrm{X} ( \mathrm{X}^{\textrm{T}} \mathrm{Q}_{\Updelta} \mathrm{X} ) ^{-1} \mathrm{X}^{\textrm{T}} {\upiota}_{n} }{n}$, $k_{0_{a}} \equiv 2 \frac{ {\upiota}_{n}^{\textrm{T}} \mathrm{X} (\mathrm{X}^{\textrm{T}}\mathrm{Q}_{\Updelta}\mathrm{X})^{-1} \mathrm{X}_{a}^{\textrm{T}}{\upiota}_{n_a} }{n} \linebreak = 2 \frac{ {\upiota}_{n_a}^{\textrm{T}} \mathrm{X}_{a} (\mathrm{X}^{\textrm{T}}\mathrm{Q}_{\Updelta}\mathrm{X})^{-1} \mathrm{X}^{\textrm{T}}{\upiota}_{n} }{n}$, with ${\upiota}_{n}$ and ${\upiota}_{n_a}$ vectors of ones of dimension $n$ and $n_a$ respectively, $\lambda _{\mu} \equiv \frac{ {\upiota}_{n}^{\textrm{T}} {\Updelta}_{\mu} {\Updelta}_{\mu}^{\textrm{T}} {\upiota}_{n} } {n} = \frac{\sum_{i=1}^{N} T_{i}^{2}} {n}$, $\lambda _{\mu_{a}} \varphi_{a}^2 \equiv \frac{ {\upiota}_{n}^{\textrm{T}} {\Updelta}_{\mu} {\Upphi} {\Updelta}_{\mu}^{\textrm{T}} \mathrm{H}_{a}^{\textrm{T}} {\upiota}_{n_a} } {n} = \frac{\sum_{i \in I_a} T_{i}^{2}} {n} \varphi_{a}^2$, $\lambda _{\nu} \equiv \frac{{\upiota}_{n}^{\textrm{T}} {\Updelta}_{\nu} {\Updelta}_{\nu}^{\textrm{T}} {\upiota}_{n} } {n} = \frac{\sum_{t=1}^{T} N_{t}^{2} } {n}$, $\lambda _{\nu_{a}} \equiv \frac{{\upiota}_{n}^{\textrm{T}} {\Updelta}_{\nu} {\Updelta}_{\nu}^{\textrm{T}} \mathrm{H}_{a}^{\textrm{T}} {\upiota}_{n_a} } {n} = \frac{\sum_{t \in J_a} N_{t}} {n}$, with $J_a$ the set of periods in which individuals belonging to stratum $a$ are observed. The estimated variances $\hat{\sigma}_{\mu}^{2}$ and $\hat{\sigma}_{\nu}^{2}$ are obtained jointly by equating $q_{N}$ and $q_{T}$ to their expected values \citep[see][]{WansbeekKapteyn(1989)}.

Simpler heteroscedastic schemes (i.e., heteroscedasticity only on the individual-specific disturbance or on the remainder error) can be obtained combining results for the general scheme with those for the homoscedastic case, although when we consider the case of heteroscedasticity only on the individual-specific disturbance the expected value of $q_{N_a}$ and the estimated variance $\hat{\varphi}_{a}^2$ are obtained differently as detailed in equations \eqref{eq.SEQUE_mu_1}-\eqref{eq.SEQUE_mu_2} in Appendix \ref{appTEC_QUE}.
%, although when we consider the case of heteroscedasticity only on the individual-specific disturbance, the expected value of $q_{N_a}$ is obtained differently as:
%\begin{equation}
%      \begin{split}
%            \textit{E} \left( q_{N_A} \right) = &
%            \left( N_a + k_{N_{a}} - k_{0_{a}} + \frac{n_a}{n} \cdot k_{0} - \frac{n_a}{n} \right) \cdot \sigma_{u}^{2} + \left( n_a - 2 \cdot \lambda_{\mu_a} \right) \cdot \varphi_{a} \\ & + \frac{n_a}{n} \cdot \lambda_{\mu} \cdot \bar{\varphi} + \left( N_a - 2 \cdot \lambda_{\nu_a} + \frac{n_a}{n} \cdot \lambda_{\nu} \right) \cdot \sigma_{\nu}^{2} ,
%      \end{split}
%\end{equation}
%and, therefore,
%\begin{equation}
%      \begin{split}
%            \hat{\varphi}_{a} =
%            &   \dfrac{q_{N_a} - \left( N_a + k_{N_{a}} - k_{0_{a}} + \frac{n_a}{n} \cdot k_{0} - \frac{n_a}{n} \right) \cdot \hat{\sigma}_{u}^{2}}{n_a - 2 \cdot \lambda_{\mu_a}} \\
%            & + \dfrac{ - \frac{n_a}{n} \cdot \lambda_{\mu} \cdot \hat{\sigma}_{\mu}^{2} - \left( N_a - 2 \cdot \lambda_{\nu_a} + \frac{n_a}{n} \cdot \lambda_{\nu} \right) \cdot \hat{\sigma}_{\nu}^{2}  }{n_a - 2 \cdot \lambda_{\mu_a}} .
%      \end{split}
%\end{equation}

\subsection{Monte Carlo experiment -- single-equation case}\label{SIMULATION_1}

In order to analyze the performances of the proposed techniques, we develop a simple simulation\footnote{The simulations have been implemented with the econometric software \textit{TSP} version 5.1.} on
\begin{equation*}
            y =\beta_{0}+\beta_1 x_{1} +\beta _2 x_{2} + \beta_3 x_{3} +\varepsilon,
\end{equation*}
where $\beta_0=10$, $\beta_1=-3$, $\beta_2=8$, and $\beta_3=-2$.

We assume unbalanced panels with a large number of individuals ($N=250$ and $N=500$) extended over a rather long time period ($T=12$). This should mimic a real world situation of a large unbalanced panel for which the two-way \textit{EC} model is the appropriate one.

Moreover, the experiment is implemented by considering as strata the deciles of the independent variable $x_2$. The homoscedastic time variance is $\sigma^2_{\nu}=6.271$, while the heteroscedastic variances have been generated with $\varphi^2_a=\sigma^2_{\mu} (1 + \lambda \bar{x}_{2_{a}})^2$, where $\sigma^2_{\mu}=6.488$, and $\psi^2_a=\sigma^2_u (1 + \lambda \bar{x}_{2_{a}})^2$, where $\sigma^2_u=6.039$; $\lambda$ is assigned values $0,1,\text{ and }2$, where $\lambda=0$ denotes the homoscedastic case and the degree of heteroscedasticity increases as the value of $\lambda$ becomes larger.\footnote{Whereas data have been generated by specifying the same parametric variance functions as in \citet{LiStengos(1994)} and \citet{Roy(2002)}, the proposed estimation method proves to be effective also in the the case of heteroscedasticity of unknown form, if the strata are identified by using a proper selection procedure, such as the \citet{Akaike(1974)} information criterion.}

Finally, the independent variables' values $x_{kit}$ ($k=1,2,3$) are generated according to a modified version of the scheme introduced by \citet{Nerlove(1971)} and used, among others, by \citet{Baltagi(1981)}, \citet{WansbeekKapteyn(1989)}, and \citet{PlatoniSckokaiMoro(2012)}:
\begin{equation*}
      x_{kit}=0.1 t + 0.5 x_{kit-1}+\omega _{kit}, \quad k=1,2,3
\end{equation*}
with $\omega _{kit}$ following the uniform distribution $[-\frac{1}{2},\frac{1}{2}]$ and $x_{ki0}=5 + 10 \omega _{ki0}$.

In order to construct the unbalanced panels, we adopt the procedure currently used for rotating panels, in which we have approximately the same number of individuals every time period: a fixed percentage of individuals ($20\%$ in our case\footnote{Also in \citet{WansbeekKapteyn(1989)} each period $20\%$ of the households in the panel is removed randomly.}) is replaced each time period, but they can re-enter the sample in later periods. Thus, if the number of individuals is $N=250$ then the number of observations is $n=1031$, if the number of individuals is $N=500$ then he number of observations is $n=2062$.

The results of a $2000$-run simulation\footnote{With $N=250$ the average numbers of observations for each stratum $a$ are $\bar{n}_{a=1}=78$, $\bar{n}_{a=2}=113$, $\bar{n}_{a=3}=134$, $\bar{n}_{a=4}=144$, $\bar{n}_{a=5}=145$, $\bar{n}_{a=6}=141$, $\bar{n}_{a=7}=116$, $\bar{n}_{a=8}=77$, $\bar{n}_{a=9}=51$, and $\bar{n}_{a=10}=32$; and with $N=500$ they are $\bar{n}_{a=1}=155$, $\bar{n}_{a=2}=226$, $\bar{n}_{a=3}=269$, $\bar{n}_{a=4}=287$, $\bar{n}_{a=5}=290$, $\bar{n}_{a=6}=282$, $\bar{n}_{a=7}=232$, $\bar{n}_{a=8}=153$, $\bar{n}_{a=9}=104$, and $\bar{n}_{a=10}=64$.}  are shown in Table \ref{Tab_SE_Var} and Table \ref{Tab_SE_SE}\footnote{As in \citet{BaltagiGriffin(1988)} and \citet{Phillips(2003)}, negative variance estimates are replaced by zero.}\footnote{Whereas data have been generated such that the individual-specific error $\mu_i$ and the time-specific error $\nu_t$ are random variables, Table \ref{Tab_SE_SE} displays also the results of the two-way \textit{FE} and robust two-way \textit{FE} estimations to check the method suggested in subsection \ref{sect:FE}. Moreover, note that the two-way \textit{FE} residuals are used in the \textit{QUE} procedure of the \textit{GLS} estimation (and both in the \textit{QUE} and \textit{WB}  procedures of the \textit{SUR} systems estimation in the following section \ref{SUR-SYSTEMS}).}.

Table \ref{Tab_SE_Var} reports the estimated variances $\hat{\psi}_{a}^2$ and $\hat{\varphi}_{a}^2$, being the latter computed on the basis of a remainder error either homoscedastic $(\hat{\sigma}_{u}^{2})$ or heteroscedastic $(\hat{\psi}_{a}^2$). As one can notice right away, if $\lambda$ is equal to $1$ or $2$ (i.e. in the heteroscedastic cases) the estimated variance $\hat{\varphi}_{a}^2\left(\hat{\psi}_{a}^2\right)$ is closer than the estimated variance $\hat{\varphi}_{a}^2\left(\hat{\sigma}_{u}^{2}\right)$ to the true value $\varphi_{a}^2$. Moreover, when $\lambda$ is equal to $0$ (homoscedastic case), the heteroscedastic procedures allow to obtain estimated variances $\hat{\psi}_{a}^2$ and $\hat{\varphi}_{a}^2$ that do not substantially vary among strata, and that are very close to the estimated values $\hat{\sigma}_{u}^{2}$ and $\hat{\sigma}_{\mu}^{2}$ obtained through the homoscedastic procedure (also reported in Table \ref{Tab_SE_Var}).

\singlespacing

\begin{table}[!ht]
\renewcommand{\tabcolsep}{0.26pc}\renewcommand{\arraystretch}{1}
\caption{Simulation results on single-equation two-way \textit{EC} model:\\estimated variances $\hat{\psi}_{a}^2$ and $\hat{\varphi}_{a}^2$\label{Tab_SE_Var}}
\vspace{-0.5em}
{\footnotesize
\begin{tabular}{r rr rrr r rr rrr}
\toprule
   &  \multicolumn{5}{c}{$N=250$, $T=12$, and $n=1031$} & & \multicolumn{5}{c}{$N=500$, $T=12$, and $n=2062$} \\
 \\
%   & \centering{  (a)} &  (b) &   (c) &  (d) &  (e) &  (f) &   (g) &   (h) &  (i) & (j) \\
 a & \multicolumn{1}{c}{$\psi_{a}^2$} & \multicolumn{1}{c}{$\hat{\psi}_{a}^2$}  & \multicolumn{1}{c }{$\varphi_{a}^2$}
   & \multicolumn{1}{c}{$\hat{\varphi}_{a}^2\left(\hat{\sigma}_{u}^{2}\right)$} & \multicolumn{1}{c }{$\hat{\varphi}_{a}^2\left(\hat{\psi}_{a}^2\right)$}
   &
   & \multicolumn{1}{c}{$\psi_{a}^2$} & \multicolumn{1}{c}{$\hat{\psi}_{a}^2$}  & \multicolumn{1}{c }{$\varphi_{a}^2$}
   & \multicolumn{1}{c}{$\hat{\varphi}_{a}^2\left(\hat{\sigma}_{u}^{2}\right)$} & \multicolumn{1}{c }{$\hat{\varphi}_{a}^2\left(\hat{\psi}_{a}^2\right)$} \\
\\
   & \multicolumn{11}{c}{\cellcolor{lightgray}$\lambda=0$} \\
 1 &   6.039 &   6.045 &   6.488 &   6.589 &   6.589  & &  6.039 &   6.019 &   6.488 &   6.560 &   6.569 \\
 2 &   6.039 &   6.043 &   6.488 &   6.509 &   6.510  & &  6.039 &   6.046 &   6.488 &   6.538 &   6.537 \\
 3 &   6.039 &   6.059 &   6.488 &   6.513 &   6.511  & &  6.039 &   6.051 &   6.488 &   6.522 &   6.521 \\
 4 &   6.039 &   6.012 &   6.488 &   6.536 &   6.542  & &  6.039 &   6.060 &   6.488 &   6.499 &   6.496 \\
 5 &   6.039 &   6.038 &   6.488 &   6.534 &   6.535  & &  6.039 &   6.032 &   6.488 &   6.525 &   6.527 \\
 6 &   6.039 &   6.050 &   6.488 &   6.478 &   6.476  & &  6.039 &   6.040 &   6.488 &   6.488 &   6.489 \\
 7 &   6.039 &   6.073 &   6.488 &   6.451 &   6.444  & &  6.039 &   6.044 &   6.488 &   6.528 &   6.528 \\
 8 &   6.039 &   6.061 &   6.488 &   6.532 &   6.527  & &  6.039 &   6.051 &   6.488 &   6.529 &   6.526 \\
 9 &   6.039 &   6.046 &   6.488 &   6.530 &   6.536  & &  6.039 &   6.042 &   6.488 &   6.527 &   6.528 \\
10 &   6.039 &   5.962 &   6.488 &   6.617 &   6.881  & &  6.039 &   5.969 &   6.488 &   6.561 &   6.652 \\
   & \multicolumn{1}{c }{\textcolor[rgb]{0.50,0.50,0.50}{$\sigma_{u}^{2}$}}
   & \multicolumn{1}{c }{\textcolor[rgb]{0.50,0.50,0.50}{$\hat{\sigma}_{u}^{2}$}}
   & \multicolumn{1}{c }{\textcolor[rgb]{0.50,0.50,0.50}{$\sigma_{\mu}^{2}$}}
   & \multicolumn{2}{c }{\textcolor[rgb]{0.50,0.50,0.50}{$\hat{\sigma}_{\mu}^{2}$}}
   &
   & \multicolumn{1}{c }{\textcolor[rgb]{0.50,0.50,0.50}{$\sigma_{u}^{2}$}}
   & \multicolumn{1}{c }{\textcolor[rgb]{0.50,0.50,0.50}{$\hat{\sigma}_{u}^{2}$}}
   & \multicolumn{1}{c }{\textcolor[rgb]{0.50,0.50,0.50}{$\sigma_{\mu}^{2}$}}
   & \multicolumn{2}{c }{\textcolor[rgb]{0.50,0.50,0.50}{$\hat{\sigma}_{\mu}^{2}$}}  \\
   & \textcolor[rgb]{0.50,0.50,0.50}{6.039} & \textcolor[rgb]{0.50,0.50,0.50}{6.044} & \textcolor[rgb]{0.50,0.50,0.50}{6.488} & \multicolumn{2}{c }{\textcolor[rgb]{0.50,0.50,0.50}{6.515}}
   &
   & \textcolor[rgb]{0.50,0.50,0.50}{6.039} & \textcolor[rgb]{0.50,0.50,0.50}{6.043} & \textcolor[rgb]{0.50,0.50,0.50}{6.488} & \multicolumn{2}{c }{\textcolor[rgb]{0.50,0.50,0.50}{6.521}} \\
   & \multicolumn{11}{c}{\cellcolor{lightgray}$\lambda=1$} \\
 1 &  12.352 &  12.947 &  13.270 &   5.400 &  13.220 & &  12.296 &  12.554 &  13.211 &   4.796 &  13.213 \\
 2 &  19.770 &  20.189 &  21.239 &  16.807 &  21.114 & &  19.760 &  19.982 &  21.229 &  16.930 &  21.242 \\
 3 &  25.800 &  26.164 &  27.718 &  25.154 &  27.650 & &  25.775 &  25.964 &  27.692 &  25.277 &  27.787 \\
 4 &  31.743 &  31.750 &  34.103 &  32.944 &  34.274 & &  31.741 &  31.922 &  34.101 &  32.748 &  34.049 \\
 5 &  38.086 &  38.108 &  40.918 &  40.964 &  41.173 & &  38.081 &  38.048 &  40.912 &  40.855 &  41.077 \\
 6 &  45.119 &  45.088 &  48.473 &  49.268 &  48.212 & &  45.104 &  45.053 &  48.458 &  49.350 &  48.321 \\
 7 &  53.775 &  53.783 &  57.773 &  60.349 &  57.110 & &  53.741 &  53.623 &  57.737 &  61.068 &  57.940 \\
 8 &  67.934 &  67.518 &  72.985 &  82.875 &  73.373 & &  67.800 &  67.596 &  72.841 &  82.462 &  73.091 \\
 9 &  94.070 &  92.953 & 101.064 & 128.090 & 101.610 & &  93.710 &  93.079 & 100.677 & 127.137 & 101.043 \\
10 & 152.259 & 147.422 & 163.580 & 255.287 & 173.531 & & 152.315 & 149.183 & 163.639 & 253.722 & 167.770 \\
   & \multicolumn{11}{c}{\cellcolor{lightgray}$\lambda=2$} \\
 1 &  20.935 &  22.737 &  22.491 &   3.298 &  22.052 & &  20.774 &  21.616 &  22.319 &   1.632 &  22.117 \\
 2 &  41.435 &  42.753 &  44.516 &  30.440 &  44.059 & &  41.396 &  42.077 &  44.474 &  30.685 &  44.367 \\
 3 &  59.331 &  60.436 &  63.743 &  55.292 &  63.446 & &  59.246 &  59.811 &  63.651 &  55.632 &  63.835 \\
 4 &  77.650 &  77.814 &  83.423 &  79.246 &  83.749 & &  77.633 &  78.145 &  83.405 &  78.762 &  83.211 \\
 5 &  97.740 &  97.849 & 105.007 & 104.667 & 105.633 & &  97.713 &  97.656 & 104.977 & 104.342 & 105.354 \\
 6 & 120.505 & 120.402 & 129.465 & 131.767 & 128.658 & & 120.449 & 120.301 & 129.404 & 131.993 & 128.970 \\
 7 & 149.079 & 148.994 & 160.163 & 168.423 & 158.219 & & 148.955 & 148.567 & 160.030 & 170.403 & 160.534 \\
 8 & 196.797 & 195.348 & 211.429 & 243.594 & 212.610 & & 196.321 & 195.606 & 210.917 & 242.139 & 211.593 \\
 9 & 287.036 & 283.267 & 308.377 & 398.811 & 309.990 & & 285.749 & 283.619 & 306.995 & 395.537 & 308.077 \\
10 & 493.850 & 477.513 & 530.568 & 847.322 & 563.254 & & 494.017 & 483.568 & 530.747 & 841.774 & 544.214 \\
\bottomrule
\end{tabular}}
\vspace{0.5em}
\begin{tablenotes}[normal,flushleft]
\item \hspace{-0.25em}\scriptsize\textit{Note}: $\psi_{a}^2$ and $\varphi_{a}^2$ are the true values of the variances, $\hat{\psi}_{a}^2$ are the estimated variances of the remainder error $u_{it}$, $\hat{\varphi}_{a}^2$ are the estimated variances of the individual-specific error $\mu_i$ computed on the basis of a remainder error either homoscedastic $(\hat{\sigma}_{u}^{2})$ or heteroscedastic $(\hat{\psi}_{a}^2$).
\end{tablenotes}
\end{table}

%\onehalfspacing

Table \ref{Tab_SE_SE} shows that the heteroscedastic procedures allow to obtain standard errors lower than those obtained through the homoscedastic procedure if $\lambda=1,2$, but higher standard errors if $\lambda=0$. However, in the latter case (i.e., the homoscedastic case) if the number of individuals (and thus the number of observations) increases, then the standard errors computed with the heteroscedastic procedures become closer to the standard errors computed with the homoscedastic procedure.

\singlespacing

\begin{table}[!ht]
\renewcommand{\tabcolsep}{0.04pc}\renewcommand{\arraystretch}{1}
\caption{Simulation results on single-equation two-way \textit{EC} model: standard errors of the estimated parameters and (average) estimated variances of the error components\label{Tab_SE_SE}}
\vspace{-0.5em}
{\scriptsize
\begin{tabular}{r rrrrrrr r rrrrrrr }
\toprule
& \multicolumn{7}{c }{$N=250$, $T=12$, and $n=1031$} & & \multicolumn{7}{c}{$N=500$, $T=12$, and $n=2062$} \\
\\
& & & & & \multicolumn{3}{c }{RE QUE} & & & & & & \multicolumn{3}{c}{RE QUE} \\
& \multicolumn{1}{c }{true} & & \multicolumn{1}{c }{FE} & \multicolumn{1}{c }{RE} & \multicolumn{3}{c }{heteroscedasticity on}
&
& \multicolumn{1}{c }{true} & & \multicolumn{1}{c }{FE} & \multicolumn{1}{c }{RE} & \multicolumn{3}{c }{heteroscedasticity on} \\
& \multicolumn{1}{c }{value} & \multicolumn{1}{c}{FE} & \multicolumn{1}{c }{robust} & \multicolumn{1}{c}{homosc.} & \multicolumn{1}{c}{$u_{it}$} & \multicolumn{1}{c}{$\mu_i$} & \multicolumn{1}{c }{$u_{it}$, $\mu_i$}
&
& \multicolumn{1}{c }{value} & \multicolumn{1}{c}{FE} & \multicolumn{1}{c }{robust} & \multicolumn{1}{c}{homosc.} & \multicolumn{1}{c}{$u_{it}$} & \multicolumn{1}{c}{$\mu_i$} & \multicolumn{1}{c }{$u_{it}$, $\mu_i$} \\
\\
& & \multicolumn{1}{c}{(a)} & \multicolumn{1}{c }{(b)} & \multicolumn{1}{c }{(c)} & \multicolumn{1}{c}{(d)} & \multicolumn{1}{c }{(e)} & \multicolumn{1}{c }{(f)} &
& & \multicolumn{1}{c}{(a)} & \multicolumn{1}{c }{(b)} & \multicolumn{1}{c }{(c)} & \multicolumn{1}{c}{(d)} & \multicolumn{1}{c }{(e)} & \multicolumn{1}{c }{(f)} \\
                    & \multicolumn{15}{c}{\cellcolor{lightgray}$\lambda=0$}  \\
$\beta_{0}$  & & & & 0.756 & 0.248 & 0.248 & 0.247 & & & & & 0.731 & 0.176 &0.176 & 0.176 \\
$\beta_{1}$  & & 0.132 & 0.129 & 0.108 & 0.122 & 0.122 & 0.121 & & & 0.093 & 0.092 & 0.076 & 0.086 & 0.086 & 0.086 \\
$\beta_{2}$ & & 0.132 & 0.129 & 0.108 & 0.121 & 0.121 & 0.120 & & & 0.092 & 0.092 & 0.076 & 0.086 & 0.086 & 0.086 \\
$\beta_{3}$  & & 0.132 & 0.129 & 0.108 & 0.122 & 0.121 & 0.121 & & & 0.093 & 0.092 & 0.076 & 0.086 & 0.086 & 0.086 \\
$\bar{\varphi}^{2}$ &  6.488 & & & 6.516 & 6.516 & 6.519 & 6.526 & & 6.488 & & & 6.521 & 6.521 & 6.521 & 6.524 \\
\textcolor[rgb]{0.50,0.50,0.50}{$\sigma_{\nu}^{2}$} & \textcolor[rgb]{0.50,0.50,0.50}{6.271} & & & \multicolumn{4}{c}{\textcolor[rgb]{0.50,0.50,0.50}{6.225}}
                                                    &
                                                    & \textcolor[rgb]{0.50,0.50,0.50}{6.271} & & & \multicolumn{4}{c }{\textcolor[rgb]{0.50,0.50,0.50}{6.234}}  \\
$\bar{\psi}^{2}$  & 6.039 & 6.044 & 6.044 & 6.044 & 6.043 & 6.044 & 6.043 & & 6.039 & 6.043 & 6.043 & 6.043 & 6.041 & 6.043 & 6.041 \\
                    & \multicolumn{15}{c}{\cellcolor{lightgray}$\lambda=1$}  \\
$\beta_{0}$ & & & & 1.009 & 0.463 & 0.499 & 0.459 & & & & & 0.871 & 0.330 & 0.355 & 0.328 \\
$\beta_{1}$ & & 0.336 & 0.320 & 0.280 & 0.228 & 0.245 & 0.223 & & & 0.236 & 0.228 & 0.197 & 0.162 & 0.174 & 0.159 \\
$\beta_{2}$ & & 0.336 & 0.357 & 0.280 & 0.244 & 0.250 & 0.244 & & & 0.236 & 0.254 & 0.197 & 0.175 & 0.177 & 0.175 \\
$\beta_{3}$ & & 0.335 & 0.320 & 0.280 & 0.228 & 0.245 & 0.223 & & & 0.236 & 0.228 & 0.197 & 0.162 & 0.174 & 0.159 \\
$\bar{\varphi}^{2}$ & 46.040 & & & 49.792 & 49.792 & 49.748 & 46.339 & & 46.071 & & & 49.784 & 49.784 & 49.755 & 46.263 \\
\textcolor[rgb]{0.50,0.50,0.50}{$\sigma_{\nu}^{2}$} & \textcolor[rgb]{0.50,0.50,0.50}{6.271} & & & \multicolumn{4}{c}{\textcolor[rgb]{0.50,0.50,0.50}{6.265}}
                                                    &
                                                    & \textcolor[rgb]{0.50,0.50,0.50}{6.271} & & & \multicolumn{4}{c }{\textcolor[rgb]{0.50,0.50,0.50}{6.259}}  \\
$\bar{\psi}^{2}$ & 42.854 & 39.282 & 39.282 & 39.282 & 42.750 & 39.282 & 42.750 & & 42.883 & 39.324 &  39.324 &  39.324 &  42.809 &  39.324 &  42.809 \\
                    & \multicolumn{15}{c}{\cellcolor{lightgray}$\lambda=2$}  \\
$\beta_{0}$ & & & & 1.376 & 0.669 & 0.772 & 0.660 & & & & & 1.095 & 0.477 & 0.549 & 0.471 \\
$\beta_{1}$ & & 0.544 & 0.517 & 0.456 & 0.329 & 0.379 & 0.316 & & & 0.383 & 0.368 & 0.321 & 0.233 & 0.269 & 0.225 \\
$\beta_{2}$ & & 0.544 & 0.591 & 0.456 & 0.364 & 0.387 & 0.361 & & & 0.382 & 0.421 & 0.321 & 0.261 & 0.274 & 0.259 \\
$\beta_{3}$ & & 0.544 & 0.517 & 0.455 & 0.329 & 0.379 & 0.316 & & & 0.383 & 0.368 & 0.321 & 0.233 & 0.269 & 0.225 \\
$\bar{\varphi}^{2}$ & 124.279 & & & 137.130 & 137.130 & 137.333 & 125.178 & & 124.377 & & & 137.055 & 137.055 & 137.330 & 124.894 \\
\textcolor[rgb]{0.50,0.50,0.50}{$\sigma_{\nu}^{2}$} & \textcolor[rgb]{0.50,0.50,0.50}{6.271} & & & \multicolumn{4}{c}{\textcolor[rgb]{0.50,0.50,0.50}{6.382}}
                                                    &
                                                    & \textcolor[rgb]{0.50,0.50,0.50}{6.271} & & & \multicolumn{4}{c }{\textcolor[rgb]{0.50,0.50,0.50}{6.323}}  \\
$\bar{\psi}^{2}$  & 115.678 & 103.328 & 103.328 & 103.328 & 115.296 & 103.328 & 115.296 & & 115.770 & 103.471 & 103.471 & 103.471 & 115.501 & 103.471 & 115.501 \\
\bottomrule
\end{tabular}}
\vspace{0.5em}
\begin{tablenotes}[normal,flushleft]
      \item \hspace{-0.25em}\scriptsize\textit{Note}: Parameters estimation based on (a-b) the estimated homoscedastic variance $\hat{\sigma}^2_{u}$; (c) the estimated homoscedastic variances $\hat{\sigma}^2_{\nu}$, $\hat{\sigma}^2_{\mu}$, and $\hat{\sigma}^2_{u}$; (d) the estimated homoscedastic variances $\hat{\sigma}^2_{\nu}$ and $\hat{\sigma}^2_{\mu}$ and heteroscedastic variances $\hat{\psi}^2_{a}$, whose the average value is $\hat{\psi}^2$; (e) the estimated homoscedastic variances $\hat{\sigma}^2_{\nu}$ and $\hat{\sigma}^2_{u}$ and heteroscedastic variances $\hat{\varphi}^2_{a}(\hat{\sigma}^2_{u})$, whose the average value is $\hat{\varphi}^2$; (f) the estimated homoscedastic variance $\hat{\sigma}^2_{\nu}$ and heteroscedastic variances $\hat{\psi}^2_{a}$ and $\hat{\varphi}^2_{a}(\hat{\psi}^2_{a})$.
\end{tablenotes}
\end{table}

%\onehalfspacing

Focusing on the heteroscedastic cases, considering heteroscedasticity only on the remainder error (columns (d)) allows to obtain standard errors that are lower than the standard errors obtained considering heteroscedasticity only on the individual-specific effect (columns (e)). In other words, misspecifying the form of heteroscedasticity can be costly when heteroscedasticity is assumed only on the individual-specific effect; this loss in efficiency is smaller when heteroscedasticity is assumed only on the remainder error. These findings confirm the conclusions in \cite{BaltagiBressonPirotte(2005)}. Obviously, the smallest standard errors are obtained implementing the estimation procedure which considers both heteroscedasticity types (columns (f)).

As in \cite{LiStengos(1994)}, \cite{Roy(2002)}, and \cite{BaltagiBressonPirotte(2005)}, we consider the relative efficiency of the different estimators, computed as the ratio of the mean square error (\emph{MSE}) of the estimator under consideration to the \emph{MSE} of the true \textit{GLS} estimator. Results are reported in Table \ref{Tab_RE_SE}.

\singlespacing

\begin{table}[!ht]
\renewcommand{\tabcolsep}{0.34pc}\renewcommand{\arraystretch}{1}
\caption{Relative efficiency of the single-equation two-way \textit{EC} model\label{Tab_RE_SE}}
\vspace{-0.5em}
{\footnotesize
\begin{tabular}{l cccc c cccc }
\toprule
     & \multicolumn{4}{c}{$N=250$, $T=12$, and $n=1031$} & & \multicolumn{4}{c}{$N=500$, $T=12$, and $n=2062$}            \\
\\
     & & \multicolumn{3}{c}{heteroscedasticity on} & & & \multicolumn{3}{c }{heteroscedasticity on} \\
       & homoscedasticity & $u_{it}$ & $\mu_i$ & $u_{it}$, $\mu_i$ & & homoscedasticity & $u_{it}$ & $\mu_i$ & $u_{it}$, $\mu_i$  \\
\\
\cline{2-10}
\\
$\lambda=0$ & 1.0025 & 1.0001 & 1.0000 & 1.0001 & & 1.0013 & 1.0000 & 1.0000 & 1.0000  \\
$\lambda=1$ & 1.0011 & 0.9997 & 0.9989 & 1.0000 & & 1.0006 & 0.9998 & 0.9994 & 1.0000  \\
$\lambda=2$ & 1.0002 & 0.9997 & 0.9982 & 1.0002 & & 1.0001 & 0.9998 & 0.9991 & 1.0000  \\
\bottomrule
\end{tabular}}
\vspace{0.5em}
\begin{tablenotes}[normal,flushleft]
      \item \hspace{-0.25em}\scriptsize\textit{Note}: Relative efficiency is defined as the ratio of the \emph{MSE} of the estimator under consideration to the \emph{MSE} of the true \textit{GLS} estimator (computed considering the true variances $\psi_{a}^2$, $\varphi_{a}^2$, and $\sigma^2_{\nu}$). Note that values of the ratio both larger and smaller than $1$ indicate a loss in efficiency: if the ratio is larger than $1$, then the absolute value of the composite error term $\varepsilon_{it}=\mu_{i}+\nu_{t}+u_{it}$ is larger than the true value; and if the ratio is smaller than $1$, then the absolute value of the composite error term $\varepsilon_{it}$ is smaller than the true value.
\end{tablenotes}
\end{table}

%\onehalfspacing

We see that there are improvements in relative \emph{MSE} numbers as the sample size increases, especially when we refer to the homoscedastic estimator. Furthermore, confirming our previous remarks, misspecifying the form of heteroscedasticity may be costly when only the individual-specific effect is considered heteroscedastic, especially if the sample size is small. Besides, as already observed in the comments to Table \ref{Tab_SE_SE}, the most efficient estimator is the one that considers both the remainder error and the individual-specific effect heteroscedastic.

\section{Heteroscedastic two-way SUR systems}\label{SUR-SYSTEMS}

When systems of equations have to be estimated, as it is the case of \textit{SUR} systems, single-equation estimation techniques are not appropriate. In order to estimate heteroscedastic two-way \textit{SUR} systems we extend the procedure in \citet{Biorn(2004)}, with individuals grouped according to the number of times they are observed.

\subsection{Model and notation}

Let $N_{p}$ denote the number of individuals observed exactly in $p$ periods, with $p=1,\ldots,T$. Hence $\sum_{p} N_{p} = N$ and $\sum_{p} \left(N_{p} p \right) = n$. Moreover, let $N_{a,p}$ denote the number of individuals belonging to stratum $a$ and observed in $p$ periods; therefore, $\sum_{a} N_{a,p} = N_{p}$ and $ \sum_{p} \sum_{a} N_{a,p} = N$.

We assume that the $T$ groups of individuals are ordered such that the $N_{p=1}$ individuals observed once come first, the $N_{p=2}$ individuals observed twice come second, etc. Hence, with $C_{p}=\sum_{h=1}^{p} N_{h}$ being the cumulated number of individuals observed at most $p$ times, the index sets of the individuals observed exactly $p$ times can be written as $I_{p} = \{ C_{p-1}+1 , \ldots , C_{p} \}$. Note that $I_{p=1}$ may be considered as a pure cross section and $I_{p}$, with $p \geq 2$, as a pseudo-balanced panel with $p$ observations for each individual. This structure allows us to use a number of results derived for the two-way \textit{SUR} systems in the balanced case.

If $k_{m}$ is the number of regressors for equation $m$, the total number of regressors for the system is $K=\sum_{m=1}^{M} k_{m}$. Stacking the $M$ equations, indexed $m=1,\ldots,M$, for the observation $( i,t )$ we have:
\begin{equation} \label{eqSUR}
      \begin{array}{l}
            \mathrm{y}_{it} = \mathrm{X}_{it} {\upbeta} + {\upmu}_{i} + {\upnu}_{t} + \mathrm{u}_{it} = \mathrm{X}_{it} {\upbeta} + {\upvarepsilon} ,
      \end{array}
\end{equation}
where the $M \times K$ matrix of explanatory variables is $\mathrm{X}_{it} = \textrm{diag} ( \mathrm{x}_{1it}^{\textrm{T}} , \ldots , \mathrm{x}_{Mit}^{\textrm{T}} )$ and the $K \times 1$ vector of parameters is ${\upbeta} = ( {\upbeta}_{1} ^{\textrm{T}} , \ldots , {\upbeta}_{M}^{\textrm{T}} ) ^{\textrm{T}}$ and where ${\upmu}_{i} \equiv ( \mu_{1i}, \ldots , \mu_{Mi} ) ^{\textrm{T}}$, ${\upnu}_{t} \equiv ( \nu_{1t}, \ldots , \nu_{Mt} ) ^{\textrm{T}}$, and $\mathrm{u}_{it} \equiv ( u_{1it} , \ldots , u_{Mit} ) ^{\textrm{T}}$. If we do not have cross-equation restrictions, we can assume $\textit{E} ( u_{mit} | \mathrm{x}_{1it}^{\textrm{T}}, \mathrm{x}_{2it}^{\textrm{T}} , \ldots , \mathrm{x}_{Mit}^{\textrm{T}} ) = 0$, and then $\textit{E} ( y_{mit} | \mathrm{x}_{1it}^{\textrm{T}}, \mathrm{x}_{2it}^{\textrm{T}} , \ldots , \mathrm{x}_{Mit}^{\textrm{T}} ) = \linebreak \textit{E} ( y_{mit} | \mathrm{x}_{mit}^{\textrm{T}} ) = \mathrm{x}_{mit}^{\textrm{T}} {\upbeta}_{m}$. On the contrary, if we have cross-equation restrictions\footnote{As \citet{Biorn(2004)} suggests, with cross-equations restrictions we can redefine ${\upbeta}$ as the complete $K \times 1$ coefficient vector (without duplication) and the $M \times K$ regression matrix as $\mathrm{X}_{it} = ( \mathrm{x}_{1it}, \mathrm{x}_{2it} , \ldots , \mathrm{x}_{Mit} ) ^{\textrm{T}} $, where the $k^\textrm{th}$ element of the $k_{m} \times 1$ vector $\mathrm{x}_{mit}$ either contains the observation on the variable in the $m^\textrm{th}$ equation which corresponds to the $k^\textrm{th}$ coefficient in ${\upbeta}$ or is zero if the $k^\textrm{th}$ coefficient does not occur in the $m^\textrm{th}$ equation.}, we can only assume $\textit{E} ( \mathrm{u}_{it} | \mathrm{x}_{it}^{\textrm{T}} ) = 0 $, where $\mathrm{x}_{it}\equiv ( \mathrm{x}_{1it}^{\textrm{T}},\mathrm{x}_{2it}^{\textrm{T}}, \ldots , \mathrm{x}_{Mit}^{\textrm{T}} )^{\textrm{T}}$.

With heteroscedasticity on both the individual-specific disturbance and the remainder error, for $i \in I_a$ across the regression equations $m$ and $j$, we assume that:
\begin{equation}
      \begin{array}{l}
            \textit{E}\left( \mu_{m i},\mu_{j i^{\prime}}\right)
           = \left\{ %\Biggl\{
            \begin{array}{ll}
                   \varphi_{a,mj} & i =    i^{\prime}    \\
                   0              & i \neq i^{\prime} ,
            \end{array}
            \right.
            \textit{E}\left(\nu_{mt},\nu_{jt^{\prime}}\right)
            =\left\{ %\Biggl\{
            \begin{array}{ll}
                  \sigma_{\nu,mj} & t=t^{\prime} \\
                  0               & t\neq t^{\prime} ,
            \end{array}
            \right.
            \\
            \textit{E}\left(u_{m i t},u_{j i^{\prime} t^{\prime}}\right)
            =\left\{ %\Biggl\{
            \begin{array}{ll}
                  \psi_{a,mj}     & i =    i^{\prime} \text{ and }    t=t^{\prime} \\
                  0               & i \neq i^{\prime} \text{ and/or } t\neq t^{\prime} .
            \end{array}
            \right.
      \end{array}
\end{equation}
Let us consider the $NM\times 1$ vector ${\upmu} \equiv ( {\upmu}_{1} ^{\textrm{T}} , \ldots , {\upmu}_{N} ^{\textrm{T}} ) ^{\textrm{T}} $, the $TM\times 1$ vector ${\upnu} \equiv ( {\upnu}_{1} ^{\textrm{T}} , \ldots , {\upnu}_{T} ^{\textrm{T}} ) ^{\textrm{T}} $, and the $nM\times 1$ vector $\mathrm{u} \equiv ( \mathrm{u}_{11}^{\textrm{T}}, \mathrm{u}_{12}^{\textrm{T}} , \ldots , \mathrm{u}_{1T_{1}}^{\textrm{T}} , \mathrm{u}_{21}^{\textrm{T}} , \ldots , \mathrm{u}_{NT_{N}}^{\textrm{T}} ) ^{\textrm{T}}$. Since the $M \times 1$ vectors $\mathrm{u}_{i t} \sim \left( 0 , {\Uppsi}_{a} \right)$, the $M \times 1$ vectors ${\upmu}_{i} \sim \left( 0 , {\Upphi}_{a} \right)$, and the $TM \times 1$ vector ${\upnu} \sim \left( 0 , {\Upsigma}_{\nu} \right)$, with the $M\times M$ matrices ${\Uppsi}_{a} = \left[\psi_{a,mj}\right]$, ${\Upphi}_{a} = \left[\varphi_{a,mj}\right]$, and ${\Upsigma}_{\nu} = [ \sigma_{\nu,mj} ]$, we can assume that the expected values of the vectors $\mathrm{u}_{i t}$, ${\upmu}_{i}$, and ${\upnu}_{t}$ are zero and their covariance matrices are equal to ${\Uppsi}_{a}$, ${\Upphi}_{a}$, and $\Upsigma_{\nu}$. It follows that $\textit{E} ( {\upvarepsilon}_{i t} {\upvarepsilon}_{i^{\prime} t^{\prime}}^{\textrm{T}} ) = \delta_{i i^{\prime}}{\Upphi}_{a} + \delta_{tt^{\prime}}\Upsigma_{\nu} + \delta_{i i^{\prime}}\delta_{tt^{\prime}}{\Uppsi}_{a}$, with $\delta_{i i^{\prime}}=1$ for $i = i^{\prime}$ and $\delta_{i i^{\prime}}=0$ for $i \neq i^{\prime}$, $\delta_{tt^{\prime}}=1$ for $t=t^{\prime}$ and $\delta_{tt^{\prime}}=0$ for $t\neq t^{\prime}$.

As in \citet{Biorn(2004)}, let us consider the $pM\times 1$ vector of independent variables $\mathrm{y}_{i(p)} \equiv ( \mathrm{y}_{i1}^{\textrm{T}} , \ldots , \mathrm{y}_{ip}^{\textrm{T}} )^{\textrm{T}}$, the $pM\times K$ matrix of explanatory variables $\mathrm{X}_{i(p)} \equiv ( \mathrm{X}_{i1}^{\textrm{T}} , \ldots , \mathrm{X}_{ip}^{\textrm{T}} )^{\textrm{T}}$, and the $pM\times 1$ vector of composite error terms ${\upvarepsilon}_{i(p)} \equiv ( {\upvarepsilon}_{i1}^{\textrm{T}} , \ldots , {\upvarepsilon}_{ip}^{\textrm{T}} )^{\textrm{T}}$ for $i \in I_{p}$. If we define the $pM\times TM$ matrix ${\Updelta}_{i(p)}$, indicating in which period
$t$ the individual $i$ of the group $p$ is observed, and if we consider the $TM\times 1$ vector ${\upnu}$, for the individual $i \in I_{p}$ we can define the $pM\times 1$ vector ${\upnu}_{i(p)} \equiv {\Updelta}_{i(p)} {\upnu}$ and write the model:
\begin{equation}
      \begin{array}{l}
            \mathrm{y}_{i \left( p \right)} = \mathrm{X}_{i \left( p \right)} {\upbeta} + ( {\upiota}_{p} \otimes {\upmu}_{i} ) + {\upnu}_{i \left( p \right)} + \mathrm{u}_{i
            \left( p \right)} = \mathrm{X}_{i \left( p \right)} {\upbeta} + {\upvarepsilon}_{i \left( p \right)} ,
      \end{array}
\end{equation}
where ${\upiota}_{p}$ is a $p \times 1$ vector of ones \citep[see][]{PlatoniSckokaiMoro(2012)}.

The $pM \times pM$ heteroscedastic variance-covariance matrix of the $pM \times 1$ composite error terms ${\upvarepsilon}_{i(a,p)}$ for the individual $i \in I_{a,p}$, with $I_{a,p} = I_a \cap I_{p}$ the set of individuals belonging to stratum $a$ and observed in $p$ periods, is given by:
\begin{equation}
      \begin{array}{l}
            {\Upomega}_{a,p} = \mathrm{E}_{p} \otimes \left({\Uppsi}_a + {\Upsigma}_{\nu}\right) + \bar{\mathrm{J}}_{p} \otimes \left( {\Uppsi}_a +
            {\Upsigma}_{\nu} + p {\Upphi}_a\right) ,
      \end{array}
\end{equation}
where $\mathrm{E}_{p} = \mathrm{I}_{p} - \bar{\mathrm{J}}_{p}$ (with $\mathrm{I}_{p}$ identity matrix of dimension $p$) and $\bar{\mathrm{J}}_{p} = \frac{\mathrm{J}_{p}} {p}$ (with $\mathrm{J}_{p}$ matrix of ones of dimension $p$). Since $\mathrm{E}_{p}$ and $\bar{\mathrm{J}}_{p}$ are symmetric, idempotent, and have orthogonal columns, the inverse of the variance-covariance matrix of the individuals belonging to stratum $a$ and
group $p$ is:
\begin{equation}
      \begin{array}{l}
            {\Upomega}_{a,p}^{-1} = \mathrm{E}_{p} \otimes \left({\Uppsi}_a + {\Upsigma}_{\nu}\right)^{-1} + \bar{\mathrm{J}}_{p} \otimes \left({\Uppsi}_a +
            {\Upsigma}_{\nu} + p {\Upphi}_a\right)^{-1}.
      \end{array}
\end{equation}
This specification nests simpler heteroscedastic schemes as well as the homoscedastic case by replacing ${\Upphi}_a$ with ${\Upsigma}_{\mu}$ and/or ${\Uppsi}_a$ with ${\Upsigma}_{u}$.

If we assume that ${\Uppsi}_a$, ${\Upphi}_a$, and ${\Upsigma}_{\nu}$ are known, then in the heteroscedastic case we can write the \textit{GLS} estimator for the $K \times 1$ vector of parameters ${\upbeta}$ as the problem of minimizing:
\begin{equation}
      \begin{array}{l}
            \underset{p=1}{\overset{T}{\textstyle\sum}} \underset{a=1}{\overset{A}{\textstyle\sum}} \underset{i \in I_{a,p}}{\textstyle\sum} {\upvarepsilon}_{i}^{\textrm{T}}
            {\Upomega}_{a,p}^{-1} {\upvarepsilon}_{i}.
      \end{array}
\end{equation}
where, for sake of simplicity and since there is no risk of ambiguity, ${\upvarepsilon}_{i}$ is used instead of ${\upvarepsilon}_{i\left( a,p \right)}$.

If we apply \textit{GLS} on the observations for the individuals observed $p$ times we obtain:
\begin{equation}
      \begin{array}{l}
            \hat{{\upbeta}}_{p}^{GLS} =  \left( \underset{a=1}{\overset{A}{\textstyle\sum}} \underset{i \in I_{a,p}}{\textstyle\sum} \mathrm{X}_{i}^{\textrm{T}} {\Upomega}_{a,p}^{-1} \mathrm{X}_{i} \right)^{-1} \underset{a=1}{\overset{A}{\textstyle\sum}} \underset{i \in I_{a,p}}{\textstyle\sum} \mathrm{X}_{i}^{\textrm{T}} {\Upomega}_{a,p}^{-1} \mathrm{y}_{i} ,
      \end{array}
\end{equation}
while the full \textit{GLS} estimator is:
\begin{equation}
      \hat{{\upbeta}}^{GLS} = \left( \underset{p=1}{\overset{T}{\textstyle\sum}} \underset{a=1}{\overset{A}{\textstyle\sum}} \underset{i \in I_{a,p}}{\textstyle\sum} \mathrm{X}_{i}^{\textrm{T}} {\Upomega}_{a,p}^{-1} \mathrm{X}_{i} \right) ^{-1} \underset{p=1}{\overset{T}{\textstyle\sum}} \underset{a=1}{\overset{A}{\textstyle\sum}}
      \underset{i \in I_{a,p}}{\textstyle\sum } \mathrm{X}_{i}^{\textrm{T}} {\Upomega}_{a,p}^{-1} \mathrm{y}_{i} ,
\end{equation}
where $\mathrm{X}_{i}$ is the $pM\times K$ matrix of explanatory variables related to individual $i \in I_{a,p}$.

\subsection{Estimation of the covariance matrices}\label{SURestimation}

The next step is to find an appropriate technique to estimate the components of the variance-covariance matrices of the two-way \textit{SUR} system ${\Uppsi}_a$, ${\Upphi}_a$, and ${\Upsigma}_{\nu}$. This can be achieved adopting either the \textit{QUE} procedure suggested by \citet{WansbeekKapteyn(1989)} for the homoscedastic single-equation case or the within-between (\textit{WB}) procedure suggested by \citet{Biorn(2004)} for the homoscedastic one-way \textit{SUR} system. In the following sub-sections we modify both procedures making them suitable for the heteroscedastic two-way \textit{SUR} system.

\subsubsection{The QUE procedure}\label{QUE}

The \textit{QUE} procedure considers the $n \times 1$ residuals $\mathrm{e}_{m} \equiv \mathrm{y}_{m} - \mathrm{X}_{m} \hat{{\upbeta}}_{m}^{W}$ from the \textit{W} estimator in (\ref{Westimator}) for the equation $m=1, \ldots ,M$, where $\mathrm{X}_{m}$ is a matrix of dimension $n \times \left(k_m-1\right)$. If we assume that the $n \times k_{m}$ matrix $\mathrm{X}_{m}$ contains a vector of ones, then we have to define the $n \times 1$ consistent centered residuals $\mathrm{f}_{m} \equiv \mathrm{E}_{n} \mathrm{e}_{m} = \mathrm{e}_{m} - \bar{e}_{m}$.

With heteroscedasticity, we can obtain the adapted \textit{QUE}s for ${\Uppsi}_{mj}$, ${\Upphi}_{mj}$, and $\sigma_{\nu,mj}$ by equating:
\begin{equation}\label{q_QUE_SUR}
      \begin{array}{lll}
            q_{n_a,mj} & \hspace{-0.5em} \equiv \mathrm{f}_{j}^{\textrm{T}} \mathrm{Q}_{\Updelta} \mathrm{H}_{a}^{\textrm{T}} \mathrm{H}_{a} \mathrm{Q}_{\Updelta} \mathrm{f}_{m} & \rightarrow
            \underset{a=1} {\overset{A} {\textstyle\sum}} q_{n_a,mj} = q_{n,mj} \equiv \mathrm{f}_{j}^{\textrm{T}} \mathrm{Q}_{\Updelta} \mathrm{f}_{m} , \\
            q_{N_a,mj} & \hspace{-0.5em} \equiv \mathrm{f}_{a_{j}}^{\textrm{T}} \bar{\mathrm{J}}_{N_a} \mathrm{f}_{a_{m}} & \rightarrow \underset{a=1} {\overset{A} {\textstyle\sum}} q_{N_a,mj} = q_{N,mj} \equiv
            \mathrm{f}_{j}^{\textrm{T}} {\Updelta}_{\mu} {\Updelta}_{N}^{-1} {\Updelta}_{\mu}^{\textrm{T}} \mathrm{f}_{m} , \\
            q_{T,mj}   & \hspace{-0.5em} \equiv \mathrm{f}_{j}^{\textrm{T}} {\Updelta}_{\nu} {\Updelta}_{T}^{-1} {\Updelta}_{\nu}^{\textrm{T}} \mathrm{f}_{m} , &
      \end{array}
\end{equation}
to their expected values. The identities in (\ref{q_QUE_SUR}) can be further detailed as already done in formula \eqref{q_QUE_details}, Appendix \ref{appTEC_QUE}, for the identities in (\ref{q_QUE}).

Hence, the estimator of $\psi_{a,mj}$ is:
\begin{equation}
      \begin{array}{l}
            \hat{\psi}_{a,mj} = \dfrac{q_{n_a,mj} + \left( k_{a,m} + k_{a,j} - k_{a,mj} \right) \hat{\sigma}_{u,mj}}{n_a - N_a - \tau_a}
      \end{array}
\end{equation}
where $ k_{a,mj} \equiv \text{tr} [ (\mathrm{X}_{m}^{\textrm{T}} \mathrm{Q}_{\Updelta} \mathrm{X}_{m})^{-1} \mathrm{X}_{m}^{\textrm{T}}\mathrm{Q}_{\Updelta}\mathrm{X}_{j} (\mathrm{X}_{j}^{\textrm{T}} \mathrm{Q}_{\Updelta} \mathrm{X}_{j})^{-1} \mathrm{X}_{j}^{\textrm{T}} \mathrm{Q}_{\Updelta} \mathrm{H}_{a}^{\textrm{T}} \mathrm{H}_{a} \mathrm{Q}_{\Updelta} \mathrm{X}_{m} ]$, with \linebreak $\sum_{a=1}^{A} k_{a,mj}= k_{mj}$ and $k_{mj} \equiv \text{tr} [ ( \mathrm{X}_{m}^{\textrm{T}} \mathrm{Q}_{\Updelta}\mathrm{X}_{m} )^{-1} \mathrm{X}_{m}^{\textrm{T}} \mathrm{Q}_{\Updelta} \mathrm{X}_{j} ( \mathrm{X}_{j}^{\textrm{T}} \mathrm{Q}_{\Updelta} \mathrm{X}_{j} )^{-1} \mathrm{X}_{j}^{\textrm{T}} \mathrm{Q}_{\Updelta} \mathrm{X}_{m} ]$. The estimated variance-covariance $\hat{\sigma}_{u,mj}$ is obtained by equating $q_{n,mj}$ to its expected value \citep[see][]{PlatoniSckokaiMoro(2012)}.
Furthermore, the estimator of $\varphi_{a,mj}$ is:
\begin{equation}
      \begin{split}
            \hat{\varphi}_{a,mj} =
            & \dfrac{q_{N_a,mj} - \left( N_a - 2 \frac{n_a}{n} \right) \hat{\psi}_{a,mj}  - \left( k_{N_a,mj} - k_{0_a,mj} + \frac{n_a}{n} k_{0,mj} + \frac{n_a}{n} \right) \hat{\sigma}_{u,mj} } {n_a - 2 \lambda_{\mu_a}} \\
            & + \dfrac{- \frac{n_a}{n} \lambda_{\mu} \hat{\sigma}_{\mu,mj} - \left( N_a - 2 \lambda_{\nu_a} + \frac{n_a}{n} \lambda_{\nu}  \right) \hat{\sigma}_{\nu,mj} } {n_a - 2 \lambda_{\mu_a}},
      \end{split}
\end{equation}
where $k_{N_a,mj} \equiv \text{tr} [ ( \mathrm{X}_{m}^{\textrm{T}}\mathrm{Q}_{\Updelta} \mathrm{X}_{m} ) ^{-1} \mathrm{X}_{m}^{\textrm{T}}\mathrm{Q}_{\Updelta}\mathrm{X}_{j} ( \mathrm{X}_{j}^{\textrm{T}}\mathrm{Q}_{\Updelta} \mathrm{X}_{j} ) ^{-1} \mathrm{X}_{a_{j}}^{\textrm{T}} \bar{\mathrm{J}}_{N_a} \mathrm{X}_{a_{m}} ]$, $k_{0_a,mj} \equiv \linebreak \frac{ {\upiota}_{N_a}^{\textrm{T}} \mathrm{X}_{a_m} (\mathrm{X}_{m}^{\textrm{T}}\mathrm{Q}_{\Updelta}\mathrm{X}_{m})^{-1} \mathrm{X}_{m}^{\textrm{T}}\mathrm{Q}_{\Updelta}\mathrm{X}_{j} (\mathrm{X}_{j}^{\textrm{T}}\mathrm{Q}_{\Updelta}\mathrm{X}_{j})^{-1} \mathrm{X}_{j}^{\textrm{T}}{\upiota}_{n}}{n} + \frac{{\upiota}_{n}^{\textrm{T}} \mathrm{X}_{m} (\mathrm{X}_{m}^{\textrm{T}}\mathrm{Q}_{\Updelta}\mathrm{X}_{m})^{-1} \mathrm{X}_{m}^{\textrm{T}}\mathrm{Q}_{\Updelta}\mathrm{X}_{j} (\mathrm{X}_{j}^{\textrm{T}}\mathrm{Q}_{\Updelta}\mathrm{X}_{j})^{-1} \mathrm{X}_{a_j}^{\textrm{T}}{\upiota}_{N_a} } {n}  $, \linebreak $k_{0,mj} \equiv \frac{ {\upiota}_{n}^{\textrm{T}} {\mathrm{X}_{m}} { ( \mathrm{X}_{m}^{\textrm{T}} \mathrm{Q}_{\Updelta} \mathrm{X}_{m} ) ^{-1} } \mathrm{X}_{m}^{\textrm{T}} \mathrm{Q}_{\Updelta} \mathrm{X}_{j} { (\mathrm{X}_{j}^{\textrm{T}} \mathrm{Q}_{\Updelta} \mathrm{X}_{j} ) ^{-1}} \mathrm{X}_{j}^{\textrm{T}} {\upiota}_{n} } {n}$. The estimated variance-covariance $\hat{\sigma}_{\mu,mj}$ is obtained jointly with $\hat{\sigma}_{\nu,mj}$ by equating $q_{N,mj}$ and $q_{T,mj}$ to their expected values \citep[see][]{PlatoniSckokaiMoro(2012)}.

As in the single-equation case, simpler heteroscedastic scheme (i.e., heteroscedasticity only on the individual-specific disturbance or on the remainder error) can be obtained combining results for the general scheme with those for the homoscedastic case, although when we consider the case of heteroscedasticity only on the individual-specific disturbance the expected value of $q_{N_a,mj}$ and the estimated variance-covariance $\hat{\varphi}_{a,mj}$ are obtained differently (see equations \eqref{eq.SURQUE_mu_1}-\eqref{eq.SURQUE_mu_2} in Appendix \ref{appTEC_QUE}).
%as:
%\begin{equation}
%      \begin{split}
%             \textit{E} \left( q_{N_a,mj} \right) = &
%             \left( N_a + k_{N_a,mj} - k_{0_a,mj} + \frac{n_a}{n} \cdot k_{0,mj} - \frac{n_a}{n} \right) \cdot \sigma_{u,mj} \\ + \left( N_a - 2 \cdot \lambda_{\mu_a} \right) & \cdot \varphi_{a,mj} + \frac{n_a}{n} \cdot \lambda_{\mu} \cdot \bar{\varphi}_{mj} + \left( N_a - 2 \cdot \lambda_{\nu_a} + \frac{n_a}{n} \cdot \lambda_{\nu} \right) \cdot \sigma_{\nu,mj}
%      \end{split}
%\end{equation}
%and, therefore,
%\begin{equation}
%      \begin{split}
%            \hat{\varphi}_{a,mj} = &
%              \dfrac{q_{N_a,mj} - \left( N_a + k_{N_a,mj} - k_{0_a,mj} + \frac{n_a}{n} \cdot k_{0,mj} - \frac{n_a}{n} \right) \cdot \hat{\sigma}_{u,mj} } {n_a - 2 \cdot \lambda_{\mu_a}} \\ &
%            + \dfrac{- \frac{n_a}{n} \cdot \lambda_{\mu} \cdot \hat{\sigma}_{\mu,mj} - \left( N_a - 2 \cdot \lambda_{\nu_a} + \frac{n_a}{n} \cdot \lambda_{\nu} \right) \cdot \hat{\sigma}_{\nu,mj} } {n_a - 2 \cdot \lambda_{\mu_a}} .
%      \end{split}
%\end{equation}

\subsubsection{The WB procedure}\label{WB}

With heteroscedastic two-way systems of equations, the $M \times M$ matrices of within individuals, between individuals, and between times (co)variations in the ${\upvarepsilon}$'s of the $M$ equations are the following:
\begin{equation}\label{EqForAppWB}
      \begin{array}{l}
            \mathrm{W}_{\varepsilon}     = \underset{a=1}{\overset{A}{\textstyle\sum}}\mathrm{W}_{\varepsilon_a}        = \underset{a=1}{\overset{A}{\textstyle\sum}} \underset{i \in I_a}{\textstyle\sum}
            \underset{t=1}{\overset{T_{i}}{\textstyle\sum}} \left( {\upvarepsilon}_{it}-{\bar{\upvarepsilon}}_{i \centerdot} - {\bar{\upvarepsilon}}_{\centerdot t} \right) \left(
            {\upvarepsilon}_{it}-{\bar{\upvarepsilon}}_{i \centerdot}-{\bar{\upvarepsilon}}_{\centerdot t} \right) ^{\textrm{T}} , \\
            \mathrm{B}_{\varepsilon}^{C} = \underset{a=1}{\overset{A}{\textstyle\sum}} \mathrm{B}_{\varepsilon_{a}}^{C} = \underset{a=1}{\overset{A}{\textstyle\sum}} \underset{i \in I_a}{\textstyle\sum} T_{i} \left(
            {\bar{\upvarepsilon}}_{i \centerdot}-{\bar{\upvarepsilon}} \right) \left( {\bar{\upvarepsilon}}_{i \centerdot}-{\bar{\upvarepsilon}} \right) ^{\textrm{T}} , \\
            \mathrm{B}_{\varepsilon}^{T} = \underset{t=1}{\overset{T}{\textstyle\sum}}N_{t} \left( {\bar{\upvarepsilon}}_{\centerdot t}-{\bar{\upvarepsilon}} \right) \left(
            {\bar{\upvarepsilon}}_{\centerdot t}-{\bar{\upvarepsilon}} \right) ^{\textrm{T}} ,
      \end{array}
\end{equation}
where for each equation $m$ we have $\bar{\varepsilon}_{m i \centerdot} = \frac{ \sum_{t=1}^{T_{i}} \varepsilon_{mit} } {T_{i}}$, $\bar{\varepsilon}_{m \centerdot t} = \frac{ \sum_{i=1}^{N_{t}} \varepsilon_{mit} } {N_{t}}$, and $\bar{\varepsilon}_{m} = \frac{ \sum_{i=1}^{N} \sum_{t=1}^{T_{i}} \varepsilon_{mit} } {n} = \frac{ \sum_{i=1}^{N} (T_{i} \bar{\varepsilon}_{m i \centerdot}) } {n}$ or $\bar{\varepsilon}_{m} = \frac{ \sum_{t=1}^{T} \sum_{i=1}^{N_{t}} \varepsilon_{m i t}} {n} = \frac{ \sum_{t=1}^{T} ( N_{t} \bar{\varepsilon}_{m \centerdot t} ) } {n}$.

Because the $\mathrm{u}_{it}$'s, the ${\upmu}_{i}$'s, and the ${\upnu}_{t}$'s are independent, from the equations in \eqref{EqForAppWB} we can write:
\begin{equation}
      \begin{array}{l}
            \textit{E} \left( \mathrm{W}_{\varepsilon_{a}}     \right) = \textit{E} \left( \mathrm{W}_{u_{a}}       \right) , \\
            \textit{E} \left( \mathrm{B}_{\varepsilon_{a}}^{C} \right) = \textit{E} \left( \mathrm{B}_{\mu_{a}}^{C} \right) + \textit{E} \left( \mathrm{B}_{u_{a}}^{C} \right) , \\
            \textit{E} \left( \mathrm{B}_{\varepsilon}^{T}     \right) = \textit{E} \left( \mathrm{B}_{\nu}^{T}     \right) + \textit{E} \left( \mathrm{B}_{u}^{T} \right) ,
      \end{array}
\end{equation}
where the within individuals (co)variation is:
\begin{equation}
      \begin{split}
            \mathrm{W}_{u_{a}} & = \underset{i \in I_a}{\textstyle\sum} \underset{t=1}{\overset{T_{i}}{\textstyle\sum }} \left( \mathrm{u}_{it} - \mathrm{\bar{u}}_{i \centerdot} - \mathrm{\bar{u}}_{\centerdot t} \right) \left(
            \mathrm{u}_{it} - \mathrm{\bar{u}}_{i \centerdot} - \mathrm{\bar{u}}_{\centerdot t} \right)^{\textrm{T}} \\
            & = \underset{i \in I_a}{\textstyle\sum} \underset{t=1}{\overset{T_{i}}{\textstyle\sum}} \mathrm{u}_{it} \mathrm{u}_{it}^{\textrm{T}} - \underset{i \in I_a}{\textstyle\sum} T_{i} \mathrm{\bar{u}}_{i \centerdot} \mathrm{\bar{u}}_{i \centerdot}^{\textrm{T}} - \underset{i \in I_a}{\textstyle\sum} \underset{t=1}{\overset{T_{i}}{\textstyle\sum}} \mathrm{\bar{u}}_{\centerdot t} \mathrm{\bar{u}}_{\centerdot t}^{\textrm{T}},
      \end{split}
\end{equation}
the between individuals (co)variations are:
\begin{equation}
      \begin{array}{l}
            \mathrm{B}_{\mu_{a}}^{C} = \underset{i \in I_a}{\textstyle\sum} T_{i} \left( {\upmu}_{i} - {\bar{\upmu}} \right) \left( {\upmu}_{i} - {\bar{\upmu}} \right)^{\textrm{T}}
            = \underset{i \in I_a}{\textstyle\sum} T_{i} {\upmu}_{i} {\upmu}_{i}^{\textrm{T}} - \underset{i \in I_a}{\textstyle\sum} T_{i} {\bar{\upmu}} {\bar{\upmu}}^{\textrm{T}} , \\
            \mathrm{B}_{u_{a}}^{C} = \underset{i \in I_a}{\textstyle\sum} T_{i} \left( \mathrm{\bar{u}}_{i \centerdot} - \mathrm{\bar{u}} \right) \left( \mathrm{\bar{u}}_{i \centerdot} - \mathrm{\bar{u}} \right)^{\textrm{T}} =
            \underset{i \in I_a}{\textstyle\sum} T_{i} \mathrm{\bar{u}}_{i \centerdot} \mathrm{\bar{u}}_{i \centerdot}^{\textrm{T}} - \underset{i \in I_a}{\textstyle\sum} T_{i} \mathrm{\bar{u}} \mathrm{\bar{u}}^{\textrm{T}} ,
      \end{array}
\end{equation}
and the between times (co)variations, as in the homoscedastic case, are:
\begin{equation}
      \begin{array}{l}
            \mathrm{B}_{\nu}^{T} = \underset{t=1}{\overset{T}{\textstyle\sum}} N_{t} \left( {\upnu}_{t} - {\bar{\upnu}} \right) \left( {\upnu}_{t} - {\bar{\upnu}}
            \right)^{\textrm{T}} = \underset{t=1}{\overset{T}{\textstyle\sum}} N_{t} {\upnu}_{t} {\upnu}_{t}^{\textrm{T}} - n {\bar{\upnu}} {\bar{\upnu}}^{\textrm{T}} , \\
            \mathrm{B}_{u}^{T} = \underset{t=1}{\overset{T}{\textstyle\sum}} N_{t} \left( \mathrm{\bar{u}}_{\centerdot t} - \mathrm{\bar{u}} \right) \left( \mathrm{\bar{u}}_{\centerdot t} - \mathrm{\bar{ u}} \right)^{\textrm{T}} =
            \underset{t=1}{\overset{T}{\textstyle\sum}} N_{t} \mathrm{\bar{u}}_{\centerdot t} \mathrm{\bar{u}}_{\centerdot t}^{\textrm{T}} - n \mathrm{\bar{u}} \mathrm{\bar{u}}^{\textrm{T}} ,
      \end{array}
\end{equation}
where $\bar{u}_{m i \centerdot} = \frac{ \sum_{t=1}^{T_{i}}u_{mit} } {T_{i}}$, $\bar{u}_{m \centerdot t} = \frac{ \sum_{i=1}^{N_{t}}u_{mit} } {N_{t}}$, $\bar{u}_{m} = \frac{ \sum_{i=1}^{N} \sum_{t=1}^{T_{i}} u_{mit}} {n} = \frac{ \sum_{i=1}^{N} ( T_{i} \bar{u}_{m i \centerdot} ) } {n}$ or $\bar{u}_{m} = \frac{ \sum_{t=1}^{T} \sum_{i=1}^{N_{t}} u_{mit}} {n} \linebreak = \frac{ \sum_{t=1}^{T} ( N_{t} \bar{u}_{m \centerdot t} )} {n}$, $\bar{\mu}_{m} = \frac{\sum_{i=1}^{N} (T_{i} \mu_{mi})} {n}$, and $\bar{\nu}_{m} = \frac{\sum_{t=1}^{T} (N_{t} \nu_{mt})} {n}$ \citep[see][]{Biorn(2004),PlatoniSckokaiMoro(2012)}.

Since for $i \in I_a$ we have $\textit{E} ( {\upvarepsilon}_{it} {\upvarepsilon}_{i^{\prime} t^{\prime}}^{\textrm{T}} ) = \delta_{ii^{\prime}} {\Upphi}_a + \delta_{tt^{\prime}} {\Upsigma}_{\nu} + \delta_{ii^{\prime}} \delta_{tt^{\prime}} {\Uppsi}_a$, where $\textit{E}(\mathrm{u}_{it}\mathrm{u}_{i^{\prime}t^{\prime}}^{\textrm{T}}) = \delta_{ii^{\prime}} \delta_{tt^{\prime}}
{\Uppsi}_a$, $\textit{E}({\upmu}_{i} {\upmu}_{i^{\prime}}^{\textrm{T}}) = \delta_{ii^{\prime}} {\Upphi}_a$, and $\textit{E}({\upnu}_{t} {\upnu}_{t^{\prime}}^{\prime}) = \delta_{tt^{\prime}} {\Upsigma}_{\nu}$, it follows that $\textit{E}(\mathrm{\bar{u}}_{i \centerdot} \mathrm{\bar{u}}_{i \centerdot}^{\textrm{T}}) = \frac{{\Uppsi}_a} {T_{i}}$, $\textit{E}(\mathrm{\bar{u}}_{\centerdot t} \mathrm{\bar{u}}_{\centerdot t}^{\textrm{T}}) = \frac{ \sum_{i \in I_t} {\Uppsi}_a } {N_{t}^{2}} \simeq \frac{\bar{\Uppsi}} {N_{t}} \approx \frac{{\Upsigma}_{u}} {N_{t}}$, with $I_t$ the set of individuals observed in period $t$, $\textit{E}(\mathrm{\bar{u}\bar{u}}^{\textrm{T}}) = \frac{\sum_{i=1}^{N} (T_{i}{\Uppsi}_a) } {n^{2}} = \frac{\bar{{\Uppsi}}} {n} \approx \frac{{\Upsigma}_{u}} {n}$, $\textit{E}({\bar{\upmu}\bar{\upmu}}^{\textrm{T}}) = \frac{\sum_{i=1}^{N} (T_{i}^{2} {\Upphi}_a )} {(\sum_{i=1}^{N} T_{i})^{2}} = \frac{\sum_{i=1}^{N} T_{i}^{2}} {n^{2}} \bar{{\Upphi}} \approx \frac{\sum_{i=1}^{N} T_{i}^{2}} {n^{2}} {\Upsigma}_{\mu}$, and $\textit{E}({\bar{\upnu}\bar{\upnu}}^{\textrm{T}}) = \frac{\sum_{t=1}^{T} N_{t}^{2}} {n^{2}} {\Upsigma}_{\nu}$.

Hence, the $M \times M$ matrices
\begin{equation}\label{SigmaPsi}
      \begin{array}{l}
            {\hat{\Uppsi}}_a = \dfrac{ \mathrm{W}_{\varepsilon_a} +  \underset{i \in I_a}{\textstyle\sum} \underset{t=1}{\overset{T_{i}}{\textstyle\sum}} \frac{1}{N_t} {\hat{\Upsigma}}_{u}}
            {n_a-N_a},
      \end{array}
\end{equation}
with $\sum_{a=1}^{A} \sum_{i \in I_a} \sum_{t=1}^{T_{i}} \frac{1}{N_t}=T$,  and
\begin{equation}\label{SigmaPhi}
      \begin{array}{l}
            {\hat{\Upphi}}_a = \dfrac{ \mathrm{B}_{\varepsilon_{a}}^{C} + \underset{i \in I_a}{\textstyle\sum} \frac{T_i}{n} \underset{j=1}{\overset{N}{\textstyle\sum}} \frac{T^2_j}{n} {\hat{\Upsigma}}_{\mu} - N_a {\hat{\Uppsi}}_a + \underset{i \in I_a}{\textstyle\sum} \frac{T_i}{n} {\hat{\Upsigma}}_{u}} {\underset{i \in I_a}{\textstyle\sum} T_i}
      \end{array}
\end{equation}
would be unbiased estimators of ${\Uppsi}_a$ and ${\Upphi}_a$ if the ${\upvarepsilon}$'s were known. Both the estimators of ${\Upsigma}_{u}$ and ${\Upsigma}_{\mu}$ and the estimator of ${\Upsigma}_{\nu}$  are derived as in the homoscedastic case:
\begin{equation}
            {\hat{\Upsigma}}_{u} = \dfrac {\mathrm{W}_{\varepsilon}} {n-N-T}, \text{ }
            {\hat{\Upsigma}}_{\mu} = \dfrac {\mathrm{B}_{\varepsilon}^{C} - \left( N-1\right) {\hat{\Upsigma}}_{u}} {n-\underset{i=1}{\overset{N}{\textstyle\sum }} \frac {T_{i}^{2}} {n}}, \text{
            and } {\hat{\Upsigma}}_{\nu} = \dfrac {\mathrm{B}_{\varepsilon}^{T} - \left( T-1\right) {\hat{\Upsigma}}_{u}} {n-\underset{t=1}{\overset{T}{\textstyle\sum }} \frac {N_{t}^{2}} {n}},
\end{equation}
that would be unbiased estimators of ${\Upsigma}_{u}$, ${\Upsigma}_{\mu}$, and ${\Upsigma}_{\nu}$ if the ${\upvarepsilon}$'s were known \citep[see][]{Biorn(2004),PlatoniSckokaiMoro(2012)}.

Again, a simpler heteroscedastic scheme (i.e., heteroscedasticity only on the individual-specific disturbance and on the remainder error) can be obtained combining results for the general scheme with those for the homoscedastic case, although when we consider the case of heteroscedasticity only on the individual-specific disturbance the estimator ${\hat{\Upphi}}_a$ is obtained differently (see equation \eqref{SigmaPhiBis} in Appendix \ref{appTEC_WB}).
%\begin{equation}\label{SigmaPhiBis}
%      \begin{array}{l}
%            {\hat{\Upphi}}_a = \dfrac{\mathrm{B}_{\varepsilon_{a}}^{C} + \underset{i \in \hat{I}_a}{\textstyle\sum} \frac{T_i}{n} \cdot \underset{j=1}{\overset{N}{\textstyle\sum}} \frac{T^2_j}{n} \cdot {\hat{\Upsigma}}_{\mu} - \left( N_a - \underset{i \in \hat{I}_a}{\textstyle\sum} \frac{T_i}{n} \right)\cdot {\hat{\Upsigma}}_{u}} {\underset{i \in \hat{I}_a}{\textstyle\sum} T_i},
%      \end{array}
%\end{equation}
%that would be an unbiased estimator of ${\Upsigma}_{\varphi_a}$ if the ${\upvarepsilon}$'s were known.%\footnote{Alternative computations of the estimators (\ref{SigmaPsi}), (\ref{SigmaPhi}), and (\ref{SigmaPhiBis}) are provided in Appendix A.\ref{app3}.}.

As \citet{Biorn(2004)} suggested, in empirical applications consistent residuals can replace ${\upvarepsilon}$'s in (\ref{EqForAppWB}) to obtain consistent estimates of ${\Uppsi}_a$, ${\Upphi}_a$, and ${\Upsigma}_{\nu}$. Since the \textit{QUE} procedure is based on the \textit{W} residuals, for coherence also in the \textit{WB} procedure we consider the $M \times 1$ residuals $\mathrm{e}_{it} \equiv \mathrm{y}_{it} - \mathrm{X}_{it} \hat{{\upbeta}}^{W}$ from the \textit{W} estimator in (\ref{Westimator}) for the individual $i$ in period $t$, where $\mathrm{X}_{it}$ is a matrix of dimension $M \times \left(K-M\right)$. As above, if we assume that the $M \times K$ matrix $\mathrm{X}_{it}$ in (\ref{eqSUR}) always contains $M$ vectors of ones (a vector of ones for each equation $m$), then we have to
define the $M \times 1$ consistent centered residuals $\mathrm{f}_{it} = \mathrm{e}_{it} - \mathrm{\bar{e}}$, where $\bar{e}_{m} = \frac{\sum_{i=1}^{N} \sum_{t=1}^{T_{i}} e_{mit}} {n} = \frac{\sum_{t=1}^{T} \sum_{i=1}^{N_{t}} e_{mit}} {n}$. Therefore, the $M \times M$ matrices of within individuals, between individuals, and between times (co)variations in the $\mathrm{f}$'s of the different $M$ equations are the following:
\begin{equation}
      \begin{array}{l}
            \mathrm{W}_{f} = \underset{a=1}{\overset{A}{\textstyle\sum}} \mathrm{W}_{f_a} = \underset{a=1} {\overset{A}{\textstyle\sum}} \underset{i \in I_a} {\textstyle\sum} \underset{t=1}
            {\overset{T_{i}}{\textstyle\sum}} \left( \mathrm{f}_{it} - \mathrm{\bar{f}}_{i \centerdot} - \mathrm{\bar{f}}_{\centerdot t} \right) \left( \mathrm{f}_{it} - \mathrm{\bar{f}}_{i \centerdot} - \mathrm{\bar{f}}_{\centerdot t} \right)
            ^{\textrm{T}} , \\
            \mathrm{B}_{f}^{C} =  \underset{a=1} {\overset{A} {\textstyle\sum}} \mathrm{B}_{f_{a}}^{C} = \underset{a=1} {\overset{A} {\textstyle\sum}} \underset{i \in I_a} {\textstyle\sum} T_{i} \left(
            \mathrm{\bar{f}}_{i \centerdot} - \mathrm{\bar{f}} \right) \left( \mathrm{\bar{f}}_{i \centerdot} - \mathrm{\bar{f}} \right) ^{\textrm{T}} ,\\
            \mathrm{B}_{f}^{T} = \underset{t=1} {\overset{T}{\textstyle\sum}} N_{t} \left( \mathrm{\bar{f}}_{\centerdot t} - \mathrm{\bar{f}} \right) \left( \mathrm{\bar{f}}_{\centerdot t} - \mathrm{\bar{f}} \right) ^{\textrm{T}} ,
      \end{array}
\end{equation}
where for each equation $m$ we have $\bar{f}_{m i \centerdot} = \frac{ \sum_{t=1}^{T_{i}} f_{m i t}} {T_{i}}$, $\bar{f}_{m \centerdot t} = \frac{ \sum_{i=1}^{N_{t}} f_{m i t} } {N_{t}}$, and $\bar{f}_{m} = \frac{ \sum_{i=1}^{N} \sum_{t=1}^{T_{i}} f_{mit} } {n} = \frac{ \sum_{i=1}^{N} (T_{i} \bar{f}_{m i \centerdot}) } {n}$ or $\bar{f}_{m} = \frac{ \sum_{t=1}^{T} \sum_{i=1}^{N_{t}} f_{mit} } {n} = \frac{ \sum_{t=1}^{T} ( N_{t} \bar{f}_{m \centerdot t} ) } {n}$. Given that:
\begin{equation}\label{WBcBt}
      \begin{array}{l}
            \textit{E} \left( \mathrm{W}_{f_a} \right) = \left( n_a-N_a \right) {\Uppsi}_{a} - \underset{i \in I_a}{\textstyle\sum} \underset{t \in J_i} {\textstyle\sum} \frac{1}{N_t} \bar{{\Uppsi}} , \\
            \textit{E}\left( \mathrm{B}_{f_{a}}^{C} \right) = \underset{i \in I_a}{\textstyle\sum} T_i {\Upphi}_a - \underset{i \in I_a}{\textstyle\sum} \frac{T_i}{n} \underset{j=1}{\overset{N}{\textstyle\sum}} \frac{T^2_j}{n} \bar{{\Upphi}} + N_a {\Uppsi}_a - \underset{i \in I_a}{\textstyle\sum} \frac{T_i}{n} \bar{{\Uppsi}} , \\
            \textit{E}\left(\mathrm{B}_{f}^{T}\right) = \left( n - \underset{t=1}{\overset{T}{\textstyle\sum}} \frac {N_{t}^{2}} {n} \right) {\Upsigma}_{\nu} + \left(T-1\right) \bar{{\Uppsi}} ,
      \end{array}
\end{equation}
where $J_i$ is the set of periods in which individual $i$ is observed and with $\bar{{\Uppsi}} \approx {\Upsigma}_{u}$ and $\bar{{\Upphi}} \approx {\Upsigma}_{\mu}$, we can conclude that the estimators in (\ref{SigmaPsi}) and (\ref{SigmaPhi}), with $\mathrm{W}_{f_a}$ instead of $\mathrm{W}_{\varepsilon_a}$ and $\mathrm{B}_{f_{a}}^{C}$ instead of $\mathrm{B}_{\varepsilon_{a}}^{C}$ respectively, are consistent estimators of ${\Uppsi}_a$ and ${\Upphi}_a$. As mentioned above, both the consistent estimators of ${\Upsigma}_{u}$ and ${\Upsigma}_{\mu}$ and the consistent estimator of ${\Upsigma}_{\nu}$ are derived as in the homoscedastic case \citep[see][]{Biorn(2004),PlatoniSckokaiMoro(2012)}. Finally, with heteroscedasticity only on the individual-specific disturbance, the expected value $\textit{E}\left(\mathrm{B}_{f_{a}}^{C}\right)$ is given by the equation \eqref{BcBis} in Appendix \ref{appTEC_WB}.

%\begin{equation}\label{BcBis}
%      \begin{array}{l}
%            \textit{E}\left(\mathrm{B}_{f_{a}}^{C}\right) = \underset{i \in I_a}{\textstyle\sum} T_i \cdot {\Upphi}_a - \underset{i \in I_a}{\textstyle\sum} \frac{T_i}{n} \cdot \underset{j=1}{\overset{N}{\textstyle\sum}} \frac{T^2_j}{n} \cdot \bar{{\Upphi}}  + \left( N_a - \underset{i \in I_a}{\textstyle\sum} \frac{T_i}{n} \right) \cdot {\Upsigma}_{u},
%      \end{array}
%\end{equation}
%and therefore the estimator in (\ref{SigmaPhiBis}), with $\mathrm{B}_{f_{a}}^{C}$ instead of $\mathrm{B}_{\varepsilon_{a}}^{C}$, is a consistent estimator of ${\Upphi}_a$.%\footnote{Alternative computations of consistent estimators are provided in Appendix A.\ref{app3}.}.

\subsection{Monte Carlo experiment -- SUR system case}\label{SIMULATION_2}

In order to analyze the performances of the proposed techniques, we develop a simple simulation on a three-equation system ($M=3$). The simulated model is:
\begin{displaymath}
      \begin{array}{llllll}
            y_{1} & =\beta _{10} & + \beta _{11} x_{1} & +\beta _{12} x_{2} & & +\varepsilon _{1}, \\
            y_{2} & =\beta _{20} & + \beta _{21} x_{1} & +\beta _{22} x_{2} & + \beta _{23} x_{3} & +\varepsilon _{2}, \\
            y_{3} & =\beta _{30} &  & + \beta _{32} x_{2} & +\beta _{33} x_{3} & +\varepsilon _{3},
      \end{array}
\end{displaymath}
where ${\upbeta}_{1} = ( 15,6,-3 )^{\textrm{T}}$, ${\upbeta}_{2} = ( 10,-3,8,-2 )^{\textrm{T}}$, and ${\upbeta}_{3} = ( 20,-2,5 )^{\textrm{T}}$.\footnote{Note that the second equation is the same equation of the single-equation case in subsection \ref{SIMULATION_1}.} Then we also allow the cross equations restrictions $\beta_{12}=\beta_{21}$ and $\beta_{23}=\beta_{32}$.

The independent variables' values $x_{kit}$ ($k=1,2,3$) have been generated and the unbalanced panel has been constructed according to the same \textit{DGP} of the single-equation case\footnote{With $N=250$ the numbers of individuals for each group $p$ are $N_{p=1}=54$, $N_{p=2}=43$, $N_{p=3}=34$, $N_{p=4}=27$, $N_{p=5}=22$, $N_{p=6}=18$, $N_{p=7}=14$, $N_{p=8}=11$, $N_{p=9}=9$, $N_{p=10}=7$, $N_{p=11}=6$, and $N_{p=12}=5$; and with $N=500$ they are $N_{p=1}=107$, $N_{p=2}=86$, $N_{p=3}=69$, $N_{p=4}=55$, $N_{p=5}=44$, $N_{p=6}=35$, $N_{p=7}=28$, $N_{p=8}=22$, $N_{p=9}=18$, $N_{p=10}=14$, $N_{p=11}=12$, and $N_{p=12}=10$.}. This should mimic a real world situation of a large unbalanced panel for which the two-way \textit{SUR} system is the appropriate model. Moreover, as in the single-equation case, the experiment is implemented by considering as strata the deciles of the independent variable $x_2$. The homoscedastic time variance-covariance matrix is:
\begin{displaymath}
      \begin{array}{c}
            {\Upsigma}_{\nu} =
            \footnotesize
            \left[
            \renewcommand{\tabcolsep}{0.2pc}
            \begin{tabular}{rrr}
                  6.429 &  0.717 & -1.107 \\
                              &  6.271 &  1.235 \\
                              &              &  9.371
            \end{tabular}
            \right] ,
            \normalsize
      \end{array}
\end{displaymath}
while the heteroscedastic variances-covariances $\varphi_{a,mj}$ and $\psi_{a,mj}$ have been generated from the matrices:
\begin{displaymath}
      \begin{array}{c}
            {\Upsigma}_{\mu} =
            \footnotesize
            \left[
            \renewcommand{\tabcolsep}{0.2pc}
            \begin{tabular}{rrr}
                  9.377 & -1.048 &  1.276 \\
                              &  6.488 &  0.710 \\
                              &              &  6.207
            \end{tabular}
            \right]
            \normalsize
            \text{and  } {\Upsigma}_{u} =
            \footnotesize
            \left[
            \renewcommand{\tabcolsep}{0.2pc}
            \begin{tabular}{rrr}
                  6.544 &  0.738 &  0.881 \\
                              &  6.039 & -1.232 \\
                              &              &  9.489
            \end{tabular}
            \right]
            \normalsize
      \end{array}
\end{displaymath}
with $\varphi_{a,mj}=\sigma_{\mu,mj} (1+\lambda \bar{x}_{2_{a}})^2$ and $\psi_{a,mj}=\sigma_{u,mj} (1+\lambda \bar{x}_{2_{a}})^2$, where $\sigma_{\mu,mj}$ and $\sigma_{u,mj}$ are elements of the matrices ${\Upsigma}_{\mu}$ and ${\Upsigma}_{u}$ respectively and $\bar{x}_{2_{a}}$ is the mean of the independent variable $x_{2}$ over the decile/stratum $a$.\footnote{The correlation among equations verifies the null hypothesis of the \citet{BreuschPagan(1979)} test at $n=1,031$, which is the number of observations when $N=250$.}

The results of a $2000$-run simulation are shown in Tables \ref{Tab_SUR_SE_QUE} and \ref{Tab_SUR_SE_WB}.\footnote{As in \citet{BaltagiGriffin(1988)} and \citet{Phillips(2003)}, negative variance estimates are replaced by zero.}\footnote{Table \ref{Tab_SUR_Var} in Appendix \ref{app_tables} displays the estimated variances-covariances for the stratum $a=5$.}

Tables \ref{Tab_SUR_SE_QUE} and \ref{Tab_SUR_SE_WB} show that, contrary to the single-equation case, the heteroscedastic procedures allow to obtain standard errors lower than those obtained through the homoscedastic procedure in all cases, i.e., not only in the heteroscedastic cases $\lambda=1,2$, but also in the homoscedastic case $\lambda=0$. However, in the homoscedastic case (i.e., with $\lambda=0$) the standard errors computed with the heteroscedastic procedures are very closed to the standard errors computed with the homoscedastic procedure.

Focusing on the heteroscedastic cases (i.e., with $\lambda=1,2$)
\begin{itemize}
      \item the smallest standard errors are obtained when the estimation procedure which considers both kinds of heteroscedasticity is implemented;
      \item though, differently from the single-equation estimation, there is not an evident difference in the loss in efficiency due to the misspecification in the form of heteroscedasticity.
\end{itemize}

Finally, comparing the standard errors obtained with the \textit{QUE} procedure (displayed in Table \ref{Tab_SUR_SE_QUE}) and those obtained with the \textit{WB} procedure (displayed in Table \ref{Tab_SUR_SE_WB}), it is possible to assert that the \textit{QUE} procedure allows to obtained lower standard errors than those obtained with the \textit{WB} procedure.

\newpage

\singlespacing

\begingroup
\renewcommand{\tabcolsep}{0.23pc}\renewcommand{\arraystretch}{0.96}
{\footnotesize
\begin{longtable}{l r rrrr r r rrrr }
\caption{\normalsize Simulation results on two-way \textit{SUR} systems - \textit{QUE} procedure:\\standard errors of the estimated parameters and (average) estimated variances and covariances of the error components\label{Tab_SUR_SE_QUE}}
\vspace*{-0.5em} \\
\toprule
                      & \multicolumn{5}{c}{$N=250$, $T=12$, and $n=1031$} & & \multicolumn{5}{c}{$N=500$, $T=12$, and $n=2062$} \\
\\
                        & \multicolumn{1}{c}{true} & & \multicolumn{3}{c}{heteroscedasticity on} & & \multicolumn{1}{c}{true} & & \multicolumn{3}{c}{heteroscedasticity on}\\
                        & \multicolumn{1}{c}{value} & \multicolumn{1}{c}{homosc.} & \multicolumn{1}{c}{$u_{it}$} & \multicolumn{1}{c}{$\mu_{it}$} & \multicolumn{1}{c}{$\mu_{it}$, $u_{it}$}
                      & & \multicolumn{1}{c}{value} & \multicolumn{1}{c}{homosc.} & \multicolumn{1}{c}{$u_{it}$} & \multicolumn{1}{c}{$\mu_{it}$} & \multicolumn{1}{c}{$\mu_{it}$, $u_{it}$} \\
\\
                        & & \multicolumn{1}{c}{(a)} & \multicolumn{1}{c}{(b)} & \multicolumn{1}{c}{(c)} & \multicolumn{1}{c}{(d)}
                      & & & \multicolumn{1}{c}{(a)} & \multicolumn{1}{c}{(b)} & \multicolumn{1}{c}{(c)} & \multicolumn{1}{c}{(d)}  \\
\\
                      & \multicolumn{11}{c}{\cellcolor{lightgray}$\lambda=0$} \\
\\
$\beta_{10}$          &         &   0.333 &   0.332 &   0.317 &   0.316 & &         &   0.235 &   0.235 &   0.230 &   0.229 \\
$\beta_{11}$          &         &   0.130 &   0.129 &   0.128 &   0.127 & &         &   0.092 &   0.092 &   0.091 &   0.091 \\
$\beta_{12}$          &         &   0.096 &   0.096 &   0.094 &   0.094 & &         &   0.068 &   0.068 &   0.067 &   0.067 \\
$\bar{\varphi}_{1}^2$ &   9.377 &   9.437 &   9.437 &   9.438 &   9.443 & &   9.377 &   9.393 &   9.393 &   9.393 &   9.394 \\
$\bar{\varphi}_{12}$  &  -1.048 &  -1.067 &  -1.067 &  -1.067 &  -1.069 & &  -1.048 &  -1.044 &  -1.044 &  -1.044 &  -1.044 \\
$\bar{\varphi}_{13}$  &   1.276 &   1.260 &   1.260 &   1.260 &   1.258 & &   1.276 &   1.291 &   1.291 &   1.291 &   1.294 \\
\textcolor[rgb]{0.50,0.50,0.50}{$\sigma_{\nu,1}^2$} &   \textcolor[rgb]{0.50,0.50,0.50}{ 6.429} & \multicolumn{4}{c}{\textcolor[rgb]{0.50,0.50,0.50}{ 6.292}}
                                                    & & \textcolor[rgb]{0.50,0.50,0.50}{ 6.429} & \multicolumn{4}{c}{\textcolor[rgb]{0.50,0.50,0.50}{ 6.289}} \\
\textcolor[rgb]{0.50,0.50,0.50}{$\sigma_{\nu,12}$}  &   \textcolor[rgb]{0.50,0.50,0.50}{ 0.717} & \multicolumn{4}{c}{\textcolor[rgb]{0.50,0.50,0.50}{ 0.669}}
                                                    & & \textcolor[rgb]{0.50,0.50,0.50}{ 0.717} & \multicolumn{4}{c}{\textcolor[rgb]{0.50,0.50,0.50}{ 0.670}} \\
\textcolor[rgb]{0.50,0.50,0.50}{$\sigma_{\nu,13}$}  &   \textcolor[rgb]{0.50,0.50,0.50}{-1.107} & \multicolumn{4}{c}{\textcolor[rgb]{0.50,0.50,0.50}{-1.124}}
                                                    & & \textcolor[rgb]{0.50,0.50,0.50}{-1.107} & \multicolumn{4}{c}{\textcolor[rgb]{0.50,0.50,0.50}{-1.113}} \\
$\bar{\psi}_{1}^2$    &   6.544 &   6.539 &   6.539 &   6.539 &   6.539 & &   6.544 &   6.550 &   6.550 &   6.550 &   6.550 \\
$\bar{\psi}_{12}$     &   0.738 &   0.738 &   0.740 &   0.738 &   0.740 & &   0.738 &   0.740 &   0.739 &   0.740 &   0.739 \\
$\bar{\psi}_{13}$     &   0.881 &   0.880 &   0.881 &   0.880 &   0.881 & &   0.881 &   0.885 &   0.883 &   0.885 &   0.883 \\
\\
$\beta_{20}$          &         &   0.316 &   0.315 &   0.302 &   0.301 & &         &   0.224 &   0.223 &   0.219 &   0.218 \\
\textcolor[rgb]{0.50,0.50,0.50}{$\beta_{21}$}   & & \textcolor[rgb]{0.50,0.50,0.50}{0.096} & \textcolor[rgb]{0.50,0.50,0.50}{0.096} & \textcolor[rgb]{0.50,0.50,0.50}{0.094} &
                                                    \textcolor[rgb]{0.50,0.50,0.50}{0.094}
                                              & & & \textcolor[rgb]{0.50,0.50,0.50}{0.068} & \textcolor[rgb]{0.50,0.50,0.50}{0.068} & \textcolor[rgb]{0.50,0.50,0.50}{0.067} &
                                                    \textcolor[rgb]{0.50,0.50,0.50}{0.067} \\
$\beta_{22}$          &         &   0.130 &   0.129 &   0.127 &   0.126 & &         &   0.092 &   0.091 &   0.091 &   0.091 \\
$\beta_{23}$          &         &   0.102 &   0.101 &   0.099 &   0.099 & &         &   0.072 &   0.072 &   0.071 &   0.071 \\
$\bar{\varphi}_{2}^2$ &   6.488 &   6.516 &   6.516 &   6.519 &   6.526 & &   6.488 &   6.521 &   6.521 &   6.521 &   6.524 \\
$\bar{\varphi}_{23}$  &   0.710 &   0.722 &   0.722 &   0.722 &   0.720 & &   0.710 &   0.708 &   0.708 &   0.708 &   0.709 \\
\textcolor[rgb]{0.50,0.50,0.50}{$\sigma_{\nu,2}^2$}   & \textcolor[rgb]{0.50,0.50,0.50}{ 6.271} & \multicolumn{4}{c}{\textcolor[rgb]{0.50,0.50,0.50}{ 6.225}}
                                                    & & \textcolor[rgb]{0.50,0.50,0.50}{ 6.271} & \multicolumn{4}{c}{\textcolor[rgb]{0.50,0.50,0.50}{ 6.234}} \\
\textcolor[rgb]{0.50,0.50,0.50}{$\sigma_{\nu,23}$}    & \textcolor[rgb]{0.50,0.50,0.50}{ 1.235} & \multicolumn{4}{c}{\textcolor[rgb]{0.50,0.50,0.50}{ 1.307}}
                                                    & & \textcolor[rgb]{0.50,0.50,0.50}{ 1.235} & \multicolumn{4}{c}{\textcolor[rgb]{0.50,0.50,0.50}{ 1.317}} \\
$\bar{\psi}_{2}^2$    &   6.039 &   6.044 &   6.043 &   6.044 &   6.043 & &   6.039 &   6.043 &   6.041 &   6.043 &   6.041 \\
$\bar{\psi}_{23}$     &  -1.232 &  -1.238 &  -1.236 &  -1.238 &  -1.236 & &  -1.232 &  -1.228 &  -1.229 &  -1.228 &  -1.229 \\
\\
$\beta_{30}$          &         &   0.340 &   0.338 &   0.324 &   0.323 & &         &   0.241 &   0.240 &   0.235 &   0.235 \\
\textcolor[rgb]{0.50,0.50,0.50}{$\beta_{32}$}   & & \textcolor[rgb]{0.50,0.50,0.50}{0.102} & \textcolor[rgb]{0.50,0.50,0.50}{0.101} & \textcolor[rgb]{0.50,0.50,0.50}{0.099} &
                                                    \textcolor[rgb]{0.50,0.50,0.50}{0.099}
                                              & & & \textcolor[rgb]{0.50,0.50,0.50}{0.072} & \textcolor[rgb]{0.50,0.50,0.50}{0.072} & \textcolor[rgb]{0.50,0.50,0.50}{0.071} & \textcolor[rgb]{0.50,0.50,0.50}{0.071} \\
$\beta_{33}$          &         &   0.148 &   0.147 &   0.145 &   0.145 & &         &   0.105 &   0.105 &   0.104 &   0.104 \\
$\bar{\varphi}_{3}^2$ &   6.207 &   6.189 &   6.189 &   6.204 &   6.223 & &   6.207 &   6.230 &   6.230 &   6.232 &   6.239 \\
\textcolor[rgb]{0.50,0.50,0.50}{$\sigma_{\nu,3}^2$}   & \textcolor[rgb]{0.50,0.50,0.50}{ 9.371} & \multicolumn{4}{c}{\textcolor[rgb]{0.50,0.50,0.50}{ 9.475}}
                                                    & & \textcolor[rgb]{0.50,0.50,0.50}{ 9.371} & \multicolumn{4}{c}{\textcolor[rgb]{0.50,0.50,0.50}{ 9.452}} \\
$\bar{\psi}_{3}^2$    &   9.489 &   9.484 &   9.482 &   9.484 &   9.482 & &   9.489 &   9.484 &   9.484 &   9.484 &   9.484 \\
\\
                      & \multicolumn{11}{c}{\cellcolor{lightgray}$\lambda=1$} \\
\\
$\beta_{10}$          &         &   0.776 &   0.764 &   0.640 &   0.630 & &         &   0.549 &   0.543 &  0.467  &   0.460 \\
$\beta_{11}$          &         &   0.271 &   0.246 &   0.263 &   0.237 & &         &   0.192 &   0.175 &  0.188  &   0.170 \\
$\beta_{12}$          &         &   0.201 &   0.194 &   0.197 &   0.187 & &         &   0.142 &   0.138 &  0.140  &   0.134 \\
$\bar{\varphi}_{1}^2$ &  66.542 &  70.815 &  70.815 &  70.707 &  66.960 & &  66.586 &  70.625 &  70.625 &  70.565 &  66.744 \\
$\bar{\varphi}_{12}$  &  -7.437 &  -7.101 &  -7.101 &  -7.096 &  -7.566 & &  -7.442 &  -6.950 &  -6.950 &  -6.947 &  -7.363 \\
$\bar{\varphi}_{13}$  &   9.055 &   9.408 &   9.408 &   9.392 &   8.847 & &   9.061 &   9.688 &   9.688 &   9.680 &   9.212 \\
\textcolor[rgb]{0.50,0.50,0.50}{$\sigma_{\nu,1}^2$}   & \textcolor[rgb]{0.50,0.50,0.50}{ 6.429} & \multicolumn{4}{c}{\textcolor[rgb]{0.50,0.50,0.50}{ 6.388}}
                                                    & & \textcolor[rgb]{0.50,0.50,0.50}{ 6.429} & \multicolumn{4}{c}{\textcolor[rgb]{0.50,0.50,0.50}{ 6.321}} \\
\textcolor[rgb]{0.50,0.50,0.50}{$\sigma_{\nu,12}$}    & \textcolor[rgb]{0.50,0.50,0.50}{ 0.717} & \multicolumn{4}{c}{\textcolor[rgb]{0.50,0.50,0.50}{ 0.665}}
                                                    & & \textcolor[rgb]{0.50,0.50,0.50}{ 0.717} & \multicolumn{4}{c}{\textcolor[rgb]{0.50,0.50,0.50}{ 0.670}} \\
\textcolor[rgb]{0.50,0.50,0.50}{$\sigma_{\nu,13}$}    & \textcolor[rgb]{0.50,0.50,0.50}{-1.107} & \multicolumn{4}{c}{\textcolor[rgb]{0.50,0.50,0.50}{-1.119}}
                                                    & & \textcolor[rgb]{0.50,0.50,0.50}{-1.107} & \multicolumn{4}{c}{\textcolor[rgb]{0.50,0.50,0.50}{-1.112}} \\
$\bar{\psi}_{1}^2$    &  46.438 &  42.498 &  46.316 &  42.498 &  46.316 & &  46.469 &  42.639 &  46.460 &  42.639 &  46.460 \\
$\bar{\psi}_{12}$     &   5.237 &   4.823 &   5.293 &   4.823 &   5.293 & &   5.241 &   4.825 &   5.241 &   4.825 &   5.241 \\
$\bar{\psi}_{13}$     &   6.252 &   5.716 &   6.259 &   5.716 &   6.259 & &   6.256 &   5.744 &   6.212 &   5.744 &   6.212 \\
\\
                        & & \multicolumn{1}{c}{(a)}    & \multicolumn{1}{c}{(b)} & \multicolumn{1}{c}{(c)} & \multicolumn{1}{c}{(d)}
                      & & & \multicolumn{1}{c}{(a)}    & \multicolumn{1}{c}{(b)} & \multicolumn{1}{c}{(c)} & \multicolumn{1}{c}{(d)}  \\
\\
$\beta_{20}$          &         &   0.714 &   0.711 &   0.581 &   0.583 & &         &   0.505 &   0.505 &   0.422 &   0.425 \\
\textcolor[rgb]{0.50,0.50,0.50}{$\beta_{21}$}   & & \textcolor[rgb]{0.50,0.50,0.50}{0.201} & \textcolor[rgb]{0.50,0.50,0.50}{0.194} & \textcolor[rgb]{0.50,0.50,0.50}{0.197} &
                                                    \textcolor[rgb]{0.50,0.50,0.50}{0.187}
                                              & & & \textcolor[rgb]{0.50,0.50,0.50}{0.142} & \textcolor[rgb]{0.50,0.50,0.50}{0.138} & \textcolor[rgb]{0.50,0.50,0.50}{0.140} &
                                                    \textcolor[rgb]{0.50,0.50,0.50}{0.134} \\
$\beta_{22}$          &         &   0.277 &   0.285 &   0.275 &   0.277 & &         &   0.195 &   0.204 &   0.196 &   0.199 \\
$\beta_{23}$          &         &   0.217 &   0.208 &   0.210 &   0.197 & &         &   0.154 &   0.148 &   0.151 &   0.142 \\
$\bar{\varphi}_{2}^2$ &  46.040 &  49.792 &  49.792 &  49.748 &  46.339 & &  46.071 &  49.784 &  49.784 &  49.755 &  46.263 \\
$\bar{\varphi}_{23}$  &   5.038 &   4.371 &   4.371 &   4.372 &   5.022 & &   5.042 &   4.255 &   4.255 &   4.256 &   4.993 \\
\textcolor[rgb]{0.50,0.50,0.50}{$\sigma_{\nu,2}^2$}   & \textcolor[rgb]{0.50,0.50,0.50}{ 6.271} & \multicolumn{4}{c}{\textcolor[rgb]{0.50,0.50,0.50}{ 6.265}}
                                                    & & \textcolor[rgb]{0.50,0.50,0.50}{ 6.271} & \multicolumn{4}{c}{\textcolor[rgb]{0.50,0.50,0.50}{ 6.259}} \\
\textcolor[rgb]{0.50,0.50,0.50}{$\sigma_{\nu,23}$}    & \textcolor[rgb]{0.50,0.50,0.50}{ 1.235} & \multicolumn{4}{c}{\textcolor[rgb]{0.50,0.50,0.50}{ 1.296}}
                                                    & & \textcolor[rgb]{0.50,0.50,0.50}{ 1.235} & \multicolumn{4}{c}{\textcolor[rgb]{0.50,0.50,0.50}{ 1.303}} \\
$\bar{\psi}_{2}^2$    &  42.854 &  39.282 &  42.750 &  39.282 &  42.750 & &  42.883 &  39.324 &  42.809 &  39.324 &  42.809 \\
$\bar{\psi}_{23}$     &  -8.743 &  -8.040 &  -8.690 &  -8.040 &  -8.690 & &  -8.748 &  -7.988 &  -8.725 &  -7.988 &  -8.725 \\
\\
$\beta_{30}$          &         &   0.763 &   0.743 &   0.637 &   0.619 & &         &   0.541 &   0.528 &   0.463 &   0.453 \\
\textcolor[rgb]{0.50,0.50,0.50}{$\beta_{32}$}   & & \textcolor[rgb]{0.50,0.50,0.50}{0.217} & \textcolor[rgb]{0.50,0.50,0.50}{0.208} & \textcolor[rgb]{0.50,0.50,0.50}{0.210} &
                                                    \textcolor[rgb]{0.50,0.50,0.50}{0.197}
                                              & & & \textcolor[rgb]{0.50,0.50,0.50}{0.154} & \textcolor[rgb]{0.50,0.50,0.50}{0.148} & \textcolor[rgb]{0.50,0.50,0.50}{0.151} &
                                                    \textcolor[rgb]{0.50,0.50,0.50}{0.142} \\
$\beta_{33}$          &         &   0.316 &   0.290 &   0.304 &   0.274 & &         &   0.224 &   0.207 &   0.217 &   0.197 \\
$\bar{\varphi}_{3}^2$ &  44.046 &  49.617 &  49.617 &  49.771 &  44.476 & &  44.076 &  49.871 &  49.871 &  50.013 &  44.386 \\
\textcolor[rgb]{0.50,0.50,0.50}{$\sigma_{\nu,3}^2$}   & \textcolor[rgb]{0.50,0.50,0.50}{ 9.371} & \multicolumn{4}{c}{\textcolor[rgb]{0.50,0.50,0.50}{ 9.615}}
                                                    & & \textcolor[rgb]{0.50,0.50,0.50}{ 9.371} & \multicolumn{4}{c}{\textcolor[rgb]{0.50,0.50,0.50}{ 9.515}} \\
$\bar{\psi}_{3}^2$    &  67.336 &  61.614 &  67.067 &  61.614 &  67.067 & &  67.382 &  61.741 &  67.287 &  61.741 &  67.287 \\
\\
                      & \multicolumn{11}{c}{\cellcolor{lightgray}$\lambda=2$} \\
\\
$\beta_{10}$          &         &   1.251 &   1.212 &   0.962 &   0.915 & &         &   0.885 &   0.860 &   0.696 &   0.669 \\
$\beta_{11}$          &         &   0.425 &   0.356 &   0.407 &   0.338 & &         &   0.301 &   0.253 &   0.290 &   0.242 \\
$\beta_{12}$          &         &   0.314 &   0.288 &   0.305 &   0.272 & &         &   0.222 &   0.205 &   0.217 &   0.195 \\
$\bar{\varphi}_{1}^2$ & 179.619 & 194.064 & 194.064 & 193.921 & 180.741 & & 179.760 & 193.620 & 193.620 & 193.610 & 180.231 \\
$\bar{\varphi}_{12}$  & -20.075 & -18.830 & -18.830 & -18.811 & -20.426 & & -20.091 & -18.425 & -18.425 & -18.416 & -19.852 \\
$\bar{\varphi}_{13}$  &  24.442 &  25.750 &  25.750 &  25.692 &  23.824 & &  24.461 &  26.554 &  26.554 &  26.524 &  24.898 \\
\textcolor[rgb]{0.50,0.50,0.50}{$\sigma_{\nu,1}^2$}   & \textcolor[rgb]{0.50,0.50,0.50}{ 6.429} & \multicolumn{4}{c}{\textcolor[rgb]{0.50,0.50,0.50}{ 6.589}}
                                                    & & \textcolor[rgb]{0.50,0.50,0.50}{ 6.429} & \multicolumn{4}{c}{\textcolor[rgb]{0.50,0.50,0.50}{ 6.402}} \\
\textcolor[rgb]{0.50,0.50,0.50}{$\sigma_{\nu,12}$}    & \textcolor[rgb]{0.50,0.50,0.50}{ 0.717} & \multicolumn{4}{c}{\textcolor[rgb]{0.50,0.50,0.50}{ 0.660}}
                                                    & & \textcolor[rgb]{0.50,0.50,0.50}{ 0.717} & \multicolumn{4}{c}{\textcolor[rgb]{0.50,0.50,0.50}{ 0.666}} \\
\textcolor[rgb]{0.50,0.50,0.50}{$\sigma_{\nu,13}$}    & \textcolor[rgb]{0.50,0.50,0.50}{-1.107} & \multicolumn{4}{c}{\textcolor[rgb]{0.50,0.50,0.50}{-1.111}}
                                                    & & \textcolor[rgb]{0.50,0.50,0.50}{-1.107} & \multicolumn{4}{c}{\textcolor[rgb]{0.50,0.50,0.50}{-1.111}} \\
$\bar{\psi}_{1}^2$    & 125.352 & 111.789 & 124.962 & 111.789 & 124.962 & & 125.451 & 112.207 & 125.386 & 112.207 & 125.386 \\
$\bar{\psi}_{12}$     &  14.137 &  12.702 &  14.317 &  12.702 &  14.317 & &  14.148 &  12.702 &  14.138 &  12.702 &  14.138 \\
$\bar{\psi}_{13}$     &  16.876 &  15.037 &  16.904 &  15.037 &  16.904 & &  16.889 &  15.098 &  16.723 &  15.098 &  16.723 \\
\\
$\beta_{20}$          &         &   1.144 &   1.125 &   0.879 &   0.845 & &         &   0.809 &   0.799 &   0.632 &   0.617 \\
\textcolor[rgb]{0.50,0.50,0.50}{$\beta_{21}$}   & & \textcolor[rgb]{0.50,0.50,0.50}{0.314} & \textcolor[rgb]{0.50,0.50,0.50}{0.288} & \textcolor[rgb]{0.50,0.50,0.50}{0.305} &
                                                    \textcolor[rgb]{0.50,0.50,0.50}{0.272}
                                              & & & \textcolor[rgb]{0.50,0.50,0.50}{0.222} & \textcolor[rgb]{0.50,0.50,0.50}{0.205} & \textcolor[rgb]{0.50,0.50,0.50}{0.217} &
                                                    \textcolor[rgb]{0.50,0.50,0.50}{0.195} \\
$\beta_{22}$          &         &   0.436 &   0.439 &   0.431 &   0.420 & &         &   0.308 &   0.315 &   0.306 &   0.302 \\
$\beta_{23}$          &         &   0.343 &   0.311 &   0.329 &   0.287 & &         &   0.243 &   0.221 &   0.235 &   0.206 \\
$\bar{\varphi}_{2}^2$ & 124.279 & 137.130 & 137.130 & 137.333 & 125.178 & & 124.377 & 137.055 & 137.055 & 137.330 & 124.894 \\
$\bar{\varphi}_{23}$  &  13.600 &  11.236 &  11.236 &  11.239 &  13.491 & &  13.611 &  10.932 &  10.932 &  10.933 &  13.471 \\
\textcolor[rgb]{0.50,0.50,0.50}{$\sigma_{\nu,2}^2$}   & \textcolor[rgb]{0.50,0.50,0.50}{ 6.271} & \multicolumn{4}{c}{\textcolor[rgb]{0.50,0.50,0.50}{ 6.382}}
                                                    & & \textcolor[rgb]{0.50,0.50,0.50}{ 6.271} & \multicolumn{4}{c}{\textcolor[rgb]{0.50,0.50,0.50}{ 6.323}} \\
\textcolor[rgb]{0.50,0.50,0.50}{$\sigma_{\nu,23}$}    & \textcolor[rgb]{0.50,0.50,0.50}{ 1.235} & \multicolumn{4}{c}{\textcolor[rgb]{0.50,0.50,0.50}{ 1.281}}
                                                    & & \textcolor[rgb]{0.50,0.50,0.50}{ 1.235} & \multicolumn{4}{c}{\textcolor[rgb]{0.50,0.50,0.50}{ 1.283}} \\
$\bar{\psi}_{2}^2$    & 115.678 & 103.328 & 115.296 & 103.328 & 115.296 & & 115.770 & 103.471 & 115.501 & 103.471 & 115.501 \\
$\bar{\psi}_{23}$     & -23.599 & -21.142 & -23.393 & -21.142 & -23.393 & & -23.618 & -21.021 & -23.558 & -21.021 & -23.558 \\
\\
$\beta_{30}$          &         &   1.222 &   1.163 &   0.977 &   0.891 & &         &   0.866 &   0.826 &   0.708 &   0.652 \\
\textcolor[rgb]{0.50,0.50,0.50}{$\beta_{32}$}   & & \textcolor[rgb]{0.50,0.50,0.50}{0.343} & \textcolor[rgb]{0.50,0.50,0.50}{0.311} & \textcolor[rgb]{0.50,0.50,0.50}{0.329} &
                                                    \textcolor[rgb]{0.50,0.50,0.50}{0.287}
                                              & & & \textcolor[rgb]{0.50,0.50,0.50}{0.243} & \textcolor[rgb]{0.50,0.50,0.50}{0.221} & \textcolor[rgb]{0.50,0.50,0.50}{0.235} &
                                                    \textcolor[rgb]{0.50,0.50,0.50}{0.206} \\
$\beta_{33}$          &         &   0.499 &   0.425 &   0.474 &   0.391 & &         &   0.354 &   0.302 &   0.339 &   0.280 \\
$\bar{\varphi}_{3}^2$ & 118.897 & 138.192 & 138.192 & 139.479 & 120.334 & & 118.990 & 138.836 & 138.836 & 140.218 & 119.892 \\
\textcolor[rgb]{0.50,0.50,0.50}{$\sigma_{\nu,3}^2$}   & \textcolor[rgb]{0.50,0.50,0.50}{ 9.371} & \multicolumn{4}{c}{\textcolor[rgb]{0.50,0.50,0.50}{ 9.843}}
                                                    & & \textcolor[rgb]{0.50,0.50,0.50}{ 9.371} & \multicolumn{4}{c}{\textcolor[rgb]{0.50,0.50,0.50}{ 9.617}} \\
$\bar{\psi}_{3}^2$    & 181.764 & 162.059 & 180.876 & 162.059 & 180.876 & & 181.907 & 162.482 & 181.607 & 162.482 & 181.607\\
\bottomrule
\end{longtable}}
\vspace*{-0.5em}
\begin{tablenotes}[normal,flushleft]
      \item \hspace{-0.25em}\scriptsize\textit{Note}: Parameters estimation based on (a) the estimated homoscedastic $\text{var}$s-$\text{Cov}$s $\hat{\sigma}_{\nu,mj}$, $\hat{\sigma}_{\mu,mj}$, $\hat{\sigma}_{u,mj}$; (b) the estimated homoscedastic $\text{var}$s-$\text{Cov}$s $\hat{\sigma}_{\nu,mj}$ and $\hat{\sigma}_{\mu,mj}$ and heteroscedastic $\text{var}$s-$\text{Cov}$s $\hat{\psi}_{a,mj}$, whose the average value is $\hat{\psi}_{mj}$; (c) the estimated homoscedastic $\text{var}$s-$\text{Cov}$s $\hat{\sigma}_{\nu,mj}$ and $\hat{\sigma}_{u,mj}$ and heteroscedastic $\text{var}$s-$\text{Cov}$s $\hat{\varphi}_{a,mj} (\hat{\sigma}_{u,mj})$, whose the average value is $\hat{\varphi}_{mj}$; (d) the estimated homoscedastic $\text{var}$s-$\text{Cov}$s $\hat{\sigma}_{\nu,mj}$ and heteroscedastic $\text{var}$s-$\text{Cov}$s $\hat{\psi}_{a,mj}$ and $\hat{\varphi}_{a,mj}(\hat{\psi}_{a,mj})$.
\end{tablenotes}
\bigskip
\endgroup

%\onehalfspacing

\newpage

\singlespacing

\begingroup
\renewcommand{\tabcolsep}{0.23pc}\renewcommand{\arraystretch}{0.96}
{\footnotesize
\begin{longtable}{l r rrrr r r rrrr }
\caption{\normalsize Simulation results on two-way \textit{SUR} systems - \textit{WB} procedure:\\standard errors of the estimated parameters and (average) estimated variances and covariances of the error components\label{Tab_SUR_SE_WB}}
\vspace*{-0.5em}
\\
\toprule
                      & \multicolumn{5}{c}{$N=250$, $T=12$, and $n=1031$} & & \multicolumn{5}{c}{$N=500$, $T=12$, and $n=2062$} \\
\\
                      & \multicolumn{1}{c}{true} & & \multicolumn{3}{c}{heteroscedasticity on} & & \multicolumn{1}{c}{true} & & \multicolumn{3}{c}{heteroscedasticity on}\\
                        & \multicolumn{1}{c}{value} & \multicolumn{1}{c}{homosc.} & \multicolumn{1}{c}{$u_{it}$} & \multicolumn{1}{c}{$\mu_{it}$} & \multicolumn{1}{c}{$\mu_{it}$, $u_{it}$}
                      & & \multicolumn{1}{c}{value} & \multicolumn{1}{c}{homosc.} & \multicolumn{1}{c}{$u_{it}$} & \multicolumn{1}{c}{$\mu_{it}$} & \multicolumn{1}{c}{$\mu_{it}$, $u_{it}$} \\
\\
                        & & \multicolumn{1}{c}{(a)}     & \multicolumn{1}{c}{(b)} & \multicolumn{1}{c}{(c)} & \multicolumn{1}{c}{(d)}
                      & & & \multicolumn{1}{c}{(a)}     & \multicolumn{1}{c}{(b)} & \multicolumn{1}{c}{(c)} & \multicolumn{1}{c}{(d)}  \\
\\
                      & \multicolumn{11}{c}{\cellcolor{lightgray}$\lambda=0$} \\
\\
$\beta_{10}$          &         &   0.348 &   0.349 &   0.334 &   0.334 & &         &   0.246 &   0.247 &   0.241 &   0.241 \\
$\beta_{11}$          &         &   0.137 &   0.137 &   0.135 &   0.135 & &         &   0.096 &   0.097 &   0.096 &   0.096 \\
$\beta_{12}$          &         &   0.101 &   0.103 &   0.100 &   0.101 & &         &   0.071 &   0.073 &   0.071 &   0.072 \\
$\bar{\varphi}_{1}^2$ &   9.377 &  10.111 &  10.111 &  10.061 &   9.655 & &   9.377 &  10.072 &  10.072 &  10.047 &   9.641 \\
$\bar{\varphi}_{12}$  &  -1.048 &  -0.990 &  -0.990 &  -0.985 &  -1.059 & &  -1.048 &  -0.968 &  -0.968 &  -0.966 &  -1.024 \\
$\bar{\varphi}_{13}$  &   1.276 &   1.135 &   1.135 &   1.130 &   1.233 & &   1.276 &   1.170 &   1.170 &   1.167 &   1.266 \\
\textcolor[rgb]{0.50,0.50,0.50}{$\sigma_{\nu,1}^2$}   & \textcolor[rgb]{0.50,0.50,0.50}{ 6.429} & \multicolumn{4}{c}{\textcolor[rgb]{0.50,0.50,0.50}{ 6.333}}
                                                    & & \textcolor[rgb]{0.50,0.50,0.50}{ 6.429} & \multicolumn{4}{c}{\textcolor[rgb]{0.50,0.50,0.50}{ 6.310}} \\
\textcolor[rgb]{0.50,0.50,0.50}{$\sigma_{\nu,12}$}    & \textcolor[rgb]{0.50,0.50,0.50}{ 0.717} & \multicolumn{4}{c}{\textcolor[rgb]{0.50,0.50,0.50}{ 0.661}}
                                                    & & \textcolor[rgb]{0.50,0.50,0.50}{ 0.717} & \multicolumn{4}{c}{\textcolor[rgb]{0.50,0.50,0.50}{ 0.666}} \\
\textcolor[rgb]{0.50,0.50,0.50}{$\sigma_{\nu,13}$}    & \textcolor[rgb]{0.50,0.50,0.50}{-1.107} & \multicolumn{4}{c}{\textcolor[rgb]{0.50,0.50,0.50}{-1.113}}
                                                    & & \textcolor[rgb]{0.50,0.50,0.50}{-1.107} & \multicolumn{4}{c}{\textcolor[rgb]{0.50,0.50,0.50}{-1.107}} \\
$\bar{\psi}_{1}^2$    &   6.544 &   7.980 &   8.627 &   7.980 &   8.627 & &   6.544 &   7.925 &   8.497 &   7.925 &   8.497 \\
$\bar{\psi}_{12}$     &   0.738 &   0.870 &   0.944 &   0.870 &   0.944 & &   0.738 &   0.871 &   0.930 &   0.871 &   0.930 \\
$\bar{\psi}_{13}$     &   0.881 &   0.655 &   0.553 &   0.655 &   0.553 & &   0.881 &   0.653 &   0.555 &   0.653 &   0.555 \\
\\
$\beta_{20}$          &         &   0.333 &   0.335 &   0.321 &   0.322 & &         &   0.235 &   0.237 &   0.231 &   0.232 \\
\textcolor[rgb]{0.50,0.50,0.50}{$\beta_{21}$}   & & \textcolor[rgb]{0.50,0.50,0.50}{0.101} & \textcolor[rgb]{0.50,0.50,0.50}{0.103} & \textcolor[rgb]{0.50,0.50,0.50}{0.100} &
                                                    \textcolor[rgb]{0.50,0.50,0.50}{0.101}
                                              & & & \textcolor[rgb]{0.50,0.50,0.50}{0.071} & \textcolor[rgb]{0.50,0.50,0.50}{0.073} & \textcolor[rgb]{0.50,0.50,0.50}{0.071} &
                                                    \textcolor[rgb]{0.50,0.50,0.50}{0.072} \\
$\beta_{22}$          &         &   0.137 &   0.141 &   0.136 &   0.139 & &         &   0.096 &   0.100 &   0.097 &   0.098 \\
$\beta_{23}$          &         &   0.107 &   0.109 &   0.106 &   0.107 & &         &   0.075 &   0.077 &   0.075 &   0.076 \\
$\bar{\varphi}_{2}^2$ &   6.488 &   7.196 &   7.196 &   7.160 &   6.815 & &   6.488 &   7.197 &   7.197 &   7.179 &   6.826 \\
$\bar{\varphi}_{23}$  &   0.710 &   0.858 &   0.858 &   0.854 &   0.726 & &   0.710 &   0.846 &   0.846 &   0.844 &   0.727 \\
\textcolor[rgb]{0.50,0.50,0.50}{$\sigma_{\nu,2}^2$}   & \textcolor[rgb]{0.50,0.50,0.50}{ 6.271} & \multicolumn{4}{c}{\textcolor[rgb]{0.50,0.50,0.50}{ 6.249}}
                                                    & & \textcolor[rgb]{0.50,0.50,0.50}{ 6.271} & \multicolumn{4}{c}{\textcolor[rgb]{0.50,0.50,0.50}{ 6.246}} \\
\textcolor[rgb]{0.50,0.50,0.50}{$\sigma_{\nu,23}$}    & \textcolor[rgb]{0.50,0.50,0.50}{ 1.235} & \multicolumn{4}{c}{\textcolor[rgb]{0.50,0.50,0.50}{ 1.308}}
                                                    & & \textcolor[rgb]{0.50,0.50,0.50}{ 1.235} & \multicolumn{4}{c}{\textcolor[rgb]{0.50,0.50,0.50}{ 1.318}} \\
$\bar{\psi}_{2}^2$    &   6.039 &   7.454 &   8.092 &   7.454 &   8.092 & &   6.039 &   7.414 &   7.979 &   7.414 &   7.979 \\
$\bar{\psi}_{23}$     &  -1.232 &  -0.949 &  -0.821 &  -0.949 &  -0.821 & &  -1.232 &  -0.937 &  -0.819 &  -0.937 &  -0.819 \\
\\
$\beta_{30}$          &         &   0.362 &   0.363 &   0.348 &   0.348 & &         &   0.256 &   0.257 &   0.250 &   0.251 \\
\textcolor[rgb]{0.50,0.50,0.50}{$\beta_{32}$}   & & \textcolor[rgb]{0.50,0.50,0.50}{0.107} & \textcolor[rgb]{0.50,0.50,0.50}{0.109} & \textcolor[rgb]{0.50,0.50,0.50}{0.106} &
                                                    \textcolor[rgb]{0.50,0.50,0.50}{0.107}
                                              & & & \textcolor[rgb]{0.50,0.50,0.50}{0.075} & \textcolor[rgb]{0.50,0.50,0.50}{0.077} & \textcolor[rgb]{0.50,0.50,0.50}{0.075} &
                                                    \textcolor[rgb]{0.50,0.50,0.50}{0.076} \\
$\beta_{33}$          &         &   0.156 &   0.158 &   0.155 &   0.155 & &         &   0.110 &   0.112 &   0.110 &   0.111 \\
$\bar{\varphi}_{3}^2$ &   6.207 &   7.220 &   7.220 &   7.185 &   6.746 & &  6.207  &   7.256 &   7.256 &   7.238 &   6.797 \\
\textcolor[rgb]{0.50,0.50,0.50}{$\sigma_{\nu,3}^2$}   & \textcolor[rgb]{0.50,0.50,0.50}{ 9.371} & \multicolumn{4}{c}{\textcolor[rgb]{0.50,0.50,0.50}{ 9.488}}
                                                    & & \textcolor[rgb]{0.50,0.50,0.50}{ 9.371} & \multicolumn{4}{c}{\textcolor[rgb]{0.50,0.50,0.50}{ 9.459}} \\
$\bar{\psi}_{3}^2$    &   9.489 &  11.586 &  12.565 &  11.586 &  12.565 & &  9.489  &  11.522 &  12.377 &  11.522 &  12.377 \\
\\
                      & \multicolumn{11}{c}{\cellcolor{lightgray}$\lambda=1$} \\
\\
$\beta_{10}$          &         &   0.787 &   0.778 &   0.659 &   0.649 & &         &   0.555 &   0.550 &   0.476 &   0.470 \\
$\beta_{11}$          &         &   0.278 &   0.256 &   0.271 &   0.247 & &         &   0.195 &   0.180 &   0.192 &   0.175 \\
$\beta_{12}$          &         &   0.205 &   0.200 &   0.202 &   0.193 & &         &   0.144 &   0.142 &   0.143 &   0.138 \\
$\bar{\varphi}_{1}^2$ &  66.542 &  71.431 &  71.431 &  71.083 &  66.640 & &  66.586 &  71.274 &  71.274 &  71.097 &  66.671 \\
$\bar{\varphi}_{12}$  &  -7.437 &  -7.005 &  -7.005 &  -6.970 &  -7.518 & &  -7.442 &  -6.863 &  -6.863 &  -6.846 &  -7.319 \\
$\bar{\varphi}_{13}$  &   9.055 &   9.269 &   9.269 &   9.223 &   8.770 & &   9.061 &   9.562 &   9.562 &   9.538 &   9.162 \\
\textcolor[rgb]{0.50,0.50,0.50}{$\sigma_{\nu,1}^2$}   & \textcolor[rgb]{0.50,0.50,0.50}{ 6.429} & \multicolumn{4}{c}{\textcolor[rgb]{0.50,0.50,0.50}{ 6.797}}
                                                    & & \textcolor[rgb]{0.50,0.50,0.50}{ 6.429} & \multicolumn{4}{c}{\textcolor[rgb]{0.50,0.50,0.50}{ 6.527}} \\
\textcolor[rgb]{0.50,0.50,0.50}{$\sigma_{\nu,12}$}    & \textcolor[rgb]{0.50,0.50,0.50}{ 0.717} & \multicolumn{4}{c}{\textcolor[rgb]{0.50,0.50,0.50}{ 0.621}}
                                                    & & \textcolor[rgb]{0.50,0.50,0.50}{ 0.717} & \multicolumn{4}{c}{\textcolor[rgb]{0.50,0.50,0.50}{ 0.648}} \\
\textcolor[rgb]{0.50,0.50,0.50}{$\sigma_{\nu,13}$}    & \textcolor[rgb]{0.50,0.50,0.50}{-1.107} & \multicolumn{4}{c}{\textcolor[rgb]{0.50,0.50,0.50}{-1.060}}
                                                    & & \textcolor[rgb]{0.50,0.50,0.50}{-1.107} & \multicolumn{4}{c}{\textcolor[rgb]{0.50,0.50,0.50}{-1.081}} \\
$\bar{\psi}_{1}^2$    &  46.438 &  44.624 &  49.196 &  44.624 &  49.196 & &  46.469 &  44.346 &  48.779 &  44.346 &  48.779 \\
$\bar{\psi}_{12}$     &   5.237 &   4.917 &   5.465 &   4.917 &   5.465 & &   5.241 &   4.933 &   5.407 &   4.933 &   5.407 \\
$\bar{\psi}_{13}$     &   6.252 &   5.579 &   6.031 &   5.579 &   6.031 & &   6.256 &   5.546 &   5.923 &   5.546 &   5.923 \\
\\
                        & & \multicolumn{1}{c}{(a)}     & \multicolumn{1}{c}{(b)} & \multicolumn{1}{c}{(c)} & \multicolumn{1}{c}{(d)}
                      & & & \multicolumn{1}{c}{(a)}     & \multicolumn{1}{c}{(b)} & \multicolumn{1}{c}{(c)} & \multicolumn{1}{c}{(d)}  \\
\\
$\beta_{20}$          &         &   0.725 &   0.724 &   0.601 &   0.603 & &         &   0.512 &   0.512 &   0.433 &   0.436 \\
\textcolor[rgb]{0.50,0.50,0.50}{$\beta_{21}$}   & & \textcolor[rgb]{0.50,0.50,0.50}{0.205} & \textcolor[rgb]{0.50,0.50,0.50}{0.200} & \textcolor[rgb]{0.50,0.50,0.50}{0.202} &
                                                    \textcolor[rgb]{0.50,0.50,0.50}{0.193}
                                              & & & \textcolor[rgb]{0.50,0.50,0.50}{0.144} & \textcolor[rgb]{0.50,0.50,0.50}{0.142} & \textcolor[rgb]{0.50,0.50,0.50}{0.143} &
                                                    \textcolor[rgb]{0.50,0.50,0.50}{0.138} \\
$\beta_{22}$          &         &   0.282 &   0.294 &   0.282 &   0.285 & &         &   0.198 &   0.208 &   0.200 &   0.203 \\
$\beta_{23}$          &         &   0.221 &   0.214 &   0.216 &   0.204 & &         &   0.156 &   0.152 &   0.154 &   0.145 \\
$\bar{\varphi}_{2}^2$ &  46.040 &  50.510 &  50.510 &  50.283 &  46.234 & &  46.071 &  50.478 &  50.478 &  50.359 &  46.290 \\
$\bar{\varphi}_{23}$  &   5.038 &   4.487 &   4.487 &   4.465 &   5.000 & &   5.042 &   4.382 &   4.382 &   4.371 &   4.992 \\
\textcolor[rgb]{0.50,0.50,0.50}{$\sigma_{\nu,2}^2$}   & \textcolor[rgb]{0.50,0.50,0.50}{ 6.271} & \multicolumn{4}{c}{\textcolor[rgb]{0.50,0.50,0.50}{ 6.549}}
                                                    & & \textcolor[rgb]{0.50,0.50,0.50}{ 6.271} & \multicolumn{4}{c}{\textcolor[rgb]{0.50,0.50,0.50}{ 6.402}} \\
\textcolor[rgb]{0.50,0.50,0.50}{$\sigma_{\nu,23}$}    & \textcolor[rgb]{0.50,0.50,0.50}{ 1.235} & \multicolumn{4}{c}{\textcolor[rgb]{0.50,0.50,0.50}{ 1.319}}
                                                    & & \textcolor[rgb]{0.50,0.50,0.50}{ 1.235} & \multicolumn{4}{c}{\textcolor[rgb]{0.50,0.50,0.50}{ 1.314}} \\
$\bar{\psi}_{2}^2$    &  42.854 &  41.156 &  45.343 &  41.156 &  45.343 & &  42.883 &  40.928 &  45.013 &  40.928 &  45.013 \\
$\bar{\psi}_{23}$     &  -8.743 &  -7.749 &  -8.283 &  -7.749 &  -8.283 & &  -8.748 &  -7.694 &  -8.315 &  -7.694 &  -8.315 \\
\\
$\beta_{30}$          &         &   0.776 &   0.760 &   0.655 &   0.643 & &         &   0.549 &   0.538 &   0.474 &   0.465 \\
\textcolor[rgb]{0.50,0.50,0.50}{$\beta_{32}$}   & & \textcolor[rgb]{0.50,0.50,0.50}{0.221} & \textcolor[rgb]{0.50,0.50,0.50}{0.214} & \textcolor[rgb]{0.50,0.50,0.50}{0.216} &
                                                    \textcolor[rgb]{0.50,0.50,0.50}{0.204}
                                              & & & \textcolor[rgb]{0.50,0.50,0.50}{0.156} & \textcolor[rgb]{0.50,0.50,0.50}{0.152} & \textcolor[rgb]{0.50,0.50,0.50}{0.154} &
                                                    \textcolor[rgb]{0.50,0.50,0.50}{0.145} \\
$\beta_{33}$          &         &   0.322 &   0.299 &   0.310 &   0.284 & &         &   0.227 &   0.211 &   0.221 &   0.202 \\
$\bar{\varphi}_{3}^2$ &  44.046 &  50.658 &  50.658 &  50.567 &  44.414 & &  44.076 &  50.902 &  50.902 &  50.891 &  44.503 \\
\textcolor[rgb]{0.50,0.50,0.50}{$\sigma_{\nu,3}^2$}   & \textcolor[rgb]{0.50,0.50,0.50}{ 9.371} & \multicolumn{4}{c}{\textcolor[rgb]{0.50,0.50,0.50}{ 9.888}}
                                                    & & \textcolor[rgb]{0.50,0.50,0.50}{ 9.371} & \multicolumn{4}{c}{\textcolor[rgb]{0.50,0.50,0.50}{ 9.654}} \\
$\bar{\psi}_{3}^2$    &  67.336 &  64.326 &  70.879 &  64.326 &  70.879 & &  67.382 &  64.079 &  70.526 &  64.079 &  70.526 \\
\\
                      & \multicolumn{11}{c}{\cellcolor{lightgray}$\lambda=2$} \\
\\
$\beta_{10}$          &         &   1.262 &   1.229 &   0.982 &   0.947 & &         &   0.890 &   0.868 &   0.705 &   0.682 \\
$\beta_{11}$          &         &   0.433 &   0.372 &   0.415 &   0.354 & &         &   0.305 &   0.261 &   0.294 &   0.249 \\
$\beta_{12}$          &         &   0.320 &   0.298 &   0.310 &   0.283 & &         &   0.225 &   0.210 &   0.220 &   0.200 \\
$\bar{\varphi}_{1}^2$ & 179.619 & 194.554 & 194.554 & 193.781 & 179.814 & & 179.760 & 194.205 & 194.205 & 193.865 & 179.872 \\
$\bar{\varphi}_{12}$  & -20.075 & -18.695 & -18.695 & -18.601 & -20.302 & & -20.091 & -18.318 & -18.318 & -18.272 & -19.761 \\
$\bar{\varphi}_{13}$  &  24.442 &  25.581 &  25.581 &  25.453 &  23.654 & &  24.461 &  26.415 &  26.415 &  26.349 &  24.807 \\
\textcolor[rgb]{0.50,0.50,0.50}{$\sigma_{\nu,1}^2$}   & \textcolor[rgb]{0.50,0.50,0.50}{ 6.429} & \multicolumn{4}{c}{\textcolor[rgb]{0.50,0.50,0.50}{ 7.736}}
                                                    & & \textcolor[rgb]{0.50,0.50,0.50}{ 6.429} & \multicolumn{4}{c}{\textcolor[rgb]{0.50,0.50,0.50}{ 6.981}} \\
\textcolor[rgb]{0.50,0.50,0.50}{$\sigma_{\nu,12}$}    & \textcolor[rgb]{0.50,0.50,0.50}{ 0.717} & \multicolumn{4}{c}{\textcolor[rgb]{0.50,0.50,0.50}{ 0.546}}
                                                    & & \textcolor[rgb]{0.50,0.50,0.50}{ 0.717} & \multicolumn{4}{c}{\textcolor[rgb]{0.50,0.50,0.50}{ 0.610}} \\
\textcolor[rgb]{0.50,0.50,0.50}{$\sigma_{\nu,13}$}    & \textcolor[rgb]{0.50,0.50,0.50}{-1.107} & \multicolumn{4}{c}{\textcolor[rgb]{0.50,0.50,0.50}{-0.954}}
                                                    & & \textcolor[rgb]{0.50,0.50,0.50}{-1.107} & \multicolumn{4}{c}{\textcolor[rgb]{0.50,0.50,0.50}{-1.029}} \\
$\bar{\psi}_{1}^2$    & 125.352 & 115.308 & 129.427 & 115.308 & 129.427 & & 125.451 & 114.597 & 128.463 & 114.597 & 128.463 \\
$\bar{\psi}_{12}$     &  14.137 &  12.720 &  14.420 &  12.720 &  14.420 & &  14.148 &  12.765 &  14.254 &  12.765 &  14.254 \\
$\bar{\psi}_{13}$     &  16.876 &  15.073 &  16.871 &  15.073 &  16.871 & &  16.889 &  14.974 &  16.516 &  14.974 &  16.516 \\
\\
$\beta_{20}$          &         &   1.156 &   1.141 &   0.895 &   0.878 & &         &   0.815 &   0.807 &   0.639 &   0.631 \\
\textcolor[rgb]{0.50,0.50,0.50}{$\beta_{21}$}   & & \textcolor[rgb]{0.50,0.50,0.50}{0.320} & \textcolor[rgb]{0.50,0.50,0.50}{0.298} & \textcolor[rgb]{0.50,0.50,0.50}{0.310} & \textcolor[rgb]{0.50,0.50,0.50}{0.283}
                                              & & & \textcolor[rgb]{0.50,0.50,0.50}{0.225} & \textcolor[rgb]{0.50,0.50,0.50}{0.210} & \textcolor[rgb]{0.50,0.50,0.50}{0.220} &
\textcolor[rgb]{0.50,0.50,0.50}{0.200} \\
$\beta_{22}$          &         &   0.442 &   0.451 &   0.438 &   0.432 & &         &   0.311 &   0.320 &   0.309 &   0.307 \\
$\beta_{23}$          &         &   0.347 &   0.320 &   0.334 &   0.297 & &         &   0.245 &   0.226 &   0.237 &   0.211 \\
$\bar{\varphi}_{2}^2$ & 124.279 & 137.910 & 137.910 & 137.607 & 124.799 & & 124.377 & 137.779 & 137.779 & 137.768 & 124.798 \\
$\bar{\varphi}_{23}$  &  13.600 &  11.314 &  11.314 &  11.258 &  13.410 & &  13.611 &  11.036 &  11.036 &  11.009 &  13.432 \\
\textcolor[rgb]{0.50,0.50,0.50}{$\sigma_{\nu,2}^2$}   & \textcolor[rgb]{0.50,0.50,0.50}{ 6.271} & \multicolumn{4}{c}{\textcolor[rgb]{0.50,0.50,0.50}{ 7.192}}
                                                    & & \textcolor[rgb]{0.50,0.50,0.50}{ 6.271} & \multicolumn{4}{c}{\textcolor[rgb]{0.50,0.50,0.50}{ 6.731}} \\
\textcolor[rgb]{0.50,0.50,0.50}{$\sigma_{\nu,23}$}    & \textcolor[rgb]{0.50,0.50,0.50}{ 1.235} & \multicolumn{4}{c}{\textcolor[rgb]{0.50,0.50,0.50}{ 1.345}}
                                                    & & \textcolor[rgb]{0.50,0.50,0.50}{ 1.235} & \multicolumn{4}{c}{\textcolor[rgb]{0.50,0.50,0.50}{ 1.314}} \\
$\bar{\psi}_{2}^2$    & 115.678 & 106.165 & 118.996 & 106.165 & 118.996 & & 115.770 & 105.556 & 118.242 & 105.556 & 118.242 \\
$\bar{\psi}_{23}$     & -23.599 & -20.845 & -22.997 & -20.845 & -22.997 & & -23.618 & -20.721 & -23.144 & -20.721 & -23.144 \\
\\
$\beta_{30}$          &         &   1.234 &   1.182 &   0.992 &   0.927 & &         &   0.873 &   0.836 &   0.715 &   0.669 \\
\textcolor[rgb]{0.50,0.50,0.50}{$\beta_{32}$}   & & \textcolor[rgb]{0.50,0.50,0.50}{0.347} & \textcolor[rgb]{0.50,0.50,0.50}{0.320} & \textcolor[rgb]{0.50,0.50,0.50}{0.334} & \textcolor[rgb]{0.50,0.50,0.50}{0.297}
                                              & & & \textcolor[rgb]{0.50,0.50,0.50}{0.245} & \textcolor[rgb]{0.50,0.50,0.50}{0.226} & \textcolor[rgb]{0.50,0.50,0.50}{0.237} &
\textcolor[rgb]{0.50,0.50,0.50}{0.211} \\
$\beta_{33}$          &         &   0.505 &   0.438 &   0.480 &   0.406 & &         &   0.357 &   0.308 &   0.342 &   0.287 \\
$\bar{\varphi}_{3}^2$ & 118.897 & 139.232 & 139.232 & 139.931 & 119.912 & & 118.990 & 139.865 & 139.865 & 140.889 & 119.846 \\
\textcolor[rgb]{0.50,0.50,0.50}{$\sigma_{\nu,3}^2$}   & \textcolor[rgb]{0.50,0.50,0.50}{ 9.371}  & \multicolumn{4}{c}{\textcolor[rgb]{0.50,0.50,0.50}{10.647}}
                                                    & & \textcolor[rgb]{0.50,0.50,0.50}{ 9.371}  & \multicolumn{4}{c}{\textcolor[rgb]{0.50,0.50,0.50}{10.025}} \\
$\bar{\psi}_{3}^2$    & 181.764 & 166.000 & 186.120 & 166.000 & 186.120 & & 181.907 & 165.423 & 185.528 & 165.423 & 185.528 \\
\bottomrule
\end{longtable}}
\vspace*{-0.5em}
\begin{tablenotes}[normal,flushleft]
      \item\hspace{-0.25em}\scriptsize\textit{Note}: Parameters estimation based on (a) the estimated homoscedastic $\text{var}$s-$\text{Cov}$s $\hat{\sigma}_{\nu,mj}$, $\hat{\sigma}_{\mu,mj}$, $\hat{\sigma}_{u,mj}$; (b) the estimated homoscedastic $\text{var}$s-$\text{Cov}$s $\hat{\sigma}_{\nu,mj}$ and $\hat{\sigma}_{\mu,mj}$ and heteroscedastic $\text{var}$s-$\text{Cov}$s $\hat{\psi}_{a,mj}$, whose the average value is $\hat{\psi}_{mj}$; (c) the estimated homoscedastic $\text{var}$s-$\text{Cov}$s $\hat{\sigma}_{\nu,mj}$ and $\hat{\sigma}_{u,mj}$ and heteroscedastic $\text{var}$s-$\text{Cov}$s $\hat{\varphi}_{a,mj} (\hat{\sigma}_{u,mj})$, whose the average value is $\hat{\varphi}_{mj}$; (d) the estimated homoscedastic $\text{var}$s-$\text{Cov}$s $\hat{\sigma}_{\nu,mj}$ and heteroscedastic $\text{var}$s-$\text{Cov}$s $\hat{\psi}_{a,mj}$ and $\hat{\varphi}_{a,mj}(\hat{\psi}_{a,mj})$.
\end{tablenotes}
\bigskip
\endgroup

%\onehalfspacing

\newpage

As Table \ref{Tab_RE_SE}, Table \ref{Tab_RE_SUR} displays the ratios of the \textit{MSE} of the estimators under consideration to the \textit{MSE} of the true \textit{GLS} estimator, i.e. it displays the measures of relative efficiency of the different estimators.

\singlespacing

\begin{table}[!ht]
\renewcommand{\tabcolsep}{0.29pc}\renewcommand{\arraystretch}{1}
\caption{Relative efficiency of two-way \textit{SUR} systems\label{Tab_RE_SUR}}
\vspace{-0.5em}
{\footnotesize
\begin{tabular}{ll cccc c cccc }
\toprule
      &      & \multicolumn{4}{c }{$N=250$, $T=12$, and $n=1031$}          & & \multicolumn{4}{c}{$N=500$, $T=12$, and $n=2062$}            \\
\\
            &       &                  & \multicolumn{3}{c }{heteroscedasticity on} & &                  & \multicolumn{3}{c }{heteroscedasticity on} \\
            &       & homoscedasticity & $u_{it}$ & $\mu_i$ & $u_{it}$, $\mu_i$     & & homoscedasticity & $u_{it}$ & $\mu_i$ & $u_{it}$, $\mu_i$     \\
\\
& & \multicolumn{9}{c}{\cellcolor{lightgray}\textit{QUE} procedure} \\
\\
$\lambda=0$ & $y_1$ &           1.0001 & 1.0001 & 1.0003 & 1.0011   & & 1.0000 & 1.0000 & 1.0001 & 1.0001  \\
            & $y_2$ &           1.0001 & 1.0001 & 1.0004 & 1.0014   & & 1.0000 & 1.0001 & 1.0001 & 1.0001  \\
            & $y_3$ &           1.0000 & 1.0001 & 1.0003 & 1.0010   & & 1.0000 & 1.0000 & 1.0001 & 1.0001  \\
\\
$\lambda=1$ & $y_1$ &           0.9995 & 0.9996 & 1.0041 & 1.0001   & & 0.9997 & 0.9998 & 1.0003 & 1.0000  \\
            & $y_2$ &           0.9996 & 0.9997 & 1.0070 & 1.0005   & & 0.9998 & 0.9998 & 1.0012 & 1.0001  \\
            & $y_3$ &           0.9993 & 0.9997 & 1.0037 & 1.0002   & & 0.9997 & 0.9998 & 1.0008 & 1.0000  \\
\\
$\lambda=2$ & $y_1$ &           0.9990 & 0.9992 & 1.0022 & 0.9999   & & 0.9995 & 0.9996 & 1.0004 & 1.0000  \\
            & $y_2$ &           0.9992 & 0.9994 & 1.0025 & 1.0001   & & 0.9996 & 0.9996 & 1.0010 & 1.0000  \\
            & $y_3$ &           0.9989 & 0.9994 & 1.0018 & 1.0000   & & 0.9995 & 0.9997 & 1.0004 & 1.0000  \\
\\
& & \multicolumn{9}{c}{\cellcolor{lightgray}\textit{WB} procedure} \\
\\
$\lambda=0$ & $y_1$ &           1.0000 & 1.0001 & 1.0002 & 1.0002   & & 1.0000 & 1.0000 & 1.0001 & 1.0001 \\
            & $y_2$ &           1.0001 & 1.0001 & 1.0003 & 1.0003   & & 1.0000 & 1.0001 & 1.0001 & 1.0001 \\
            & $y_3$ &           1.0001 & 1.0001 & 1.0002 & 1.0003   & & 1.0000 & 1.0001 & 1.0001 & 1.0001 \\
\\
$\lambda=1$ & $y_1$ &           0.9994 & 0.9995 & 1.0008 & 0.9999   & & 0.9997 & 0.9997 & 0.9999 & 1.0000 \\
            & $y_2$ &           0.9996 & 0.9996 & 1.0016 & 1.0002   & & 0.9998 & 0.9998 & 1.0007 & 1.0000 \\
            & $y_3$ &           0.9993 & 0.9996 & 1.0007 & 1.0001   & & 0.9997 & 0.9998 & 1.0004 & 1.0000 \\
\\
$\lambda=2$ & $y_1$ &           0.9990 & 0.9991 & 1.0012 & 0.9999   & & 0.9995 & 0.9995 & 1.0002 & 0.9999 \\
            & $y_2$ &           0.9992 & 0.9992 & 1.0024 & 1.0007   & & 0.9996 & 0.9996 & 1.0009 & 1.0000 \\
            & $y_3$ &           0.9998 & 0.9993 & 1.0012 & 0.9999   & & 0.9995 & 0.9996 & 1.0003 & 0.9999 \\
\bottomrule
\end{tabular}}
\vspace{0.5em}
\begin{tablenotes}[normal,flushleft]
      \item \hspace{-0.25em}\scriptsize\textit{Note}: Relative efficiency is defined as the ratio of the \emph{MSE} of the estimator under consideration to the \emph{MSE} of the true \textit{GLS} estimator (computed considering the true $\text{var}$s-$\text{Cov}$s $\psi_{a,mj}$, $\varphi_{a,mj}$, and $\sigma_{\nu,mj}$). Note that values of the ratio both larger and smaller than $1$ indicate a loss in efficiency: if the ratio is larger than $1$, then the absolute value of the composite error term $\varepsilon_{mit}=\mu_{mi}+\nu_{mt}+u_{mit}$ is larger than the true value; and if the ratio is smaller than $1$, then the absolute value of the composite error term $\varepsilon_{mit}$ is smaller than the true value.
\end{tablenotes}
\end{table}

%\onehalfspacing

This table highlights that, as expected, with $\lambda=0$ the most efficient procedure is the homoscedastic one, whereas with $\lambda=1,2$ the most efficient procedure is the one that considers both the remainder error and the individual-specific effect heteroscedastic. Note also that if the heteroscedasticity is low (i.e., with $\lambda=1$) the \emph{WB} procedure is more efficient than the \emph{QUE} procedure, whereas if the heteroscedasticity is high (i.e., with $\lambda=2$) the \emph{QUE} procedure is more efficient than the \emph{WB} procedure.

\section{Conclusion}\label{CONCLUSIONS}

The use of panel data is becoming very popular in applied econometrics, since large data sets including many individuals observed for several periods are increasingly accessible and manageable. Most of these data sets are unbalanced panels, since very often not all the individuals are observed over the whole time period. In estimating single-equation or system of equations \textit{EC} models on these data, the heteroscedasticity problem may be very common, especially when individuals differ in size.

In this paper, we have derived suitable \textit{EC} model estimators for heteroscedastic two-way single equations and \textit{SUR} systems (with cross-equations restrictions) on unbalanced panel data. Our simulations show that such estimators substantially improve estimation efficiency as compared to the case where heteroscedasticity is not taken into account, especially when both the individual-specific and remainder error components are heteroscedastic.

\begin{appendices}

\titleformat{\section}{\sc\center}{\sc{Appendix }\thesection:}{1em}{}
\titleformat{\subsection}{\it\raggedright}{\normalfont\thesection.\thesubsection}{1em}{}
%\titleformat{\subsubsection}{\it\raggedright}{}{0em}{}
\renewcommand{\thesection}{\Alph{section}}
\renewcommand{\thesubsection}{\arabic{subsection}}

\section{Fixed effects estimation assumptions}\label{appFE}

In the FE estimation the following assumptions are made\footnote{Details on the assumptions FE.1 and FE.2 can be found in Appendix A of \citet{PlatoniSckokaiMoro(2012)}.}.
\begin{description}[font=\normalfont\sffamily]
      \item [\sc fe.1] {\sc Strict exogeneity} The set of $\left(k-1\right) T_{i}$ explanatory variables for each individual $\mathrm{x}_{i \circ }\equiv \left( \mathrm{x}_{i1}, \mathrm{x}_{i2} , \ldots , \mathrm{x}_{iT_{i}} \right)$ is uncorrelated with the idiosyncratic error $u_{it}$ and the set of $\left(k-1\right) N_{t}$ explanatory variables in each time period $\mathrm{x}_{\circ t  } \equiv \left( \mathrm{x}_{1t},\mathrm{x}_{2t} , \ldots , \mathrm{x}_{N_{t}t} \right)$ is also uncorrelated with the same idiosyncratic error $u_{it}$:
          \begin{equation*}
                \begin{array}{r}
                        \textit{E} \left( u_{it} \left\vert \mathrm{x} \right. ,\mu_{i} , \nu_{t} \right)
                      = \textit{E} \left( u_{it} \left\vert \mathrm{x}_{i \circ } \right. , \mu_{i} , \nu_{t} \right)
                      = \textit{E} \left( u_{it} \left\vert \mathrm{x}_{\circ t} \right. , \mu_{i} , \nu_{t} \right) = 0,
                \end{array}
          \end{equation*}
          with $\mathrm{x} \equiv \left( \mathrm{x}_{11} , \ldots , \mathrm{x}_{1T_{1}},\mathrm{x}_{21}, \ldots ,\mathrm{x}_{2T_{2}} , \ldots , \mathrm{x}_{N1}, \ldots ,\mathrm{x}_{NT_{N}} \right)$.
      \item [\sc fe.2] {\sc Consistency} The \textit{W} estimator in (\ref{Westimator}) is asymptotically well behaved, in the sense that the \textquotedblleft adjusted\textquotedblright\ $\left(k-1\right) \times \left(k-1\right)$ outer product matrix $\mathrm{X}^{\textrm{T}} \mathrm{Q}_{\left[ \Updelta \right]} \mathrm{X}$ has the appropriate rank:
          \begin{equation*}
                \begin{array}{l}
                      \text{rank} \left( \mathrm{X}^{\textrm{T}} \mathrm{Q}_{ \left[ \Updelta \right] } \mathrm{X} \right) = k - 1.
                \end{array}
          \end{equation*}
      \item [\sc fe.3] {\sc No serial correlation} For each stratum $a$ the conditional variance-covariance matrix of the idiosyncratic error terms $u_{it}$ coincides with the unconditional one, and it is characterized by constant variances and zero covariances:
               \begin{equation*}
                \begin{array}{l}
                      \textit{E}\left(\mathrm{u}_a\mathrm{u}^{\textrm{T}}_a \left\vert \mathrm{x}_a,\mu_{i\left(a\right)}, \nu_t \right.\right) = \psi_{a}^2 \mathrm{I}_{n_a}.
                \end{array}
               \end{equation*}
          Hence given the ${A\times 1}$ vector ${\uppsi} = (\psi_{1}^2,\psi_{2}^2,\dots,\psi_{A}^2)^{\textrm{T}}$ we can define the $n\times n$ matrix ${\Uppsi} = \textrm{diag} \left({\Updelta}_\mu {\Updelta}_{AN}^{\textrm{T}} {\uppsi} \right)$ and the conditional variance-covariance matrix of $u_{it}$ is
               \begin{equation*}
                \begin{array}{l}
                      \textit{E}\left(\mathrm{u}\mathrm{u}^{\textrm{T}} \left\vert \mathrm{x},\mu_i, \nu_t \right.\right) = {\Uppsi}.
                \end{array}
               \end{equation*}
\end{description}

\section{Random effects estimation assumptions}\label{appRE}

In the RE estimation the following assumptions are made\footnote{Details on the assumptions RE.1 and RE.2 can be found in Appendix B of \citet{PlatoniSckokaiMoro(2012)}.}.
\begin{description}[font=\normalfont\sffamily]
      \item [\sc re.1.a] {\sc Strict exogeneity} The set of $k T_{i}$ explanatory variables for each individual
          $\mathrm{x}_{i\circ}\equiv \left( \mathrm{x}_{i1}, \mathrm{x}_{i2} , \ldots , \mathrm{x}_{iT_{i}} \right)$ is uncorrelated with the idiosyncratic error $u_{it}$ and the set of $k N_{t}$ explanatory
          variables in each time period $\mathrm{x}_{\circ t } \equiv \left( \mathrm{x}_{1t},\mathrm{x}_{2t} , \ldots , \mathrm{x}_{N_{t}t} \right)$ is also uncorrelated with the same idiosyncratic error $u_{it}$:
          \begin{equation*}
                \begin{array}{r}
                        \textit{E} \left( u_{it} \left\vert \mathrm{x} \right. ,\mu_{i} , \nu_{t} \right)
                      = \textit{E} \left( u_{it} \left\vert \mathrm{x}_{i\circ} \right. , \mu_{i} , \nu_{t} \right)
                      = \textit{E} \left( u_{it} \left\vert \mathrm{x}_{\circ t} \right. , \mu_{i} , \nu_{t} \right) = 0,
                \end{array}
          \end{equation*}
          with $\mathrm{x} \equiv \left( \mathrm{x}_{11} , \ldots , \mathrm{x}_{1T_{1}},\mathrm{x}_{21}, \ldots ,\mathrm{x}_{2T_{2}} , \ldots , \mathrm{x}_{N1}, \ldots ,\mathrm{x}_{NT_{N}} \right)$.
      \item [\sc re.1.b and re.1.c] {\sc Orthogonality conditions} Both $\mu_{i}$ and $\nu_{t}$ are orthogonal to the corresponding sets of explanatory variables, that is the $k T_{i}$ explanatory variables for each individual $\mathrm{x}_{i\circ}$ and the $k N_{t}$ explanatory variables in each time period $\mathrm{x}_{\circ t}$:
          \begin{equation*}
                \begin{array}{l}
                      \textit{E} \left( \mu_{i} \left\vert \mathrm{x}_{i\circ} \right. \right) = \textit{E} \left( \mu_{i} \right) = 0 \text{ and }
                      \textit{E} \left( \nu_{t} \left\vert \mathrm{x}_{\circ t} \right. \right) = \textit{E} \left( \nu_{t} \right) = 0.
                \end{array}
          \end{equation*}
      \item [\sc re.2] {\sc Rank condition} The $k \times k$ weighted outer product matrix $ \mathrm{X}^{\textrm{T}} {\Upomega}^{-1} \mathrm{X}$ has the appropriate rank, ensuring the \textit{GLS} estimator in (\ref{GLSestimator}) is consistent:
          \begin{equation*}
                \begin{array}{l}
                      \text{rank} \left( \mathrm{X}^{\textrm{T}} {\Upomega}^{-1} \mathrm{X} \right) = k.
                \end{array}
          \end{equation*}
      \item [\sc re.3] {\sc No serial correlation} For each stratum $a$ the conditional variance-covariance matrix of the idiosyncratic error terms $u_{it}$ is characterized by constant variances and zero covariances; in addition, whereas the variance of the time-specific effect $\nu_t$ is constant across strata, the variance of the individual-specific effect $\mu_i$ is constant within each stratum $a$:
          \begin{itemize}
          \item[a.] $\textit{E}\left(\mathrm{u}_a\mathrm{u}^{\textrm{T}}_a \left\vert \mathrm{x}_a,\mu_{i\left(a\right)}, \nu_t \right.\right) = \psi_{a}^2 \mathrm{I}_{n_a}$,
          \item[b.] $\textit{E}\left(\mu_{i\left(a\right)}^2\left\vert \mathrm{x}_{i\left(a\right)}\right.\right) = \varphi_{a}^2$,
          \item[c.] $\textit{E}\left(\nu_t^2\left\vert \mathrm{x}_t\right.\right) = \sigma^2_{\nu}$.
        \end{itemize}
\end{description}

%\newpage

\section{Alternative robust standard errors}\label{appFEROB}

Let us re-index the individuals belonging to stratum $a$ as $i_a=1_a,\ldots,N_a$, so that $T_{i_a}$ refers to the number of times the individual $i$ of the stratum $a$ is observed.

Since $u_{it} \sim \left( 0 , \psi^2_{a} \right)$, it is possible to obtain robust standard errors also by stacking the observations for each stratum $a$, and then by writing:
\begin{equation}
      \begin{array}{rl}
            \widetilde{\mathrm{y}}_{a} = & \hspace{-0.5em}
            \left[ \textrm{diag}\left(\mathrm{E}_{T_{i_a}}\right) - \left( \mathrm{E}_{T_{1_a}}\mathrm{D}_{1_a}, \dots, \mathrm{E}_{T_{N_a}} \mathrm{D}_{N_a} \right)^{\textrm{T}} \mathrm{Q}^{-} \left( \mathrm{D}^{\textrm{T}}_{1_a} \mathrm{E}_{T_{1_a}} , \dots, \mathrm{D}^{\textrm{T}}_{N_a} \mathrm{E}_{T_{N_a}} \right) \right] \\ & \mathrm{y}_{a } , \\
            \widetilde{\mathrm{X}}_{a} = & \hspace{-0.5em}
            \left[ \textrm{diag}\left(\mathrm{E}_{T_{i_a}}\right) - \left( \mathrm{E}_{T_{1_a}}\mathrm{D}_{1_a}, \dots, \mathrm{E}_{T_{N_a}} \mathrm{D}_{N_a} \right)^{\textrm{T}} \mathrm{Q}^{-} \left( \mathrm{D}^{\textrm{T}}_{1_a} \mathrm{E}_{T_{1_a}} , \dots, \mathrm{D}^{\textrm{T}}_{N_a} \mathrm{E}_{T_{N_a}} \right) \right] \\ & \mathrm{X}_{a}.
      \end{array}
\end{equation}
Therefore, we can compute the $N_a \times 1$ vector $\widetilde{\mathrm{e}}_{a} = \widetilde{\mathrm{y}}_{a} - \widetilde{\mathrm{X}}_{a} \hat{{\upbeta}}^{W}$ and the robust asymptotic variance-covariance matrix of $\hat{{\upbeta}}^{W}$ is estimated by:
\begin{equation}
      \begin{array}{l}
            \text{var}\left(\hat{{\upbeta}}^{W}\right) =
            \left( \mathrm{X}^{\textrm{T}} \mathrm{Q}_{\Updelta} \mathrm{X} \right)^{-1}
            \underset{a=1}{\overset{A}{\textstyle\sum}} \Bigl( \widetilde{\mathrm{X}}^{\textrm{T}}_{a} \widetilde{\mathrm{e}}_{a}
            \widetilde{\mathrm{e}}^{\textrm{T}}_{a} \widetilde{\mathrm{X}}_{a} \Bigr)
            \left( \mathrm{X}^{\textrm{T}} \mathrm{Q}_{\Updelta} \mathrm{X} \right)^{-1} .
      \end{array}
\end{equation}

\section{Technical appendix on QUE procedures}\label{appTEC_QUE}

\subsection{Adapted \textit{QUE}s in (\ref{q_QUE})}\label{app2}

The identities in (\ref{q_QUE}) can be further detailed as:
\begin{equation}\label{q_QUE_details}
      %\scriptsize
      \begin{array}{rl}
            q_{n_a} \equiv & \hspace{-0.5em} \left[\mathrm{f}_a^{\textrm{T}} -
                                        \bar{\mathrm{f}}_{N \centerdot}^{\textrm{T}} {\Updelta}_{\mu_{a}}^{\textrm{T}} - \left( \bar{\mathrm{f}}_{\centerdot T}^{\textrm{T}} {\Updelta}_{T} - \bar{\mathrm{f}}_{N
                                        \centerdot}^{\textrm{T}} {\Updelta}_{TN}^{\textrm{T}} \right) \mathrm{Q}^{-} \left( {\Updelta}_{\nu_{a}} - {\Updelta}_{\mu_{a}} {\Updelta}_{N}^{-1} {\Updelta}_{TN}^{\textrm{T}} \right)^{\textrm{T}} \right] \\
                           & \hspace{-0.5em}  \left[\mathrm{f}_a - {\Updelta}_{\mu_{a}} \bar{\mathrm{f}}_{N \centerdot} - \left( {\Updelta}_{\nu_{a}} - {\Updelta}_{\mu_{a}} {\Updelta}_{N}^{-1} {\Updelta}_{TN}^{\textrm{T}} \right) \mathrm{Q}^{-} \left( \bar{\mathrm{f}}_{\centerdot T}^{\textrm{T}} {\Updelta}_{T} - \bar{\mathrm{f}}_{N \centerdot}^{\textrm{T}} {\Updelta}_{TN}^{\textrm{T}} \right)^{\textrm{T}} \right] , \\
            q_{n} \equiv   & \hspace{-0.5em}  \underset{1\times n}{\mathrm{f}^{\textrm{T}}}\underset{n\times 1}{\mathrm{f}} -
                                        \underset{1\times N}{\bar{\mathrm{f}}_{N \centerdot}^{\textrm{T}}} \underset{N\times N}{{\Updelta}_{N}} \underset{N\times 1}{\bar{\mathrm{f}}_{N \centerdot}}
                           \\ & - \left( \underset{1\times T}{\bar{\mathrm{f}}_{\centerdot T}^{\textrm{T}}} \underset{T\times T}{{\Updelta}_{T}} - \underset{1\times N}{\bar{\mathrm{f}}_{N \centerdot}^{\textrm{T}}}
                                        \underset{N\times T}{{\Updelta}_{TN}^{\textrm{T}}} \right) \underset{T \times T}{\mathrm{Q}^{-}} \left( \underset{1\times T}{\bar{\mathrm{f}}_{\centerdot T}^{\textrm{T}}} \underset{T\times T}{{\Updelta}_{T}} - \underset{1\times N}{\bar{\mathrm{f}}_{N \centerdot}^{\textrm{T}}} \underset{N\times T}{{\Updelta}_{TN}^{\textrm{T}}}  \right)^{\textrm{T}} , \\
            q_{N_a} \equiv & \hspace{-0.5em}  \underset{i \in I_a} {\textstyle\sum} T_{i} \bar{f}_{i \centerdot}^{2} , \\
            q_{N} \equiv   & \hspace{-0.5em}  \underset{1\times N}{\bar{\mathrm{f}}_{N \centerdot}^{\textrm{T}}}
                                        \underset{N\times N}{{\Updelta}_{N}} \underset{N\times 1}{\bar{\mathrm{f}}_{N \centerdot}} =
                                        \underset{i=1} {\overset{N} {\textstyle\sum}} T_i \bar{f}_{i \centerdot}^{2}  = \underset{a=1} {\overset{A} {\textstyle\sum}} \underset{i \in I_a} {\textstyle\sum}  T_{i}
                                        \bar{f}_{i \centerdot}^{2} , \\
            q_{T} \equiv   & \hspace{-0.5em}  \underset{1\times T}{\bar{\mathrm{f}}_{\centerdot T}^{\textrm{T}}} \underset{T\times T} {{\Updelta}_{T}}
                                        \underset{T\times 1}{\bar{\mathrm{f}}_{\centerdot T}} = \underset{t=1} {\overset{T} {\textstyle\sum}} N_t \bar{f}_{\centerdot t}^{2} ,
      \end{array}
\end{equation}
where the elements of the $N\times 1$ matrix $\bar{\mathrm{f}}_{N \centerdot}$ are $\bar{f}_{i \centerdot} = \frac{\sum_{t=1}^{T_{i}} f_{it}} {T_{i}}$, the elements of the $T\times 1$ matrix $\bar{\mathrm{f}}_{\centerdot T}$ are $\bar{f}_{\centerdot t} = \frac{ \sum_{i=1}^{N_{t}} f_{it}} {N_{t}}$, ${\Updelta}_{\mu_{a}} = \mathrm{H}_{a} {\Updelta}_{\mu}$, and ${\Updelta}_{\nu_{a}} = \mathrm{H}_{a} {\Updelta}_{\nu}$.

%\subsection{Technical appendix on QUE procedures}\label{appTEC_QUE}

\subsection{Expected values in the single-equation case}

Referring to the identities in \eqref{q_QUE}, and considering the $n \times n$ matrix $\mathrm{M} \equiv \mathrm{I}_n - \mathrm{X} (\mathrm{X}^{\textrm{T}} \mathrm{Q}_{\Updelta} \mathrm{X})^{-1} \linebreak \mathrm{X}^{\textrm{T}} \mathrm{Q}_{\Updelta}$ (and then by definition $\mathrm{e}=\mathrm{M}\mathrm{y}=\mathrm{M}{\upvarepsilon}$ and $\mathrm{f}\mathrm{f}^{\textrm{T}} = \mathrm{E}_{n} \mathrm{e} \mathrm{e}^{\textrm{T}} \mathrm{E}_{n} = \mathrm{E}_{n} \mathrm{M} {\Upomega} \mathrm{M}^{\textrm{T}} \mathrm{E}_{n} $), the expected value of $q_{n_a}$ is:
\begin{equation}\label{Eqin}
      \textit{E} \left( q_{n_a} \right)
      = \text{tr} \left( \mathrm{H}_{a} \mathrm{Q}_{\Updelta} \mathrm{E}_{n} \mathrm{M} {\Upomega} \mathrm{M}^{\textrm{T}} \mathrm{E}_{n} \mathrm{Q}_{\Updelta} \mathrm{H}_{a}^{\textrm{T}} \right) = \left( n_a - N_a - \tau_a \right) \psi_{a}^2 - k_{a} \bar{\psi}^2 ,
\end{equation}
where $\tau_a \equiv n_a-N_a-\text{tr} (\mathrm{H}_{a} \mathrm{Q}_{\Updelta} \mathrm{H}_{a}^{\textrm{T}})$, $k_{a} \equiv \text{tr} [ ( \mathrm{X}^{\textrm{T}}\mathrm{Q}_{\Updelta}\mathrm{X})^{-1} \mathrm{X}^{\textrm{T}} \mathrm{Q}_{\Updelta} \mathrm{H}_{a}^{\textrm{T}} \mathrm{H}_{a} \mathrm{Q}_{\Updelta} \mathrm{X} ]$, and $\bar{\psi}^2 \approx \sigma_{u}^{2}$ is obtained by equating $q_{n}$ to its expected value
\citep[see][]{WansbeekKapteyn(1989),Davis(2002)}, that is:
\begin{equation}\label{Eqn}
      \begin{array}{l}
            \textit{E} \left( q_{n} \right) = \left[ n-N-\left(T-1\right)-\left(k-1\right) \right] \sigma_{u}^{2} .
      \end{array}
\end{equation}
Moreover, the expected value of $q_{N_a}$ is:
\begin{equation}
      \begin{split}
            \textit{E} \left( q_{N_a} \right) = & \text{ tr} \left( \bar{\mathrm{J}}_{n_a} \mathrm{H}_{a} \mathrm{E}_{n} \mathrm{M} {\Upomega} \mathrm{M}^{\textrm{T}} \mathrm{E}_{n} \mathrm{H}_{a}^{\textrm{T}} \right) \\
            = & \left( N_a - 2 \frac{n_a}{n} \right) \psi_{a}^2 + \left( k_{N_{a}} - k_{0_{a}} + \frac{n_a}{n} k_{0} + \frac{n_a}{n} \right) \bar{\psi}^2 \\
            & + \left( n_a - 2 \lambda_{\mu_a} \right) \varphi_{a}^2 + \frac{n_a}{n} \lambda_{\mu} \bar{\varphi}^2 + \left( N_a - 2 \lambda_{\nu_a} + \frac{n_a}{n} \lambda_{\nu} \right) \sigma_{\nu}^{2} ,
      \end{split}
\end{equation}
where $\bar{\varphi}^2 \approx \sigma_{\mu}^{2}$ is obtained jointly with $\sigma_{\nu}^{2}$ by equating $q_{N}$ and $q_{T}$ to their expected values, that is:
\begin{equation}
      \begin{array}{rl}
            \textit{E} \left( q_{N} \right) & = \left( N+k_{N}-k_{0} - 1 \right) \sigma_{u}^{2} + \left( n - \lambda_{\mu} \right) \sigma_{\mu}^{2} + \left( N - \lambda_{\nu} \right) \sigma_{\nu}^{2} , \\
            \textit{E} \left( q_{T} \right) & = \left( T+k_{T}-k_{0} - 1 \right) \sigma_{u}^{2} + \left( T - \lambda_{\mu} \right) \sigma_{\mu}^{2} + \left( n - \lambda_{\nu} \right) \sigma_{\nu}^{2} ,
      \end{array}
\end{equation}
with $k_{N} \equiv \text{tr} [( \mathrm{X}^{\textrm{T}}\mathrm{Q}_{\Updelta}\mathrm{X})^{-1} \mathrm{X}^{\textrm{T}} {\Updelta}_{\mu} {\Updelta}_{N} {\Updelta}_{\mu}^{\textrm{T}} \mathrm{X} ]$ and $k_{T} \equiv \text{tr} [( \mathrm{X}^{\textrm{T}}\mathrm{Q}_{\Updelta}\mathrm{X})^{-1} \mathrm{X}^{\textrm{T}} {\Updelta}_{\nu} {\Updelta}_{T} {\Updelta}^{\textrm{T}}_{\nu} \mathrm{X} ]$.

In the case heteroscedasticity is only on the individual-specific disturbance the expected value of $q_{N_a}$ is obtained as follows:
\begin{equation}\label{eq.SEQUE_mu_1}
      \begin{split}
            \textit{E} \left( q_{N_a} \right) = &
            \left( N_a + k_{N_{a}} - k_{0_{a}} + \frac{n_a}{n} k_{0} - \frac{n_a}{n} \right) \sigma_{u}^{2} + \left( n_a - 2 \lambda_{\mu_a} \right) \varphi_{a}^2 + \frac{n_a}{n} \lambda_{\mu} \bar{\varphi}^2 \\
        & + \left( N_a - 2 \lambda_{\nu_a} + \frac{n_a}{n} \lambda_{\nu} \right) \sigma_{\nu}^{2} ,
      \end{split}
\end{equation}
and, therefore,
\begin{equation}\label{eq.SEQUE_mu_2}
      \begin{split}
            \hat{\varphi}_{a}^2 =
            & \dfrac{q_{N_a} - \left( N_a + k_{N_{a}} - k_{0_{a}} + \frac{n_a}{n} k_{0} - \frac{n_a}{n} \right) \hat{\sigma}_{u}^{2} - \frac{n_a}{n} \lambda_{\mu} \hat{\sigma}_{\mu}^{2}}{n_a - 2 \lambda_{\mu_a}} \\
            & + \dfrac{ - \left( N_a - 2 \lambda_{\nu_a} + \frac{n_a}{n} \lambda_{\nu} \right) \hat{\sigma}_{\nu}^{2} }{n_a - 2 \lambda_{\mu_a}} .
      \end{split}
\end{equation}

\subsection{Expected values in the SUR systems case}

Referring to the identities in \eqref{q_QUE_SUR}, and considering the $n \times n$ matrix $\mathrm{M}_m \equiv \mathrm{I}_n - \mathrm{X}_m \linebreak (\mathrm{X}_m^{\textrm{T}}\mathrm{Q}_{\Updelta}\mathrm{X}_m)^{-1} \mathrm{X}_m^{\textrm{T}} \mathrm{Q}_{\Updelta}$ (and then by definition $\mathrm{e}_m = \mathrm{M}_m \mathrm{y}_m = \mathrm{M}_m {\upvarepsilon}_m$ and $\mathrm{f}_m\mathrm{f}_j^{\textrm{T}} = \mathrm{E}_{n} \mathrm{e}_m \mathrm{e}_j^{\textrm{T}} \mathrm{E}_{n} = \mathrm{E}_{n} \mathrm{M}_m {\Upomega}_{mj} \mathrm{M}_j^{\textrm{T}} \mathrm{E}_{n} $), the expected value of $q_{n_a,mj}$ is:
\begin{equation}\label{EqinSUR}
      \begin{split}
            \textit{E} \left( q_{n_a,mj} \right)
            & = \text{tr} \left( \mathrm{H}_{a} \mathrm{Q}_{\Updelta} \mathrm{E}_{n} \mathrm{M}_m {\Upomega}_{mj} \mathrm{M}_j^{\textrm{T}} \mathrm{E}_{n} \mathrm{Q}_{\Updelta}
            \mathrm{H}_{a}^{\textrm{T}} \right) \\
            & = \left( n_a - N_a - \tau_a \right) \psi_{a,mj} - \left( k_{a,m} + k_{a,j} - k_{a,mj} \right) \bar{\psi}_{mj},
      \end{split}
\end{equation}
where $ k_{a,mj} \equiv \text{tr} [ (\mathrm{X}_{m}^{\textrm{T}} \mathrm{Q}_{\Updelta} \mathrm{X}_{m})^{-1} \mathrm{X}_{m}^{\textrm{T}}\mathrm{Q}_{\Updelta}\mathrm{X}_{j} (\mathrm{X}_{j}^{\textrm{T}} \mathrm{Q}_{\Updelta} \mathrm{X}_{j})^{-1} \mathrm{X}_{j}^{\textrm{T}} \mathrm{Q}_{\Updelta} \mathrm{H}_{a}^{\textrm{T}} \mathrm{H}_{a} \mathrm{Q}_{\Updelta} \mathrm{X}_{m} ]$ and $k_{mj} \equiv \text{tr} [ ( \mathrm{X}_{m}^{\textrm{T}} \mathrm{Q}_{\Updelta}\mathrm{X}_{m} )^{-1} \mathrm{X}_{m}^{\textrm{T}} \mathrm{Q}_{\Updelta} \mathrm{X}_{j} ( \mathrm{X}_{j}^{\textrm{T}} \mathrm{Q}_{\Updelta} \mathrm{X}_{j} )^{-1} \mathrm{X}_{j}^{\textrm{T}} \mathrm{Q}_{\Updelta} \mathrm{X}_{m} ]$, and $\bar{\psi}_{mj} \approx \sigma_{u,mj}$ is obtained by equating $q_{n,mj}$ to its expected value \citep[see][]{PlatoniSckokaiMoro(2012)}:
\begin{equation}\label{EqnSUR}
      \begin{array}{l}
            \textit{E} \left( q_{n,mj} \right) = \left[ n-N-\left(T-1\right)-\left(k_m-1\right)-\left(k_j-1\right)+k_{mj} \right] \sigma_{u,mj}.
      \end{array}
\end{equation}
Moreover, the expected value of $q_{N_a,mj}$ is:
\begin{equation}
      \begin{split}
            \textit{E} \left( q_{N_a,mj}\right)
              = & \text{ tr} \left( \bar{\mathrm{J}}_{N_a} \mathrm{H}_{a} \mathrm{E}_{n} \mathrm{M}_m {\Upomega}_{mj} \mathrm{M}_j^{\textrm{T}} \mathrm{E}_{n} \mathrm{H}_{a}^{\textrm{T}} \right) \\
              = & \left( N_a - 2 \frac{n_a}{n} \right) \psi_{a,mj} + \left( k_{N_a,mj} - k_{0_a,mj} + \frac{n_a}{n} k_{0,mj} + \frac{n_a}{n} \right) \bar{\psi}_{mj} \\
              & + \left( n_a - 2 \lambda_{\mu_a} \right) \varphi_{a,mj} + \frac{n_a}{n} \lambda_{\mu} \bar{\varphi}_{mj} + \left( N_a - 2 \lambda_{\nu_a} + \frac{n_a}{n} \lambda_{\nu} \right) \sigma_{\nu,mj} ,
      \end{split}
\end{equation}
where $k_{N_a,mj} \equiv \text{tr} [ ( \mathrm{X}_{m}^{\textrm{T}}\mathrm{Q}_{\Updelta}\mathrm{X}_{m} ) ^{-1} \mathrm{X}_{m}^{\textrm{T}}\mathrm{Q}_{\Updelta} \mathrm{X}_{j} ( \mathrm{X}_{j}^{\textrm{T}}\mathrm{Q}_{\Updelta}\mathrm{X}_{j} ) ^{-1} \mathrm{X}_{a_{j}}^{\textrm{T}} \bar{\mathrm{J}}_{N_a} \mathrm{X}_{a_{m}} ]$, $k_{0_a,mj} \equiv \linebreak \frac{ {\upiota}_{N_a}^{\textrm{T}} \mathrm{X}_{a_m} (\mathrm{X}_{m}^{\textrm{T}}\mathrm{Q}_{\Updelta}\mathrm{X}_{m})^{-1} \mathrm{X}_{m}^{\textrm{T}}\mathrm{Q}_{\Updelta}\mathrm{X}_{j} (\mathrm{X}_{j}^{\textrm{T}}\mathrm{Q}_{\Updelta}\mathrm{X}_{j})^{-1} \mathrm{X}_{j}^{\textrm{T}}{\upiota}_{n} +  {\upiota}_{n}^{\textrm{T}} \mathrm{X}_{m} (\mathrm{X}_{m}^{\textrm{T}}\mathrm{Q}_{\Updelta}\mathrm{X}_{m})^{-1} \mathrm{X}_{m}^{\textrm{T}}\mathrm{Q}_{\Updelta}\mathrm{X}_{j} (\mathrm{X}_{j}^{\textrm{T}}\mathrm{Q}_{\Updelta}\mathrm{X}_{j})^{-1} \mathrm{X}_{a_j}^{\textrm{T}}{\upiota}_{N_a} } {n}  $, $k_{0,mj} \linebreak \equiv \frac{ {\upiota}_{n}^{\textrm{T}} {\mathrm{X}_{m}} { ( \mathrm{X}_{m}^{\textrm{T}} \mathrm{Q}_{\Updelta} \mathrm{X}_{m} ) ^{-1} } \mathrm{X}_{m}^{\textrm{T}} \mathrm{Q}_{\Updelta} \mathrm{X}_{j} { ( \mathrm{X}_{j}^{\textrm{T}} \mathrm{Q}_{\Updelta} \mathrm{X}_{j} ) ^{-1}} \mathrm{X}_{j}^{\textrm{T}} {\upiota}_{n} } {n}$, and $\bar{\varphi}_{mj} \approx \sigma_{\mu,mj}$ is obtained jointly with $\sigma_{\nu,mj}$ by equating $q_{N,mj}$ and $q_{T,mj}$ to their expected values \citep[see][]{PlatoniSckokaiMoro(2012)}:
\begin{equation}
      \begin{array}{rl}
            \textit{E} \left( q_{N,mj} \right) = & \hspace{-0.5em} \left( N+k_{N,mj}-k_{0,mj} - 1 \right) \sigma_{u,mj} + \left( n - \lambda_{\mu} \right) \sigma_{\mu,mj} \\ & + \left( N -
            \lambda_{\nu} \right) \sigma_{\nu,mj} , \\
            \textit{E} \left( q_{T,mj} \right) = & \hspace{-0.5em} \left( T+k_{T,mj}-k_{0,mj} - 1 \right) \sigma_{u,mj} + \left( T - \lambda_{\mu} \right) \sigma_{\mu,mj} \\ & + \left( n -
            \lambda_{\nu} \right) \sigma_{\nu,mj} ,
      \end{array}
\end{equation}
with $k_{N,mj} \equiv \text{tr} [( \mathrm{X}_j^{\textrm{T}}\mathrm{Q}_{\Updelta}\mathrm{X}_j)^{-1} \mathrm{X}_j^{\textrm{T}}\mathrm{Q}_{\Updelta}\mathrm{X}_m ( \mathrm{X}_m^{\textrm{T}}\mathrm{Q}_{\Updelta}\mathrm{X}_m)^{-1} \mathrm{X}_m^{\textrm{T}} {\Updelta}_{\mu} {\Updelta}_{N} {\Updelta}_{\mu}^{\textrm{T}} \mathrm{X}_j ]$ and $k_{T,mj} \equiv \linebreak \text{tr} [ ( \mathrm{X}_j^{\textrm{T}}\mathrm{Q}_{\Updelta}\mathrm{X}_j)^{-1} \mathrm{X}_j^{\textrm{T}}\mathrm{Q}_{\Updelta}\mathrm{X}_m ( \mathrm{X}_m^{\textrm{T}}\mathrm{Q}_{\Updelta}\mathrm{X}_m)^{-1} \mathrm{X}_m^{\textrm{T}} {\Updelta}_{\nu} {\Updelta}_{T} {\Updelta}_{\nu}^{\textrm{T}} \mathrm{X}_j ]$.

In the case heteroscedasticity is only on the individual-specific disturbance, the expected value of $q_{N_a,mj}$ is obtained differently as:
\begin{equation}\label{eq.SURQUE_mu_1}
      \begin{split}
             \textit{E} \left( q_{N_a,mj} \right) = &
             \left( N_a + k_{N_a,mj} - k_{0_a,mj} + \frac{n_a}{n} k_{0,mj} - \frac{n_a}{n} \right) \sigma_{u,mj} \\
         & + \left( n_a - 2 \lambda_{\mu_a} \right) \varphi_{a,mj} + \frac{n_a}{n} \lambda_{\mu} \bar{\varphi}_{mj} + \left( N_a - 2 \lambda_{\nu_a} + \frac{n_a}{n} \lambda_{\nu} \right) \sigma_{\nu,mj}
      \end{split}
\end{equation}
and, therefore,
\begin{equation}\label{eq.SURQUE_mu_2}
      \begin{split}
            \hat{\varphi}_{a,mj} = &
              \dfrac{q_{N_a,mj} - \left( N_a + k_{N_a,mj} - k_{0_a,mj} + \frac{n_a}{n} k_{0,mj} - \frac{n_a}{n} \right) \hat{\sigma}_{u,mj} } {n_a - 2 \lambda_{\mu_a}} \\ &
            + \dfrac{- \frac{n_a}{n} \lambda_{\mu} \hat{\sigma}_{\mu,mj} - \left( N_a - 2 \lambda_{\nu_a} + \frac{n_a}{n} \lambda_{\nu} \right) \hat{\sigma}_{\nu,mj} } {n_a - 2 \lambda_{\mu_a}} .
      \end{split}
\end{equation}

\section{Technical appendix on WB procedure}\label{appTEC_WB}

In case of heteroscedasticity only on the individual-specific disturbance the estimator is:
\begin{equation}\label{SigmaPhiBis}
      \begin{array}{l}
            {\hat{\Upphi}}_a = \dfrac{\mathrm{B}_{\varepsilon_{a}}^{C} + \underset{i \in I_a}{\textstyle\sum} \frac{T_i}{n} \underset{j=1}{\overset{N}{\textstyle\sum}} \frac{T^2_j}{n} {\hat{\Upsigma}}_{\mu} - \left( N_a - \underset{i \in I_a}{\textstyle\sum} \frac{T_i}{n} \right) {\hat{\Upsigma}}_{u}} {\underset{i \in I_a}{\textstyle\sum} T_i},
      \end{array}
\end{equation}
that would be an unbiased estimator of ${\Upphi}_a$ if the ${\upvarepsilon}$'s were known.%\footnote{Alternative computations of the estimators (\ref{SigmaPsi}), (\ref{SigmaPhi}), and (\ref{SigmaPhiBis}) are provided in Appendix A.\ref{app3}.}.

Using the centered residuals from the $W$ estimation, the expected value of the between individuals (co)variations is:
\begin{equation}\label{BcBis}
      \begin{array}{l}
            \textit{E}\left(\mathrm{B}_{f_{a}}^{C}\right) = \underset{i \in I_a}{\textstyle\sum} T_i {\Upphi}_a - \underset{i \in I_a}{\textstyle\sum} \frac{T_i}{n} \underset{j=1}{\overset{N}{\textstyle\sum}} \frac{T^2_j}{n} \bar{{\Upphi}}  + \left( N_a - \underset{i \in I_a}{\textstyle\sum} \frac{T_i}{n} \right) {\Upsigma}_{u},
      \end{array}
\end{equation}
and therefore the estimator in (\ref{SigmaPhiBis}), with $\mathrm{B}_{f_{a}}^{C}$ instead of $\mathrm{B}_{\varepsilon_{a}}^{C}$, is a consistent estimator of ${\Upphi}_a$.%\footnote{Alternative computations of consistent estimators are provided in Appendix A.\ref{app3}.}.

\section{Additional tables} \label{app_tables}

Due to the space limit it would be impossible (and unnecessary) to display $480$ variance-covariance matrices as done in Table \ref{Tab_SE_Var} for the single-equation case. Table \ref{Tab_SUR_Var} displays the estimated variances-covariances for the stratum $a=5$.\\

\singlespacing

\begin{landscape}
\begin{table}
\renewcommand{\tabcolsep}{0.29pc}\renewcommand{\arraystretch}{1}
\caption{Simulation results on two-way \textit{SUR} systems: estimated variances-covariances $\hat{\psi}_{5,mj}$ and $\hat{\varphi}_{5,mj}$\label{Tab_SUR_Var}}
\vspace{-0.5em}
{\footnotesize
\begin{tabular}{r rr rrr rrr  r rr rrr  rrr}
\toprule
     & \multicolumn{8}{c}{$N=250$, $T=12$, and $n=1031$} & \multicolumn{8}{c}{$N=500$, $T=12$, and $n=2062$} \\
\\
       & & & \multicolumn{3}{c}{QUE procedure} & \multicolumn{3}{c}{WB procedure}
     & & & & \multicolumn{3}{c}{QUE procedure} & \multicolumn{3}{c}{WB procedure}  \\
       & \multicolumn{2}{c}{true values} & & \multicolumn{2}{c}{$\hat{\varphi}_{5,mj}$ on} & & \multicolumn{2}{c}{$\hat{\varphi}_{5,mj}$ on}
     & & \multicolumn{2}{c}{true values} & & \multicolumn{2}{c}{$\hat{\varphi}_{5,mj}$ on} & & \multicolumn{2}{c}{$\hat{\varphi}_{5,mj}$ on}  \\
$mj$   & $\psi_{5,mj}$ & $\varphi_{5,mj}$ & $\hat{\psi}_{5,mj}$ & $\hat{\sigma}_{u,mj}$ & $\hat{\psi}_{5,mj}$ & $\hat{\psi}_{5,mj}$ & $\hat{\sigma}_{u,mj}$ & $\hat{\psi}_{5,mj}$
     & & $\psi_{5,mj}$ & $\varphi_{5,mj}$ & $\hat{\psi}_{5,mj}$ & $\hat{\sigma}_{u,mj}$ & $\hat{\psi}_{5,mj}$ & $\hat{\psi}_{5,mj}$ & $\hat{\sigma}_{u,mj}$ & $\hat{\psi}_{5,mj}$  \\
\\
     & \multicolumn{17}{c}{\cellcolor{lightgray}$\lambda=0$} \\
$11$ &  6.544 &  9.377 &  6.547 &  9.373 &  9.372 &  7.338 &  9.625 &  9.735 & &  6.544 &  9.377 &  6.563 &  9.414 &  9.412 &  7.293 &
 9.708 &   9.816 \\
$12$ &  0.738 & -1.048 &  0.741 & -1.079 & -1.079 &  0.805 & -1.031 & -1.019 & &  0.738 & -1.048 &  0.735 & -1.045 & -1.044 &  0.806 &  -1.003 &  -0.992 \\
$13$ &  0.881 &  1.276 &  0.872 &  1.212 &  1.214 &  0.758 &  1.141 &  1.124 & &  0.881 &  1.276 &  0.884 &  1.315 &  1.315 &  0.765 &
 1.249 &   1.230 \\
$22$ &  6.039 &  6.488 &  6.038 &  6.534 &  6.535 &  6.802 &  6.813 &  6.924 & &  6.039 &  6.488 &  6.032 &  6.525 &  6.527 &  6.754 &
 6.829 &   6.942 \\
$23$ & -1.232 &  0.710 & -1.235 &  0.730 &  0.729 & -1.072 &  0.791 &  0.812 & & -1.232 &  0.710 & -1.213 &  0.746 &  0.743 & -1.060 &
 0.810 &   0.831 \\
$33$ &  9.489 &  6.207 &  9.434 &  6.156 &  6.166 & 10.570 &  6.610 &  6.783 & &  9.489 &  6.207 &  9.490 &  6.249 &  6.248 & 10.557 &
 6.725 &   6.890 \\
     & \multicolumn{17}{c}{\cellcolor{lightgray}$\lambda=1$} \\
$11$ & 41.271 &  59.138 &  41.325 &  58.882 &  59.093 &  42.697 &  58.738 &  59.075 & &  41.265 &  59.129 &  41.401 &  59.116 &  59.331 &  42.418 &  59.208 &  59.541 \\
$12$ &  4.654 &  -6.609 &   4.681 &  -6.784 &  -6.756 &   4.713 &  -6.682 &  -6.644 & &   4.654 &  -6.608 &   4.638 &  -6.630 &  -6.597 &   4.684 &   -6.561 &  -6.517 \\
$13$ &  5.556 &   8.047 &   5.499 &   7.606 &   7.644 &   5.454 &   7.480 &   7.504 & &   5.555 &   8.046 &   5.573 &   8.170 &   8.198 &   5.480 &    8.076 &   8.085 \\
$22$ & 38.086 &  40.918 &  38.108 &  40.964 &  41.173 &  39.251 &  41.011 &  41.343 & &  38.081 &  40.912 &  38.048 &  40.855 &  41.077 &  38.962 &  41.042 &  41.382 \\
$23$ & -7.770 &   4.478 &  -7.794 &   4.643 &   4.599 &  -7.609 &   4.663 &   4.638 & &  -7.769 &   4.477 &  -7.655 &   4.683 &   4.624 &  -7.489 &   4.725 &   4.688 \\
$33$ & 59.844 &  39.146 &  59.545 &  38.501 &  38.873 &  61.184 &  38.723 &  39.274 & &  59.836 &  39.140 &  59.857 &  39.049 &  39.377 &  61.157 &  39.405 &  39.910 \\
     & \multicolumn{17}{c}{\cellcolor{lightgray}$\lambda=2$} \\
$11$ & 105.914 & 151.765 & 106.110 & 150.610 & 151.619 & 108.699 & 149.776 & 150.938 & & 105.884 & 151.722 & 106.262 & 151.181 & 152.214 & 107.884 & 150.922 & 152.084 \\
$12$ &  11.944 & -16.962 &  12.021 & -17.445 & -17.319 &  11.993 & -17.245 & -17.112 & &  11.941 & -16.957 &  11.903 & -17.074 & -16.934 &
 11.903 & -16.954 & -16.804 \\
$13$ &  14.259 &  20.652 &  14.120 &  19.452 &  19.614 &  14.219 &  19.232 &  19.384 & &  14.255 &  20.646 &  14.300 &  20.841 &  20.974 &
 14.268 &  20.697 &  20.814 \\
$22$ &  97.740 & 105.007 &  97.849 & 104.667 & 105.633 &  99.808 & 104.325 & 105.437 & &  97.713 & 104.977 &  97.656 & 104.342 & 105.354 &
 98.982 & 104.333 & 105.473 \\
$23$ & -19.940 &  11.491 & -20.012 &  12.023 &  11.821 & -19.786 &  11.966 &  11.779 & & -19.934 &  11.488 & -19.647 &  12.075 &  11.835 &
-19.464 &  12.077 &  11.857 \\
$33$ & 153.578 & 100.459 & 152.893 &  98.094 &  99.722 & 155.594 &  97.926 &  99.757 & & 153.534 & 100.431 & 153.627 &  99.466 & 101.008 &
155.420 &  99.621 & 101.356 \\
\bottomrule
\end{tabular}}
\vspace{0.5em}
\begin{tablenotes}[normal,flushleft]
      \item \hspace{-0.25em}\scriptsize\textit{Note}: $\psi_{5,mj}$ and $\varphi_{5,mj}$ are the true values of the $\text{var}$s-$\text{Cov}$s, $\hat{\psi}_{5,mj}$ are the estimated
      $\text{var}$s-$\text{Cov}$s of the remainder error $u_{it}$, $\hat{\varphi}_{5,mj}$ are the estimated $\text{var}$s-$\text{Cov}$s of
      the individual-specific error $\mu_i$ computed on the basis of a remainder error either homoscedastic $(\hat{\sigma}_{u,mj})$ or
      heteroscedastic $(\hat{\psi}_{5,mj})$.
\end{tablenotes}
\end{table}
\end{landscape}

%\onehalfspacing

\end{appendices}

%\section*{References}


\begin{thebibliography}{7}
\expandafter\ifx\csname natexlab\endcsname\relax\def\natexlab#1{#1}\fi

\bibitem[\protect\citeauthoryear{Akaike}{1974}]{Akaike(1974)} \text{Akaike, H.} (1974), ``A new look at the statistical model identification.'' \textit{IEEE Transactions on Automatic Control}, \text{19} (6), 716-723. %IEEE Trans. Auto. Contr.
\bibitem[\protect\citeauthoryear{Arellano}{1987}]{Arellano(1987)} \text{Arellano, M.} (1987), ``Computing robust standard errors for within groups estimators.'' \textit{Oxford Bulletin of Economics and Statistics}, \text{49} (4), 431-434. %Oxford Bull. Econ. Statist.
\bibitem[\protect\citeauthoryear{Baltagi}{1980}]{Baltagi(1980)} \text{Baltagi, B.~H.} (1980), ``On seemingly unrelated regressions with error components.'' \textit{Econometrica}, \text{48} (6), 1547-1551.
\bibitem[\protect\citeauthoryear{Baltagi}{1981}]{Baltagi(1981)} \text{Baltagi, B.~H.} (1981), ``Pooling: an experimental study of alternative testing and estimation procedures in a two-way error component model.'' \textit{Journal of Econometrics}, \text{17} (1), 21-49. %J. Econometrics
\bibitem[\protect\citeauthoryear{Baltagi}{1985}]{Baltagi(1985)}  \text{Baltagi, B.~H.} (1985), ``Pooling cross-sections with unequal time series lengths.'' \textit{Economics Letters}, \text{18} (2-3), 133-136. % Econ. Letters
\bibitem[\protect\citeauthoryear{Baltagi}{1988}]{Baltagi(1988)} \text{Baltagi, B.~H.} (1988), ``An alternative heteroscedastic error components model (problem 88.2.2.).'' \textit{Econometric Theory}, \text{4} (2), 349-350. %Economet. Theor.
%\bibitem[\protect\citeauthoryear{Baltagi}{2005}]{Baltagi(2005)} Baltagi, B.H., 2005. \textit{Econometric Analysis of Panel Data}, 3$^\text{rd}$ edition. Chichester, UK: Wiley and Sons.
\bibitem[\protect\citeauthoryear{Baltagi}{2013}]{Baltagi(2013)} \text{Baltagi, B.~H.} (2013). \textit{Econometric Analysis of Panel Data}, 5$^\text{th}$ edition. Wiley and Sons, Chichester (UK).
\bibitem[\protect\citeauthoryear{Baltagi et al.}{2005}]{BaltagiBressonPirotte(2005)} \text{Baltagi, B.~H., G. Bresson, and A. Pirotte} (2005), ``Adaptive estimation of heteroskedastic error component models.'' \textit{Econometric Reviews}, \text{24} (1), 39-58. %Economet. Rev.
\bibitem[\protect\citeauthoryear{Baltagi et al.}{2006}]{BaltagiBressonPirotte(2006)} \text{Baltagi, B.~H., G. Bresson, and A. Pirotte} (2006), ``Joint LM test for homoskedasticity in a one-way error component model.'' \textit{Journal of Econometrics}, \text{134} (2), 401-417.
\bibitem[\protect\citeauthoryear{Baltagi and Griffin}{1988}]{BaltagiGriffin(1988)} \text{Baltagi, B.~H. and J.~M. Griffin} (1988), ``A generalized error component model with heteroscedastic disturbances.'' \textit{International Economic Review}, \text{29} (4), 745-753. % Int. Econ. Rev.
\bibitem[\protect\citeauthoryear{Bester and Hansen}{2016}]{BesterHansen(2016)} \text{Bester, C.~A. and C.~B. Hansen} (2016), ``Grouped effects estimators in fixed effects models.'' \textit{Journal of Econometrics}, \text{190} (1), 197-208.
\bibitem[\protect\citeauthoryear{Bi{\o}rn}{1981}]{Biorn(1981)} \text{Bi{\o}rn, E.} (1981), ``Estimating economic relations from incomplete cross-section/time-series data.'' \textit{Journal of Econometrics}, \text{16} (2), 221-236.
\bibitem[\protect\citeauthoryear{Bi{\o}rn}{2004}]{Biorn(2004)} \text{Bi{\o}rn, E.} (2004), ``Regression systems for unbalanced panel data: a stepwise maximum likelihood procedure.'' \textit{Journal of Econometrics}, \text{122} (2), 281-291.
\bibitem[\protect\citeauthoryear{Bresson et al.}{2006}]{BressonHsiaoPirotte(2006)} \text{Bresson, G., C. Hsiao, and A. Pirotte} (2006), ``Heteroskedasticity and random coefficient model on panel data.'' \text{Working Papers ERMES}, No. 0601, 51 p.
\bibitem[\protect\citeauthoryear{Bresson et al.}{2011}]{BressonHsiaoPirotte(2011)} \text{Bresson, G., C. Hsiao, and A. Pirotte} (2011), ``Assessing the contribution of R\&D to total factor productivity: a Bayesian approach to account for heterogeneity and heteroskedasticity.'' \textit{Advances in Statistical Analysis}, \text{95} (4), 435-452. %Adv. Statist. Anal.
\bibitem[\protect\citeauthoryear{Breusch and Pagan}{1979}]{BreuschPagan(1979)} \text{Breusch, T.~S. and A.~R. Pagan} (1979), ``A simple test for heteroscedasticity and random coefficient variation.'' \textit{Econometrica}, \text{47} (5), 1287-1294.
\bibitem[\protect\citeauthoryear{Chib}{2008}]{Chib(2008)} \text{Chib, S.} (2008), ``Panel data modeling and inference: a Bayesian primer.''  In  \textit{The Econometrics of Panel Data} (M\'{a}ty\'{a}s, L. and P. Sevestre, eds.), book series \textit{Advanced Studies in Theoretical and Applied Econometrics}, Vol. 46, Chapter 15, 479-515, 3$^\text{rd}$ edition. Springer-Verlag, Berlin (GE).
\bibitem[\protect\citeauthoryear{Davis}{2002}]{Davis(2002)} \text{Davis, P.} (2002), ``Estimating multi-way error components models with unbalanced data structures.'' \textit{Journal of Econometrics}, \text{106} (1), 67-95.
%\bibitem[\protect\citeauthoryear{Hausman}{1978}]{Hausman(1978)} \textsc{Hausman, J.~A.} (1978). Specification tests in econometrics. \textit{Econometrica} \textbf{46}(6), 1251-1271.
\bibitem[\protect\citeauthoryear{Hsiao and Pesaran}{2004}]{HsiaoPesaran(2004)} \text{Hsiao, C. and M.~H. Pesaran} (2004), ``Random coefficient panel data models.'' \text{IZA Discussion Paper Series}, No. 1236, 39 p.
\bibitem[\protect\citeauthoryear{Lejeune}{1996}]{Lejeune(1996)} \text{Lejeune, B.} (1996), ``A full heteroscedastic one-way error components model for incomplete panel: Maximum likelihood estimation and Lagrange multiplier testing.'' \text{CORE Discussion Paper, Universit\'{e} Catholique de Louvain}, No. 1996/006, 28 p.
\bibitem[\protect\citeauthoryear{Lejeune}{2004}]{Lejeune(2004)} \text{Lejeune, B.} (2004), ``A full heteroscedastic one-way error components model allowing for unbalanced panel: pseudo-maximum likelihood estimation and specification testing.'' \text{CORE Discussion Paper, Universit\'{e} Catholique de Louvain}, No. 2004/76, 37 p.
\bibitem[\protect\citeauthoryear{Li and Stengos}{1994}]{LiStengos(1994)} \text{Li, Q. and T. Stengos} (1994), ``Adaptive estimation in the panel data error component model with heteroskedasticity of unknown form.'' \textit{International Economic Review}, \text{35} (4), 981-1000.
\bibitem[\protect\citeauthoryear{Magnus}{1982}]{Magnus(1982)} \text{Magnus, J.~R.} (1982), ``Multivariate error components analysis of linear and non-linear regression models by maximum likelihood.'' \textit{Journal of Econometrics}, \text{19} (2-3), 239-285.
\bibitem[\protect\citeauthoryear{Mazodier and Trognon}{1978}]{MazodierTrognon(1978)} \text{Mazodier, P. and A. Trognon} (1978), ``Heteroscedasticity and stratification in error components models.'' \textit{Annales de l'INSEE}, \text{30-31}, 451-482.
\bibitem[\protect\citeauthoryear{Nerlove}{1971}]{Nerlove(1971)} \text{Nerlove, M.} (1971), ``Further evidence on the estimation of dynamic economic relations from a time series of cross sections.'' \textit{Econometrica}, \text{39} (2), 359-382.
\bibitem[\protect\citeauthoryear{Neyman and Scott}{1948}]{NeymanScott(1948)} \text{Neyman, J. and E.~L. Scott} (1948), ``Consistent estimates based on partially consistent observations.'' \textit{Econometrica}, \text{16} (1), 1-32.
\bibitem[\protect\citeauthoryear{Phillips}{2003}]{Phillips(2003)} \text{Phillips, R.~F.} (2003), ``Estimation of a stratified error-components model.'' \textit{International Economic Review}, \text{44} (2), 501-521.
\bibitem[\protect\citeauthoryear{Platoni et al.}{2012}]{PlatoniSckokaiMoro(2012)} \text{Platoni, S., P. Sckokai, and  D. Moro} (2012), ``A note on two-way ECM estimation of SUR systems on unbalanced panel data.'' \textit{Econometric Reviews}, \text{31} (2), 119-141.
\bibitem[\protect\citeauthoryear{Randolph}{1988}]{Randolph(1988)} \text{Randolph, W.~C.} (1988), ``A transformation for heteroscedastic error components regression models.'' \textit{Economics Letters}, \text{27} (4), 349-354.
\bibitem[\protect\citeauthoryear{Rao et al.}{1981}]{RaoKaplanCochran(1981)} \text{Rao, P.~S.~R.~S., J. Kaplan, and W.~C. Cochran} (1981), ``Estimators for the one-way random effects model with unequal error variances.'' \textit{Journal of the American Statistical Association}, \text{76} (373), 89-97. % J. Am. Statist. Assoc.
\bibitem[\protect\citeauthoryear{Roy}{2002}]{Roy(2002)} \text{Roy, N.} (2002), ``Is adaptive estimation useful for panel models with heteroskedasticity in the individual specific error component? Some Monte Carlo evidence.'' \textit{Econometric Reviews}, \text{21} (2), 189-203.
\bibitem[\protect\citeauthoryear{Verbon}{1980}]{Verbon(1980)} \text{Verbon, H.~A.~A.} (1980), ``Testing for heteroscedasticity in a model of seemingly unrelated regression equations with variance component.'' \textit{Economics Letters}, \text{5} (2), 149-153.
\bibitem[\protect\citeauthoryear{Wang and Ho}{2010}]{WangHo(2010)} \text{Wang, H.-J. and C.-W. Ho} (2010), ``Estimating fixed-effect panel stochastic frontier models by model transformation.'' \textit{Journal of Econometrics}, \text{157} (2), 286-296.
\bibitem[\protect\citeauthoryear{Wansbeek}{1989}]{Wansbeek(1989)} \text{Wansbeek, T.} (1989), ``An alternative heteroscedastic error components model (problem 88.2.2.).'' \textit{Econometric Theory}, \text{5} (2), 326.
\bibitem[\protect\citeauthoryear{Wansbeek and Kapteyn}{1989}]{WansbeekKapteyn(1989)} \text{Wansbeek, T. and A. Kapteyn} (1989), ``Estimation of the error-components model with incomplete panels.'' \textit{Journal of Econometrics}, \text{41} (3), 341-361.
\bibitem[\protect\citeauthoryear{Wooldridge}{2010}]{Wooldridge(2010)} \text{Wooldridge, J.~M.} (2010). \textit{Econometric Analysis of Cross Section and Panel Data}, 2$^\text{nd}$ edition. The MIT Press, Cambridge, Massachusset (US) - London, England (UK).
\end{thebibliography}
\end{document}